\documentclass{article}
\usepackage{graphicx}  

\usepackage{amsmath}   
\usepackage[compress]{cite}
\usepackage{amssymb}   
\usepackage{bm} 
\usepackage{dcolumn}
\usepackage{color}
\usepackage{mathrsfs}
\usepackage{amsfonts}
\usepackage{varioref}
\usepackage{color}
\usepackage{float}
\usepackage{subfig}

\usepackage{placeins}
\usepackage{caption}

\RequirePackage[colorlinks,citecolor=blue,urlcolor=magenta,linkcolor=blue]{hyperref}
\def\LL{Lanczos-Lovelock }

\def\gr{general relativity}
\def\RN{Reissner-Nordstr\"{o}m }
\def\RNV{Reissner-Nordstr\"{o}m-Vaidya }

\def\PS{photon sphere}
\def\EH{event horizon}
\def\DS{de Sitter }
\def\AH{apparent horizon}
\def\CS{critical surface}
\def\NEC{null energy condition}
\def\SD{Schwarzschild-de Sitter }

\labelformat{section}{Section #1} 
\labelformat{subsection}{Section #1} 
\labelformat{subsubsection}{Section #1}
\labelformat{subsubsubsection}{Section #1}
\labelformat{equation}{Eq.~(#1)} \labelformat{figure}{Fig.~#1} 
\labelformat{subfigure}{Fig.~\thefigure#1} 
\labelformat{table}{Tab.~#1} 
\labelformat{appendix}{Appendix #1}
\numberwithin{equation}{section}
\DeclareMathOperator{\sech}{sech}
\title{Understanding photon sphere and black hole shadow in dynamically evolving spacetimes}
\author{Akash K Mishra\footnote{akash.mishra@iitgn.ac.in}$~^{1}$, 
Sumanta Chakraborty\footnote{sumantac.physics@gmail.com}$~^{2}$ and 
Sudipta Sarkar\footnote{sudiptas@iitgn.ac.in}$~^{1}$\\
{\small{$^{1}$ Indian Institute of Technology, Gandhinagar-382355, Gujarat, India}}\\
{\small{$^{2}$ School of Physical Sciences, Indian Association for the Cultivation of Science, Kolkata-700032, India}}}
\begin{document}
  
\maketitle
\begin{abstract}

We have derived the differential equation governing the evolution of the photon sphere for dynamical black hole spacetimes with or without spherical symmetry. Numerical solution of the same depicting evolution of the photon sphere has been presented for Vaidya, Reissner-Nordstr\"{o}m-Vaidya and de-Sitter Vaidya spacetimes. It has been pointed out that evolution of the photon sphere depends crucially on the validity of the null energy condition by the in-falling matter and may present an observational window to even test it through black hole shadow. We have also presented the evolution of the photon sphere for slowly rotating Kerr-Vaidya spacetime and associated structure of black hole shadow. Finally, the effective graviton metric for Einstein-Gauss-Bonnet gravity has been presented, and the graviton sphere has been contrasted with the photon sphere in this context. 

\end{abstract}
\section{Introduction and Motivation}

Black holes are one of the fascinating and inevitable consequences of General relativity and have often provided profound insights into the fundamental nature of spacetime at both classical and quantum level. Ever since the pioneering work of Bekenstein and Hawking, black hole physics has received a tremendous amount of attention and theoretical success over the last few decades\cite{bardeen1973,hawking1972,hawking1975,hawking_ellis_1973,Bekenstein:1972tm,Bekenstein:1973ur,Wald:1995yp}. However, the observational evidence for black holes remain elusive till the recent detection of gravitational waves, which is best described by the merger of two black holes \cite{Abbott:2016blz,TheLIGOScientific:2016wfe,Abbott:2016nmj,Abbott:2017vtc}. Besides, in recent years strong evidence from a wide range of astrophysical data have come up for the existence of super-massive black holes at the center of most of the galaxies \cite{Ghez:2008ms,Schodel:2002vg,Kormendy:2013dxa}, which also hints toward the presence of black holes. 
However all these tests including the detection of gravitational waves provided indirect evidence for black holes, while a direct detection of a black hole would correspond to the observation of black hole shadow, a direct probe of the photon sphere around the black hole \cite{Vazquez:2003zm,Shaikh:2018lcc,Hou:2018bar,Cunha:2018gql,Tsukamoto:2017fxq,Repin:2018anv,Abdujabbarov:2016hnw,Kumar:2018ple,Cunha:2018cof,Ayzenberg:2018jip,Rahman:2018fgy,Atamurotov:2013sca,Grenzebach:2015oea,Atamurotov:2015xfa,Mars:2017jkk}. Loosely speaking the black hole shadow is due to strong gravitational lensing effect\cite{Einstein:1956zz,Bartelmann:2010fz,Cunha:2018acu,Kochanek2006,PhysRev.133.B835,PhysRevLett.105.251101,PhysRevD.71.064004,PhysRevD.69.022002,Perlick2004,Bozza:2009yw,PhysRevD.59.124001,Chakraborty:2016lxo}, near the photon sphere, which is defined as the set of directions in the observer's sky, from which no signal from distant source reaches the observer. 
This effect can in principle be observed from Earth, providing a definitive test of existence of black holes and for this very purpose, the Event Horizon telescope is being designed to observe the shadow like structure around the supermassive object at the center of our galaxy\cite{Doeleman:2008qh,Dokuchaev:2018kzk,Doeleman:2012zc,Guo:2018kis}. Various other interesting aspects of photon sphere and shadow has been extensively studied by numerous authors\cite{Claudel:2000yi, Hod:2012ax, Khoo:2016xqv, Baldiotti:2014pca, Decanini:2010fz,Shoom:2017ril,Cederbaum:2015fra,Johannsen:2015qca,Teo2003,Gallo:2015bda,Bhattacharya:2016naa,Chakraborty:2012sd,Rahman:2018oso,Mukherjee:2018dmm,Abdujabbarov:2015xqa,Younsi:2016azx,Goddi:2017pfy}. However, black holes are in general not stationary since they continuously accrete matter and grow in size. Therefore it is very much desirable to understand how the photon sphere evolves when one goes beyond the stationary consideration\cite{Claudel:2000yi}. This corresponds to a nontrivial generalization of the notion of the photon sphere, which rather than being determined by an algebraic equation turns out to be governed by a second order differential equation. In this paper, we will study the evolution of the photon sphere for rotating and non-rotating black holes by solving the associated differential equation for various dynamical black hole models, leading to several non-trivial results.

Further insights can be gained when one considers theories beyond \gr. Although \gr\ seems to be tremendously successful in macroscopic length scale, it is reasonable to believe that, it is only an effective theory of a more general theory. Such a theory is expected to contain higher curvature corrections to the Einstein-Hilbert action, and the Lovelock theory represents one such unique generalization with the field equations containing at most second derivative of the metric\cite{Lovelock:1971yv,Zwiebach:1985uq,Boulware:1985wk,Zumino:1985dp,Padmanabhan:2013xyr}. The causal structure of Lovelock theories are different from that of \gr\cite{Aragone:1987jm,ChoquetBruhat:1988dw,Izumi:2014loa,Reall:2014pwa,Papallo:2015rna,Reall:2014sla,Brustein:2017iet,Andrade:2016yzc}. This is because the characteristics hypersurfaces determining the causal structure of a system are null surfaces for the case of Einstein's Equation, however in Lovelock theories, the background metric receives correction, and the characteristics hypersurfaces turn out to be null with respect to a different effective metric. Thus one can safely say that gravity propagates at the speed of light in \gr\, but in Lovelock theories, gravity propagates at speed different than light. Therefore one can ask, how the graviton circular null geodesics moving in the effective metric for Lovelock theories are different from that of the photon. This difference can be used as another probe of the presence of higher curvature terms over and above \gr. In this work, we consider the Einstein-Gauss-Bonnet gravity, which incorporates the first order correction to the gravity Lagrangian over and above \gr\ and present the evolution of photon and graviton sphere in a dynamical black hole spacetime by explicitly calculating the effective graviton metric. 

The paper is organized as follows: In \ref{Section_2} we provide a derivation of the evolution equation for the radius of photon sphere for a dynamical spherically symmetric black hole. In \ref{Section_3}, as an illustration of the analytical method, we solve this equation numerically in three different settings --- (a) Schwarzschild Vaidya black hole, (b) \RNV black hole and finally (c) Schwarzschild de-Sitter Vaidya black hole for various suitable choices of the mass and charge functions. Using appropriate mass and charge profiles, we find a novel relationship between the evolution of the photon sphere and \NEC\ of the inflowing matter to the black hole. Our result shows that not only the event horizon but the photon sphere and hence shadow are sensitive to the null energy condition. This is a significant result because of several reasons --- (a) unlike the event horizon, there is no reason for the evolution of the photon sphere to be somehow related to \NEC\ and (b) the black hole shadow is observable by a distant observer and hence provides observational evidence for the violation of null energy condition. Subsequently, in \ref{Section_4} we present the evolution for the shadow of a dynamical spherically symmetric black hole and present the evolution by plotting the shadow at various instance of time. In \ref{Section_7} we extend our analysis beyond spherical symmetry, i.e., for a rotating black hole, namely the Kerr-Vaidya black hole in slow rotation limit. The absence of spherical symmetry in the solution makes the problem considerably challenging, but in the slow rotation limit, we have derived the differential equation governing the photon sphere. Evolution of black hole shadow has also been studied. Finally, in \ref{Section_6} we generalize our analysis to Einstein-Gauss-Bonnet theory and provide a derivation of the effective graviton metric in the dynamical context and have studied the evolution of the graviton sphere and corresponding shadow, which has been contrasted with the photon sphere. 

\emph{Notations and Conventions:} We have set the fundamental constants $c$ and $\hbar$ to unity, and we will work with mostly positive signature convention. As per our notation, a `prime' will denote derivative with respect to the radial coordinate `$r$' while `dot' over a quantity implies derivative with respect to `$v$'. All the derivatives with respect to the affine parameter `$\lambda$' along the null geodesic will be displayed explicitly.
\section{Photon Sphere in a Spherically Symmetric Dynamical Spacetime}\label{Section_2}

In this section, we will explicitly derive a second-order differential equation governing the dynamical evolution of the photon sphere in a general static and spherically symmetric spacetime. However before going into the gory detail of the derivation it is instructive to recall the derivation of the photon sphere for static spacetimes as a warm up exercise. Even though the location of the photon sphere can be determined in numerous possible ways (see, e.g., \cite{Shaikh:2018lcc,Claudel:2000yi}), in what follows we will adopt a procedure which can be straightforwardly generalized to the dynamical context. In the context of static spacetime, we write down the metric using in-going null coordinate $v$, leading to
\begin{equation}
ds^2 = -f(r)dv^2 + 2 dv dr +r^2d\Omega_2^2\label{sch}
\end{equation}
The spherical symmetry allows us to choose a particular plane in the spacetime, which for convenience is chosen to be the equatorial plane with $\theta=\pi/2$. In this spacetime, there exists one null geodesic on the equatorial plane, which is circular. This is essentially the photon sphere, projected on the equatorial plane, yielding a circle, known as photon circular orbit. Since the trajectory of null geodesics (equivalently, photons) are circular in nature, we can set $r=\textrm{constant}\equiv r_{\rm ph}$. Being null geodesic, additionally we have $ds^2=0$, which from \ref{sch} gives rise to,
\begin{equation}\label{1}
\left(\frac{d\phi}{dv}\right)^2 = \frac{1}{r_{\rm ph}^{2}} f(r_{\rm ph}).
\end{equation}
Similarly, starting from the metric in \ref{sch}, one can write down the radial geodesic equation for null trajectories, which reads,
\begin{equation}\label{radial_geo_eqn}
\frac{d^2r}{d\lambda^2}-\frac{\partial f}{\partial r}\left(\frac{dr}{d\lambda}\right)\left(\frac{dv}{d\lambda}\right) + \frac{1}{2} f \frac{\partial f}{\partial r} \left(\frac{dv}{d\lambda}\right)^2 -rf\left(\frac{d\phi}{d\lambda}\right)^2=0
\end{equation}
However our interest is mainly in understanding the circular null geodesic and hence we may use the fact that $r=r_{\rm ph}=\textrm{constant}$, leading to both $\dot{r}=0$ and $\ddot{r}=0$. Thus with these results taken into account, \ref{radial_geo_eqn} for circular null geodesics become,
\begin{equation}\label{2}
\left(\frac{d\phi}{dv}\right)^2 = \frac{1}{2r_{\rm ph}} \frac{\partial f(r)}{\partial r}\Big|_{r_{\rm ph}}
\end{equation}
Thus one can immediately equate \ref{1} and \ref{2}, leading to an algebraic equation for the radial coordiante $r_{\rm ph}$, which reads, 
\begin{equation}\label{st ph sp}
r_{\rm ph}\frac{\partial f(r)}{\partial r}\Big|_{r_{\rm ph}} = 2 f(r_{\rm ph})~.
\end{equation}
As evident, \ref{st ph sp} represents well-known equation for the radius of photon sphere in a static and spherically symmetric spacetime. As a cross verification of this result, one may resort to Schwarzschild spacetime and hence substituting $f(r)=1-(2M/r)$ one easily obtains $r_{\rm ph}=3M$. This sets the stage for our subsequent discussion regarding photon sphere for dynamical black holes. 

A physical scenario where a dynamical black hole may exist correspond to a situation when the black hole is fed by accretion disk surrounding it or the black hole is radiating matter, possibly evaporating black hole. The metric ansatz associated with such a dynamical black hole is an obvious generalization of \ref{sch}, which reads, 
\begin{equation}\label{Dynamical_in}
ds^2 = -f(r,v)dv^2 + 2 dv dr +r^2d\Omega^2~.
\end{equation}
Even though the structure of the metric is very much similar to the one presented in \ref{sch}, the photon orbits will be completely different. This is because the spacetime is no longer static and as a consequence the radius of the photon sphere cannot taken to be constant and it must change with time (or the ingoing null coordinate $v$). So we can model the radius of the photon sphere to be a function of in-going time $(v)$ for accreeting matter and out-going time $(u)$ for radiating matter. Note that by virtue of spherical symmetry the photon sphere can not depend on other coordinates. Let us start with the in-going case first, which can be trivially generalized to the radiating case. As described earlier, here we have $r_{\rm ph} = r_{\rm ph}(v)$. In the dynamical case as well it is possible to follow an identical route as that of the static case, e.g., one first writes down the equation for $ds^{2}=0$ and couples it with radial null geodesic equation. This results into the desired evolution equation for the radius of the photon circular orbit in the equatorial plane, which reads (for a derivation see \ref{Appendix_A}),

\begin{align}\label{Photon_Sp_Eqn_in}
\ddot{r}_{\rm ph}(v)&+\dot{r}_{\rm ph}(v)\left[\frac{3}{r_{\rm ph}(v)}f(r_{\rm ph}(v),v) 
-\frac{3}{2}\frac{\partial f}{\partial r}\Big|_{r_{\rm ph}(v),v} \right] 
-\frac{2}{r_{\rm ph}(v)}\left\{\dot{r}_{\rm ph}(v)\right\}^2 
\nonumber
\\
&+\frac{1}{2}\left(f(r_{\rm ph}(v),v)\frac{\partial f}{\partial r}\Big|_{r_{\rm ph}(v),v} 
-\frac{\partial f}{\partial v}\Big|_{r_{\rm ph}(v),v}\right)
-\frac{1}{r_{\rm ph}(v)}f(r_{\rm ph}(v),v)^2 =0
\end{align}

On the other hand for a radiating black hole spacetime, the metric is best described in terms of the out-going null coordinate $u$, in terms of which the spacetime metric takes the following form,
\begin{equation}\label{Dynamical_out}
ds^2 = -f(r,u)du^2 - 2 du dr +r^2d\Omega^2~.
\end{equation}
In this context as well the photon circular orbit is not located at a fixed radial distance, rather it varies with the out-going null coordinate $u$. Thus in this context, $r=r_{\rm ph}(u)$. Following the path laid down in the context of accreting black hole it is straightforward to determine the differential equation governing $r_{\rm ph}(u)$ for radiating black hole as well. The corresponding equation for the evolution of the photon sphere becomes,
\begin{align}\label{Photon Sp Eqn_out}
\ddot{r}_{\rm ph}(u)&-\dot{r}_{\rm ph}(u)\left[\frac{3}{r_{\rm ph}(u)} f(r_{\rm ph}(u),u) 
-\frac{3}{2}\frac{\partial f}{\partial r}\Big|_{r_{\rm ph}(u),u}\right] 
-\frac{2}{r_{\rm ph}(u)}\left\{\dot{r}_{\rm ph}(u)\right\}^2 
\nonumber
\\
&+\frac{1}{2} \left\{f(r_{\rm ph}(u),u)\frac{\partial f}{\partial r}\Big|_{r_{\rm ph}(u),u} 
+\frac{\partial f}{\partial u}\Big|_{r_{\rm ph}(u),u}\right\}
-\frac{1}{r_{\rm ph}(u)} f(r_{\rm ph}(u),u)^2=0~.
\end{align}
Thus \ref{Photon_Sp_Eqn_in} and \ref{Photon Sp Eqn_out} represents the general equation governing the evolution of the radius of photon sphere around a dynamically evolving black hole, either accreting or radiating. An entirely different approach has been taken in Ref.\cite{Claudel:2000yi} to arrive at the same second order differential equation.
As evident, unless we specify the form of the metric function $f(r,v)$ it is not possible to solve the above differential equation and hence determine the location of the photon circular orbit $r_{\rm ph}(u)$. Our aim, in the subsequent sections, will be to study the behavior of $r_{\rm ph}(v)$ (or, $r_{\rm ph}(u)$)  evolution for various choice of $f(r,v)$. 

As an aside, let us point out two more radii of significant interest in the dynamical black hole spacetime under consideration. The first one correspond to the apparent horizon, whose location can be determined by solving the equation $f(r,v)=0$, leading to $r_{\rm ah}=r_{\rm ah}(v)$. While the event horizon being a null surface, satisfies a differential equation, namely $(dr/dv)=(1/2)f(r,v)$\cite{poisson_2004,Nielsen:2010gm}. Thus given a particular spacetime, with a certain $f(r,v)$, one can immediately determine the location of the event and apparent horizon, besides the photon circular orbit. We will explore these results as well in the next sections. 
\section{Application: Photon Sphere in Vaidya and Reissner-Nordstr\"{o}m-Vaidya Spacetimes}
\label{Section_3}

In the previous section, we have elaborated on the location of the circular photon orbit as well as event and apparent horizon in a dynamical black hole spacetime. In this section, we apply the formalism developed above in the context of two well known dynamical black hole spacetimes, namely the Vaidya spacetime and Reissner-Nordstr\"{o}m-Vaidya spacetime. In the case of Vaidya spacetime, the black hole mass changes with time, while for Reissner-Nordstr\"{o}m-Vaidya both mass and charge of the black hole changes with time. We will first discuss the case of Vaidya spacetime and hence determine the circular photon orbit along with event and apparent horizon in it, before taking up the Reissner-Nordstr\"{o}m-Vaidya spacetime. Finally, we will comment on the possible modifications pertaining to the presence of positive cosmological constant. 
\subsection{Photon Sphere in Vaidya Space time}\label{Photon_Vaidya}

As a first illustration of the method developed above, let us consider the case of Vaidya spacetime, which is basically a black hole spacetime accreting null fluid. This, in turn, demands the black hole mass to be changing with time. Since we are primarily interested in an accreting black hole, we can write down the Vaidya spacetime in the in-going null coordinate as\cite{PhysRev.83.10},
\begin{equation}\label{Vaidya_in}
ds^2 = -\left(1-\frac{2M(v)}{r}\right) dv^2 +2 dv dr +r^2 d\Omega^2~.
\end{equation}
If the above metric is supposed to be a solution of Einstein gravity, one can determine the associated energy momentum tensor by computing the Einstein tensor. Since Einstein tensor depends on derivatives of the metric, it immediately follows that the energy-momentum tensor associated with the Vaidya spacetime corresponding to \ref{Vaidya_in} takes the form,
\begin{equation}
T_{ab} = \frac{1}{4\pi r^2} \frac{dM(v)}{dv} \delta_{va} \delta_{vb}~.
\end{equation}
Interestingly, if we demand the above energy-momentum tensor to satisfy the null energy condition, it follows that $dM(v)/dv\geq 0$ constraint must hold. This is expected, as the flow of matter satisfying energy condition is supposed to increase the black hole mass. Thus one can immediately write down the differential equation governing the evolution of the photon sphere in this spacetime following \ref{Photon_Sp_Eqn_in}. As anticipated, this equation has no analytical solution possible and must resort to numerical techniques. However, solving the second order differential equation requires two boundary data, and we may use future boundary conditions for the accreting scenario. In particular, throughout this work we will assume that at late times the black hole settles down to a stationary configuration and we may use the following boundary conditions: (a) $r_{\rm ph}(v_0)= 3M(v_0)$ and (b) $\dot{r}_{\rm ph}(v_0)=0$, where $v_0$ is some future time where the mass function approaches a constant value, i.e., $\dot{M}(v_0)=0$. Moreover, the location of the apparent horizon in this spacetime is straightforward to work out and corresponds to $r_{\rm ap}=2M(v)$, while event horizon can be determined by solving $(dr/dv)=(1/2)\{1-(2M(v)/r)\}$ with appropriate future boundary conditions\cite{Nielsen:2010gm}. 

\begin{figure}[h]
\centering
    \subfloat{{\includegraphics[scale=0.375]{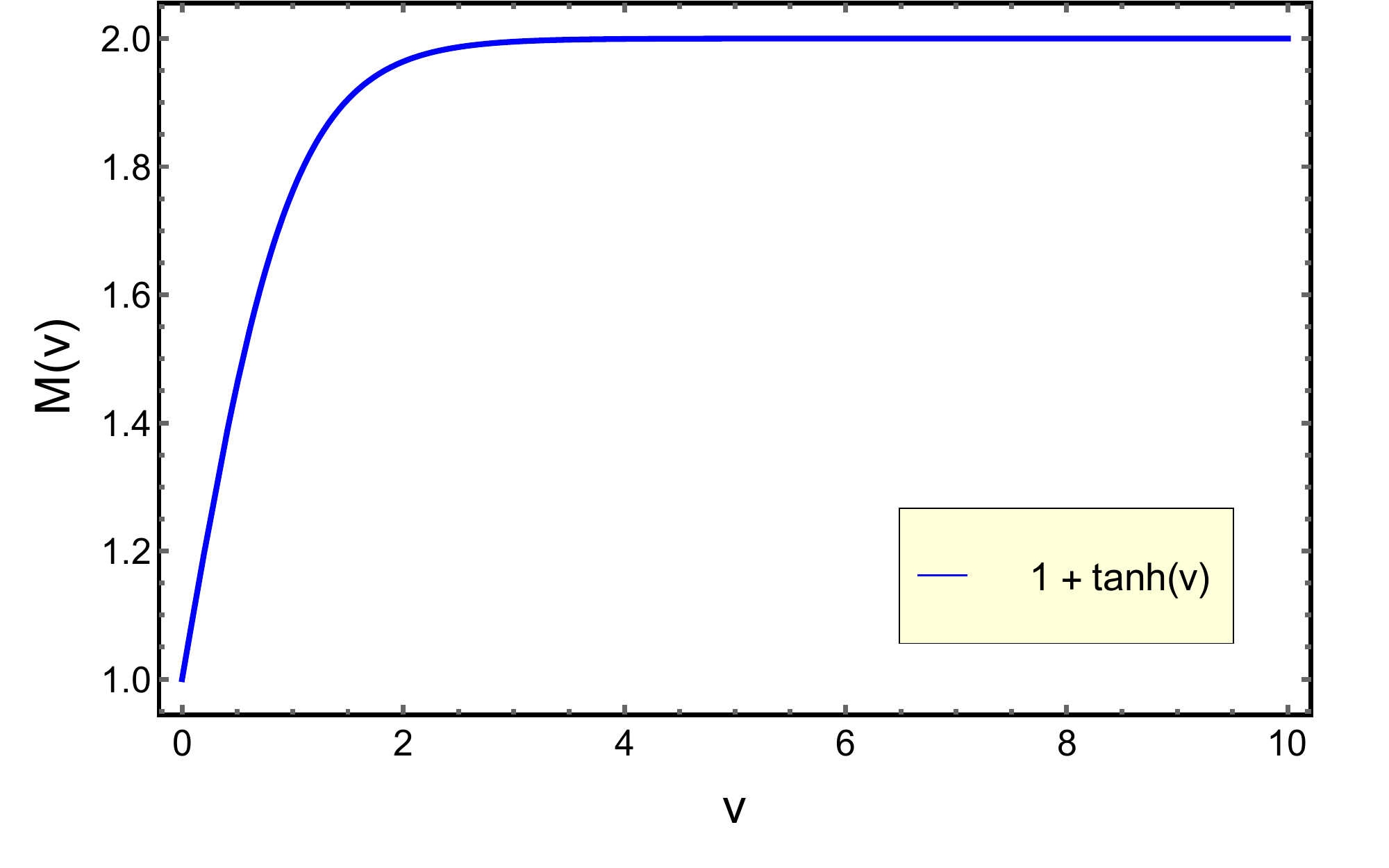} }}
    \qquad
   \subfloat{{\includegraphics[scale=0.375]{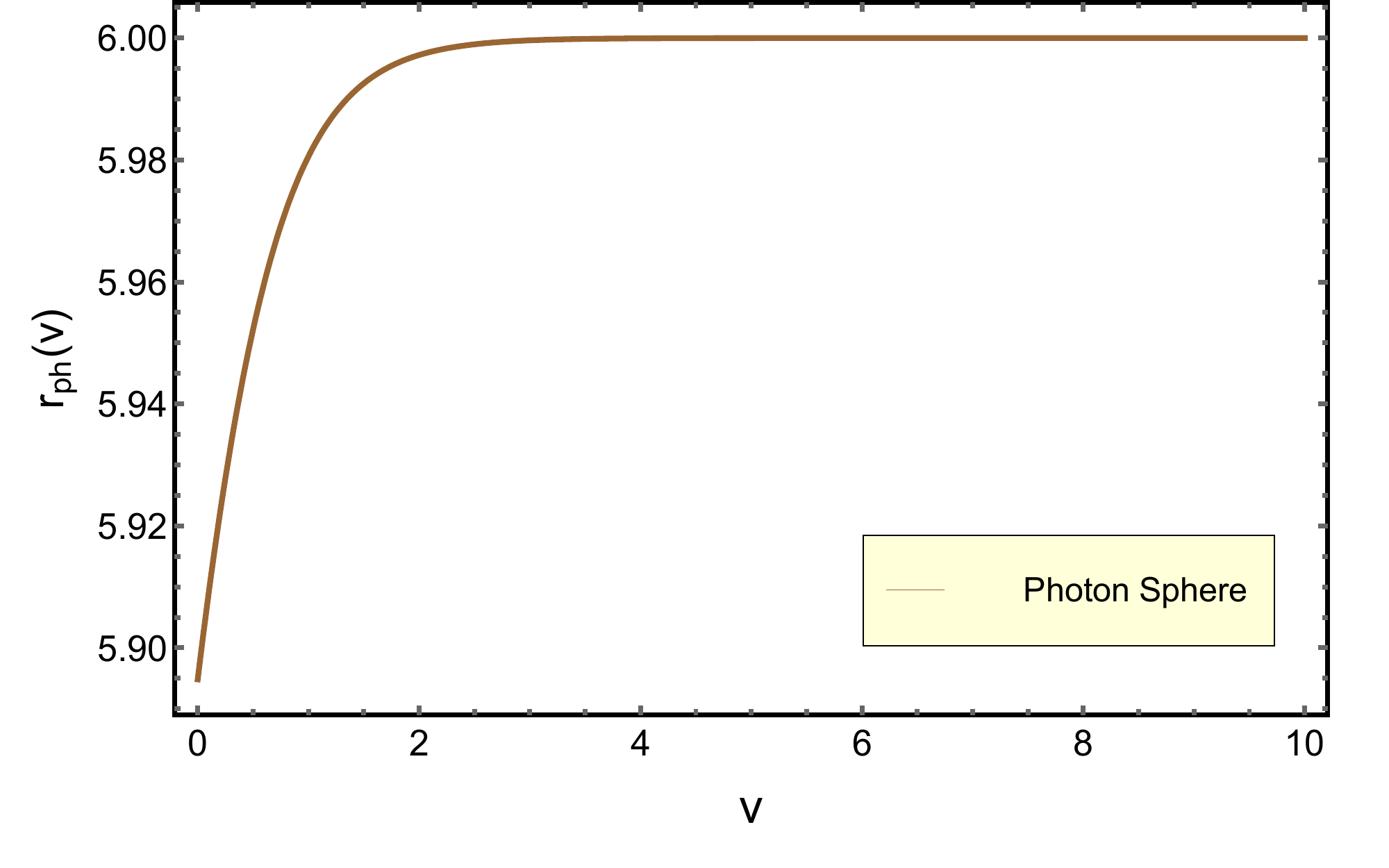} }}
    \quad
    \subfloat{{\includegraphics[scale=0.375]{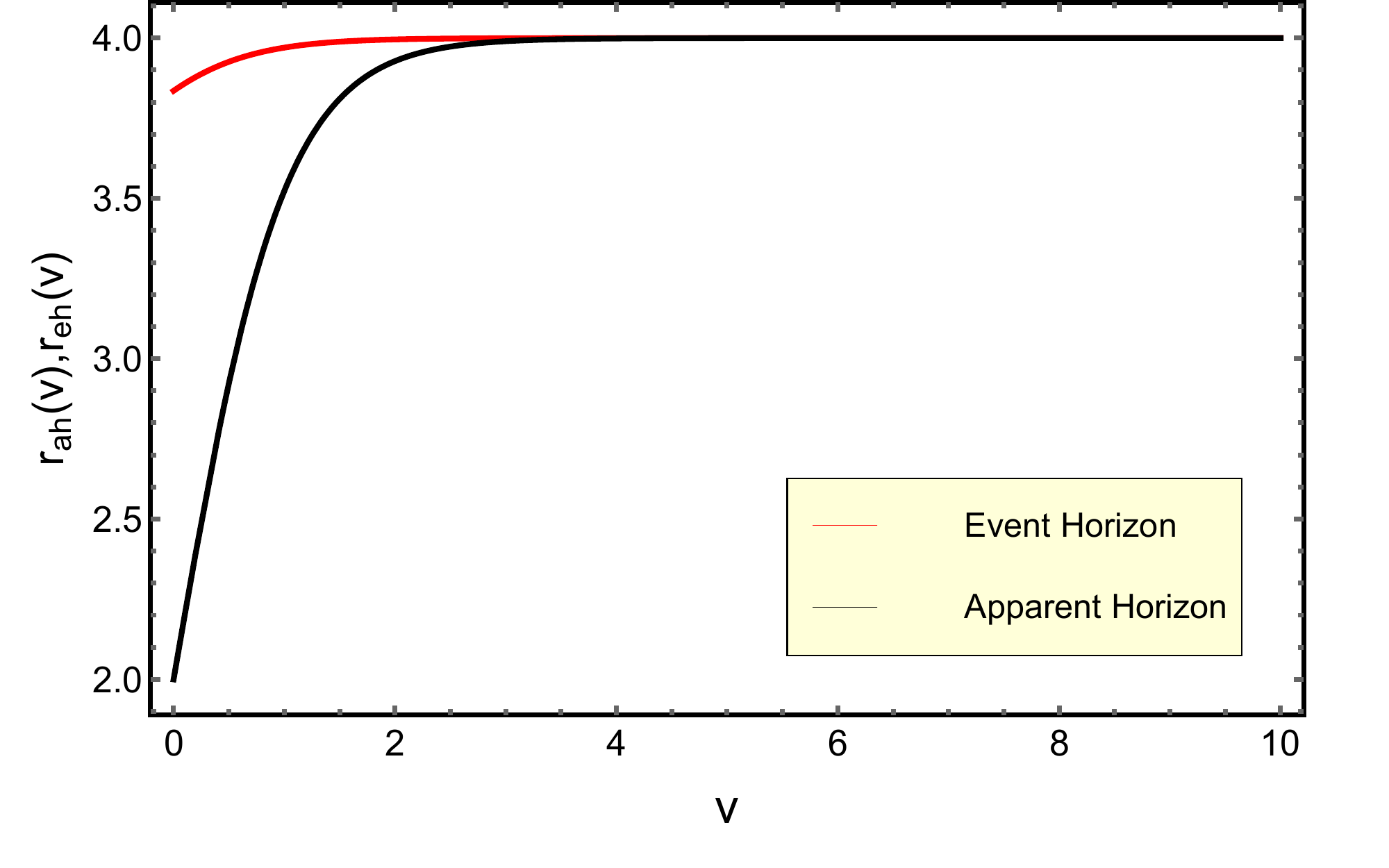} }}
 \caption{The evolution of the radius of the photon sphere, event, and apparent horizon has been presented for the mass function written down in \ref{inc_mass}. The figure on the top left panel shows the variation of this mass function with the advanced null coordinate $v$, while the top right panel shows the evolution of the radius of photon sphere projected on the equatorial plane. On the other hand, the evolution of both the event and apparent horizon has been presented in the bottom panel. In all these cases the respective radii asymptote to the static values ($M_0$ has been set to unity).}\label{tahh_mass_fn}
\end{figure}

\begin{figure}[h]
    \centering
    \subfloat{{\includegraphics[scale=0.375]{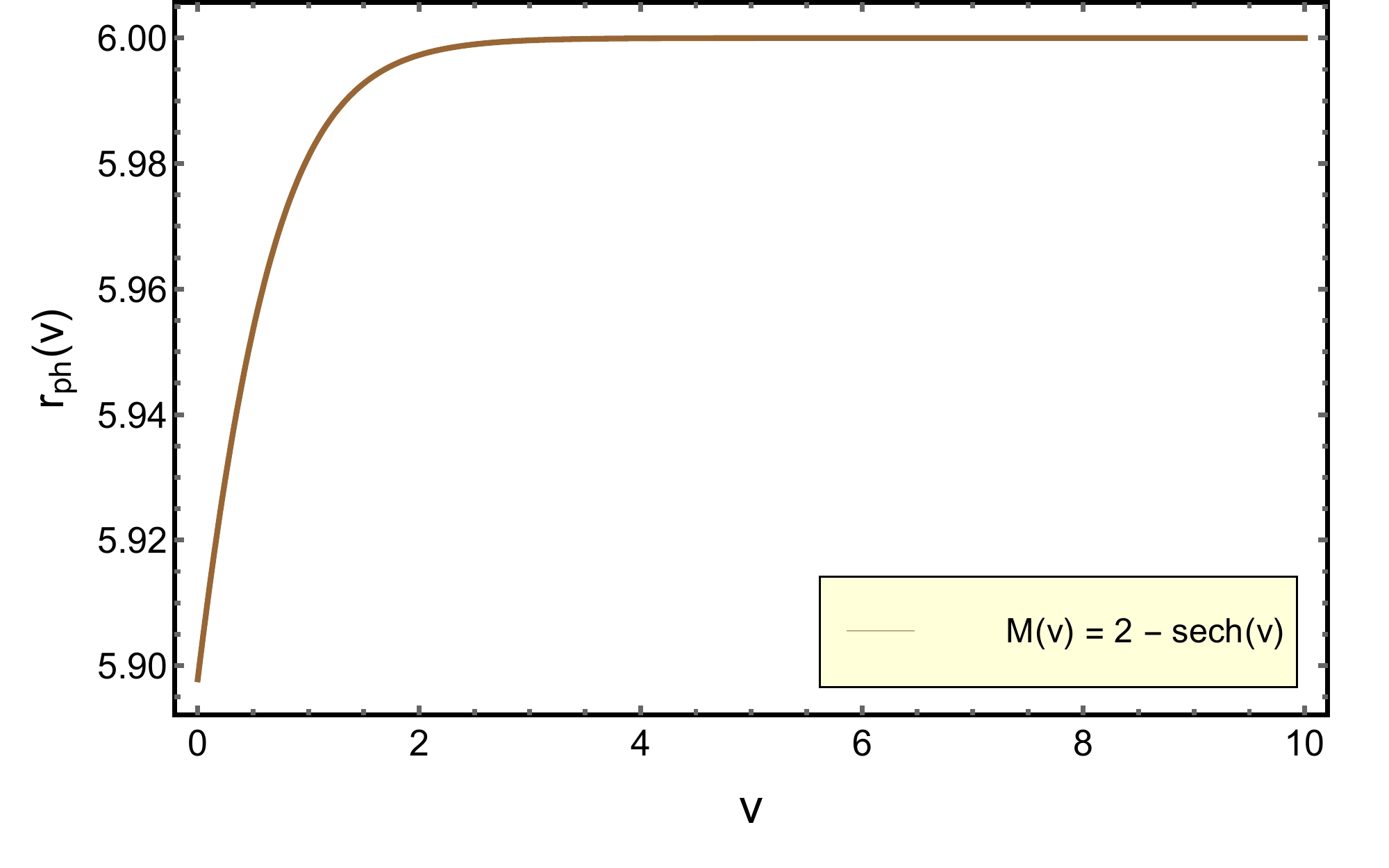} }}
    \qquad
    \subfloat{{\includegraphics[scale=0.375]{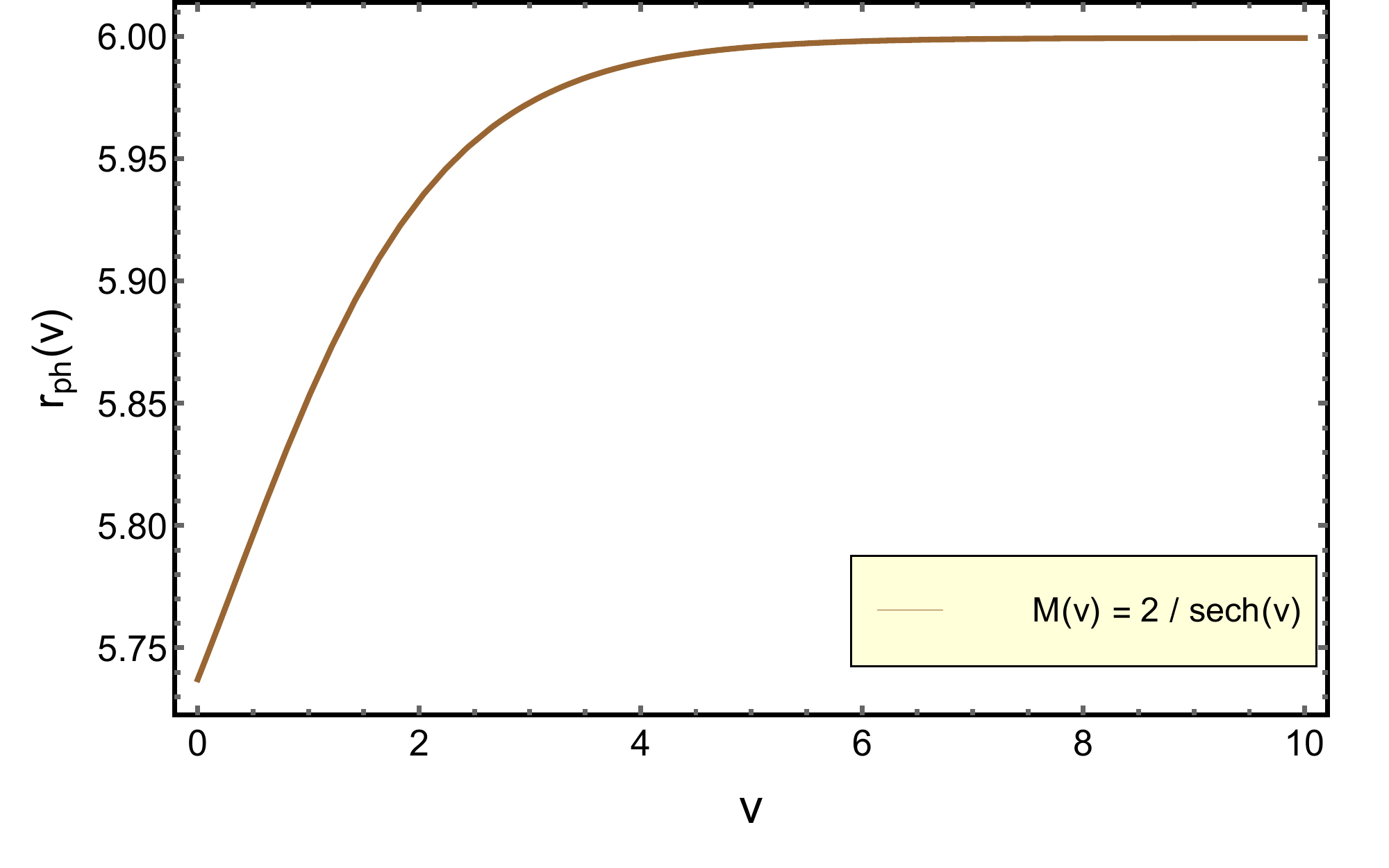} }}%
    \quad
	\subfloat{{\includegraphics[scale=0.375]{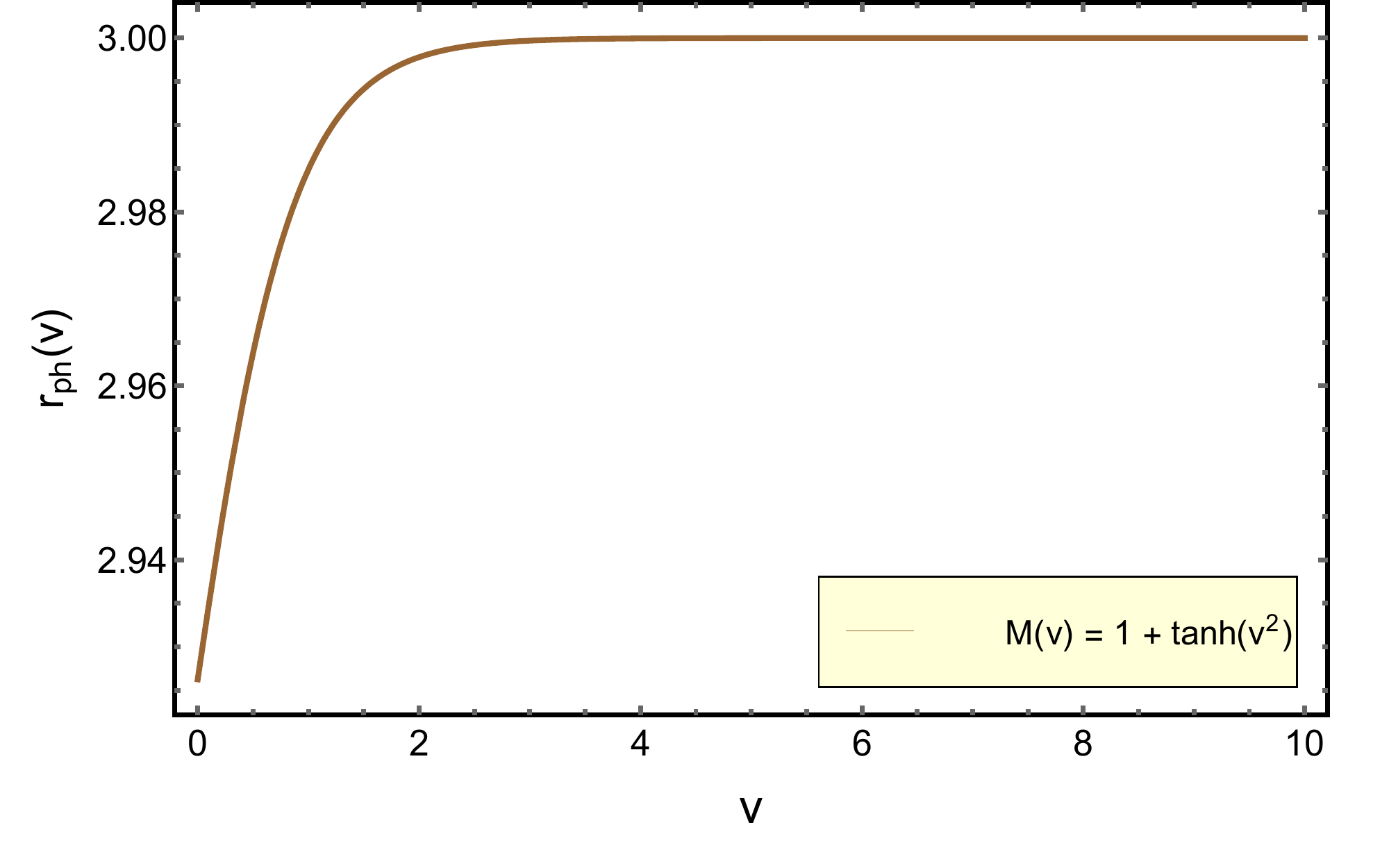} }}
    \qquad
    \subfloat{{\includegraphics[scale=0.375]{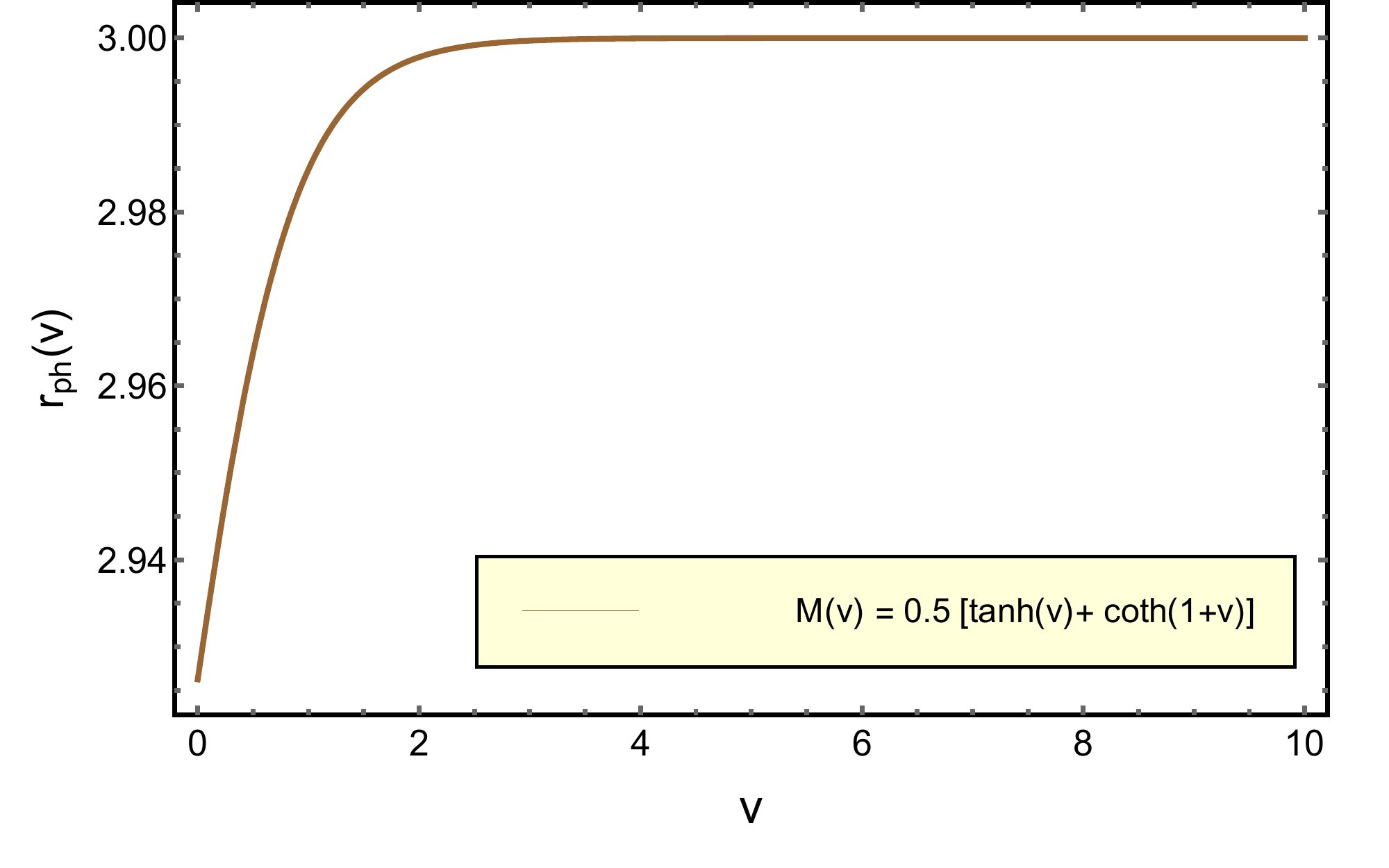} }}%
    \caption{This figure also presents the variation of the radius of the photon sphere projected on the equatorial plane. The top left panel shows the evolution of the photon sphere for mass function $M(v) = M_0\{2-\sech(v)\}$ with the choice of $M_0 =1$. The rest of the plots starting from top right and then to bottom left and bottom right panel shows the evolution of the photon sphere for mass functions $M(v)=2/\{1+\sech(v)\}$, $M(v)=1+\tanh(v^2)$ and $M(v)= 0.5\{\tanh(v)+\coth(1+v)\}$ respectively.}\label{more mass fn}
\end{figure}

\begin{figure}[h]
    \centering
    \subfloat{{\includegraphics[scale=0.375]{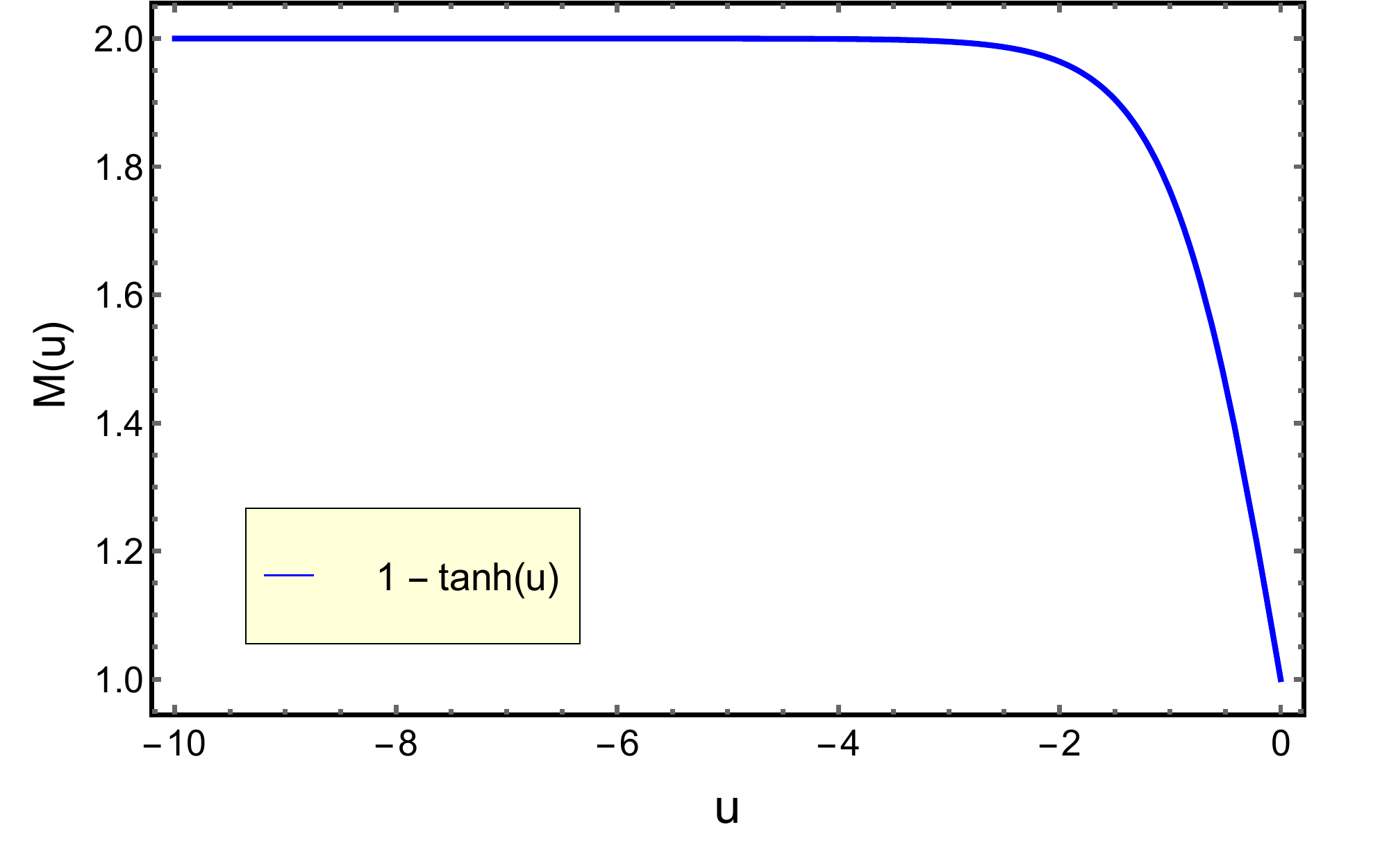} }}
    \qquad
    \subfloat{{\includegraphics[scale=0.375]{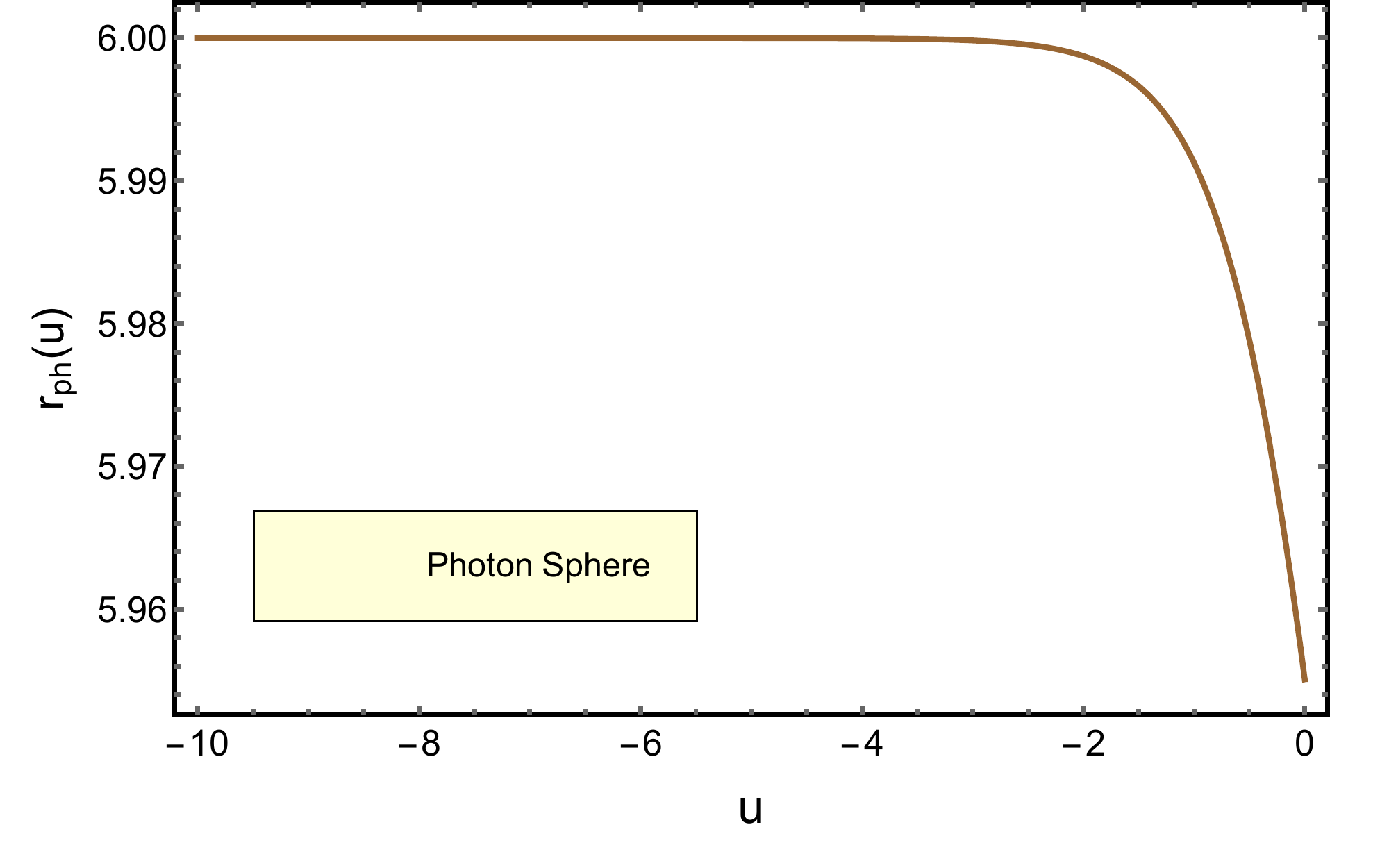} }}%
    \quad
	\subfloat{{\includegraphics[scale=0.375]{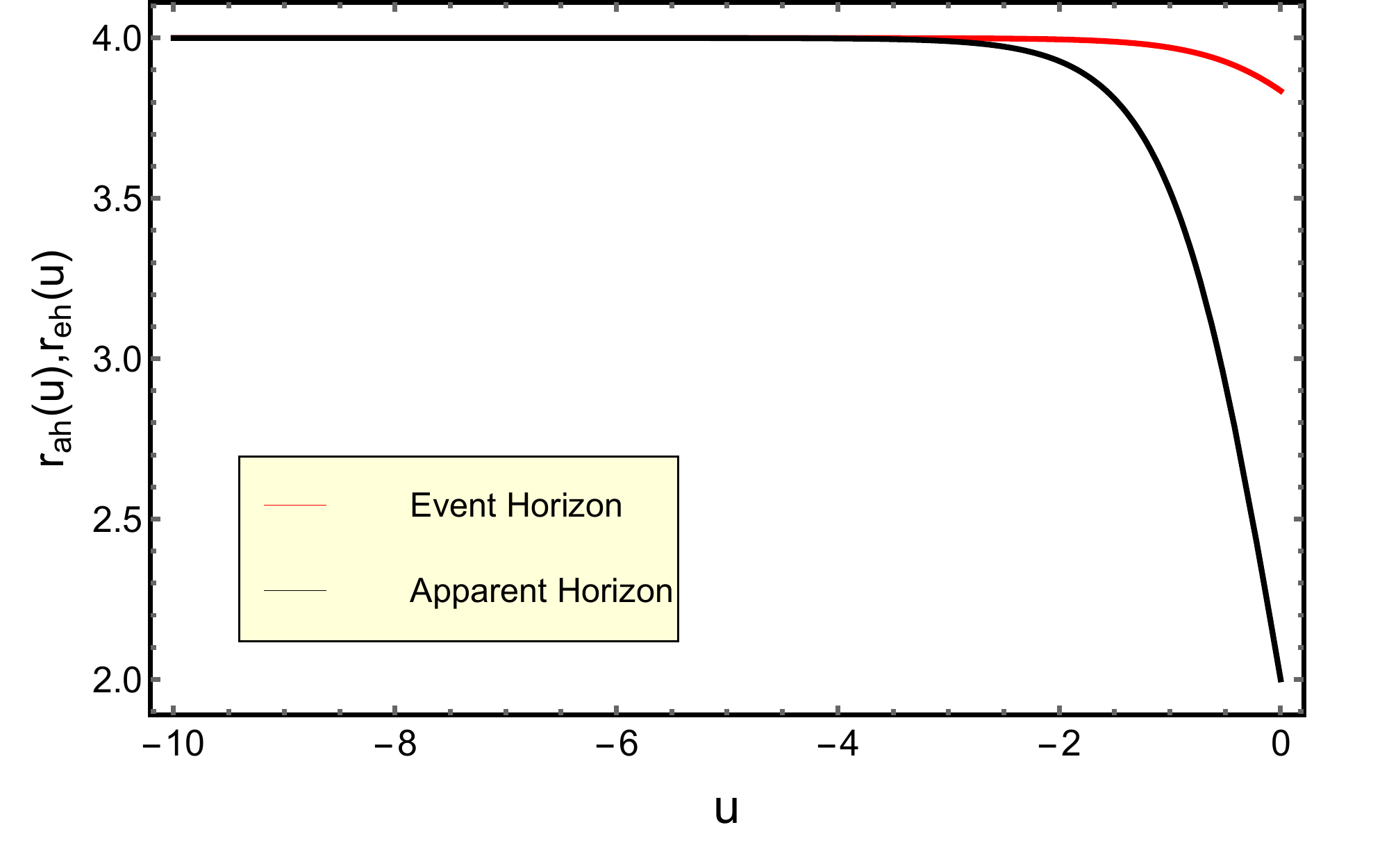} }}
	\caption{This figure demonstrates the evolution of the photon sphere with retarded null coordinate $u$ for radiating black holes, whose masses are decreasing with time. The top left panel shows the variation of the mass function with $u$, while that on the top right panel shows the evolution of the radius of the photon sphere. On the other hand, on the bottom left panel, we have plotted the evolution of the event and the apparent horizon.  As expected, they all started from a constant value and gradually decreased as the black hole mass decreases. Here also $M_0$ has been set to be unity.}\label{dec_mass_PS_vaidya}
\end{figure}

The above presents the theoretical framework necessary to discuss the evolution of the photon sphere in Vaidya spacetime. To illustrate the evolution in an explicit manner, we focus our attention particularly to smoothly varying mass function. For that matter, we start with the following choice of the mass function,
\begin{equation}\label{inc_mass}
M(v) = \frac{M_0}{2}\left\{1+\tanh(v)\right\}~,
\end{equation}
which has the nice property that, it approaches to a constant value $M_0$, in the asymptotic future (i.e., $v\rightarrow \infty$) and allows one to impose future boundary conditions, i.e., $r_{\rm ph}(v\rightarrow \infty)=3\,M_0$ and $\dot{r}_{\rm ph}(v\rightarrow \infty)=0$, to obtain the evolution of the photon sphere. Identically one can also study the behavior of the apparent horizon and the event horizon in the Vaidya spacetime. The result of such an analysis due to the mass function, written down in \ref{inc_mass}, is depicted in \ref{tahh_mass_fn}. As evident the photon circular orbit along with event and apparent horizon asymptote to constant values. In particular, the apparent horizon, in the dynamical context, lies within the event horizon and ultimately coincides with the event horizon.
\FloatBarrier
To further grasp the theoretical result derived earlier, we have considered a few other examples of smoothly increasing mass functions, e.g., $M(v) = (M_0/2)\{2 - \sech(v)\}$, which asymptotically approaches $M_0$. In a similar manner, by imposing the future boundary conditions $r_{\rm ph}(v\rightarrow \infty)=3\,M_0$ and $\dot{r}_{\rm ph}(v\rightarrow \infty)=0$ in \ref{Photon_Sp_Eqn_in}, one can obtain the corresponding evolution of photon sphere. We illustrate the result for the mass function presented above along with some other mass functions in \ref{more mass fn}. As expected, in all of them the photon sphere ultimately settles down to a radius $3M_{0}$. 
\FloatBarrier
As promised, at this stage let us consider the case of a black hole that is radiating matter, which can be modeled by a smoothly decreasing mass function. We would like to emphasize that, although such a process doesn't occur classically, it does occur quantum mechanically and nothing prevents us from modeling the black hole by a radiating mass function without worrying about the underlying phenomenon. The evolution of the photon sphere in case of a radiating black hole spacetime can be obtained by solving \ref{Photon Sp Eqn_out} with the past boundary condition, i.e., one assumes the black hole to be static to start with. One such radiating mass function takes the following form
\begin{equation}
M(u) = \left(\frac{M_0}{2}\right)\left[1-\tanh(u)\right]\label{dec mass}
\end{equation}
As evident, the black hole starts with a constant value of mass, $M_0$ in the far past (denoted by $u\rightarrow -\infty$) and allows one to impose past boundary conditions $r_{\rm ph}(u\rightarrow -\infty)=3\,M_0$ and $\dot{r}_{\rm ph}(u\rightarrow -\infty)=0$, to obtain the evolution of the photon sphere. We illustrate this result along with the evolution of the event horizon and apparent horizon in \ref{dec_mass_PS_vaidya}. As evident, all the radii start from their static values and decrease dynamically as the black hole evaporates by radiating matter. This is expected, as the size of the photon sphere must decrease as the mass function decreases. In the next section, we will extend this result to a \RNV spacetime, involving a time-dependent mass and charge function.
\FloatBarrier

\subsection{\RNV Space-time}\label{RNV_Photon}

Having understood the evolution of the photon sphere in the case of Vaidya spacetime, we shall now take over the case of a black hole with a time-dependent charge and mass, thus depicting \RNV spacetime. A physical scenario where this may arise is in the case of a black hole accreting both mass and charge. Hence in this context, it is more suitable to describe the dynamical black hole in the in-going null coordinate $v$, in which the spacetime geometry is given by \ref{Dynamical_in}, with the following identifications, 
\begin{equation}\label{RN_in}
f(r,v) = 1-\frac{2M(v)}{r} + \frac{Q(v)^2}{r^2}, \quad A_v = \frac{Q(v)}{r}~.
\end{equation}
Here $M(v)$ and $Q(v)$ are the respective mass and charge functions, with $A_v$ being the electromagnetic gauge field. Unlike the Vaidya solution, which for static case is a solution of vacuum Einstein's equations, the static case for \RNV solution requires support from Maxwell stress tensor. In particular, the total action of the static scenario, besides the Einstein-Hilbert term also has the $F_{\mu\nu}F^{\mu \nu}$ coupling. However in the dynamical situation, besides the Maxwell field, we also need some additional contribution from the matter sector, which takes the following form as Einstein's equations are assumed to hold,
\begin{equation}
8\pi\,T_{\mu\nu}^{\textrm{ext}} = \frac{1}{r^3}\left\{2r \,\dot{M}(v) - 2Q(v) \dot{Q}(v)\right\}\delta_{\mu}^{v} \delta_{\nu}^{v}~.
\end{equation}
As evident the above energy-momentum tensor is associated with some sort of null fluid and it obeys the null energy condition, i.e., $T_{\mu\nu}^{\rm ext} k^\mu k^\nu\geq 0$ for the null vector $k^{\mu}=(\partial/\partial v)^{\mu}$ if,
\begin{equation}\label{NECRNV}
2r \,\dot{M}(v) - 2Q(v) \dot{Q}(v) \geq 0~,
\end{equation}
holds \cite{Caceres:2013dma}. Clearly the \NEC\ is obeyed for all $r\geq (Q\dot{Q}/\dot{M})\equiv r_{\rm cs}$, where $r_{\rm cs}$ denotes the critical surface within which the Null Energy Condition may get violated. Thus for such spacetime, where \ref{NECRNV} is not satisfied, there exist regions where the \NEC\ is violated as well.

From the Hawking's area theorem \cite{hawking_ellis_1973}, we know that the radius of the \EH\ can decrease for infalling matter, which admits violation of the \NEC. Therefore, in the presence of a \CS\ $r_{\rm cs}$, this essentially boils down to the question that, whether the \EH\ lies inside or outside the \CS\ and the evolution of the \EH\ would behave accordingly. One can also choose the mass and charge profile in such a way that the \CS\ crosses the \EH\ at some value of the in-going null coordinate. In such a case, one would expect the \EH\ first to increase and then decrease as the \CS\ crosses it. Interestingly, it turns out that the evolution of the photon sphere is also affected by the violation of the \NEC, which is counter-intuitive. Since unlike the \EH\, there is apriori no reason for the \PS\ to be somehow related to the \NEC. This curious phenomenon has been explicitly demonstrated in this section, i.e., we have shown that the evolution of the \PS\ is related to the location of the \CS. Therefore, since the photon spheres can be probed by an external observer, it may provide an observational evidence to the violation of the \NEC. 

\begin{figure}[h]
	\centering
    \subfloat{{\includegraphics[scale=0.375]{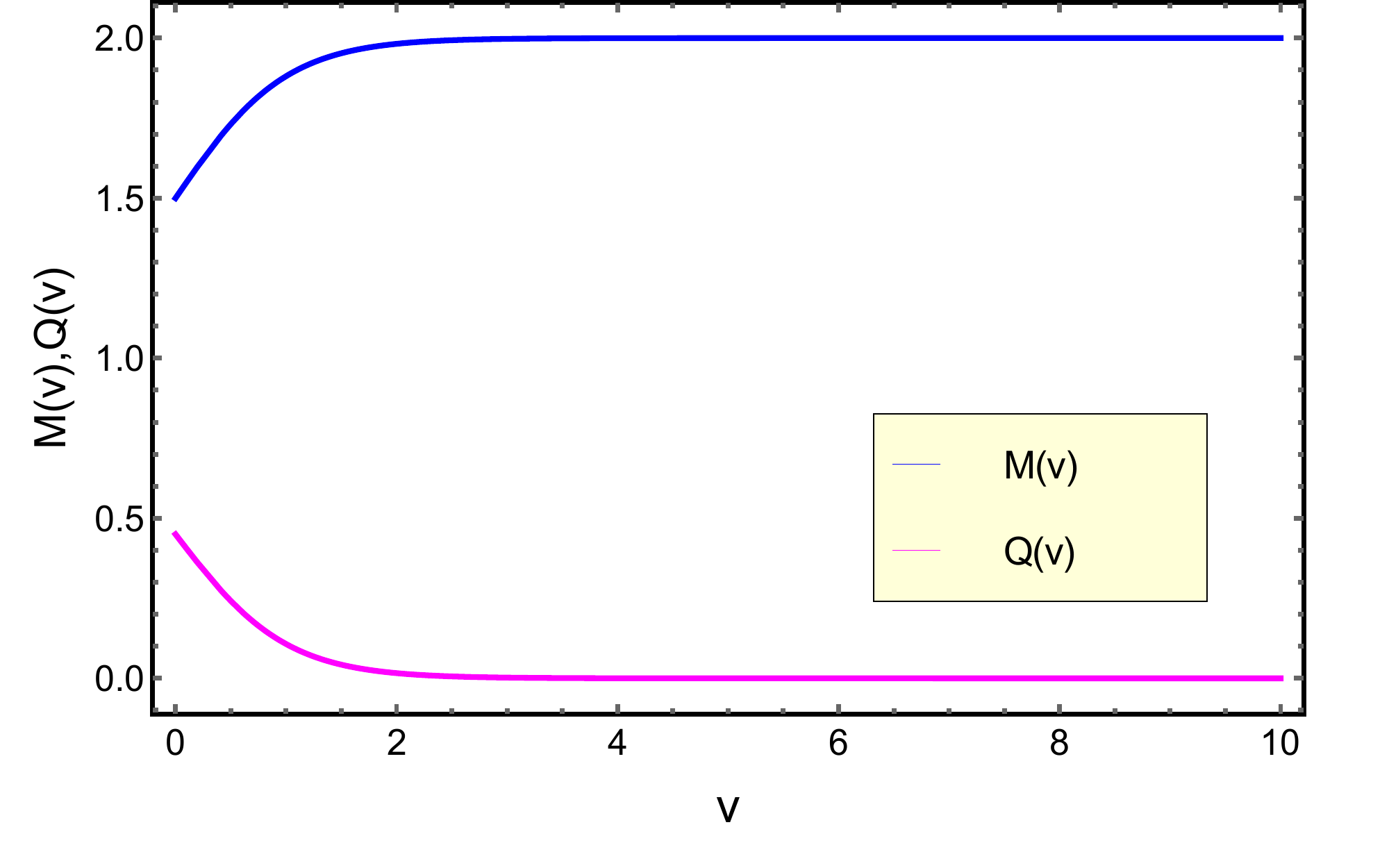} }}
    \qquad
    \subfloat{{\includegraphics[scale=0.375]{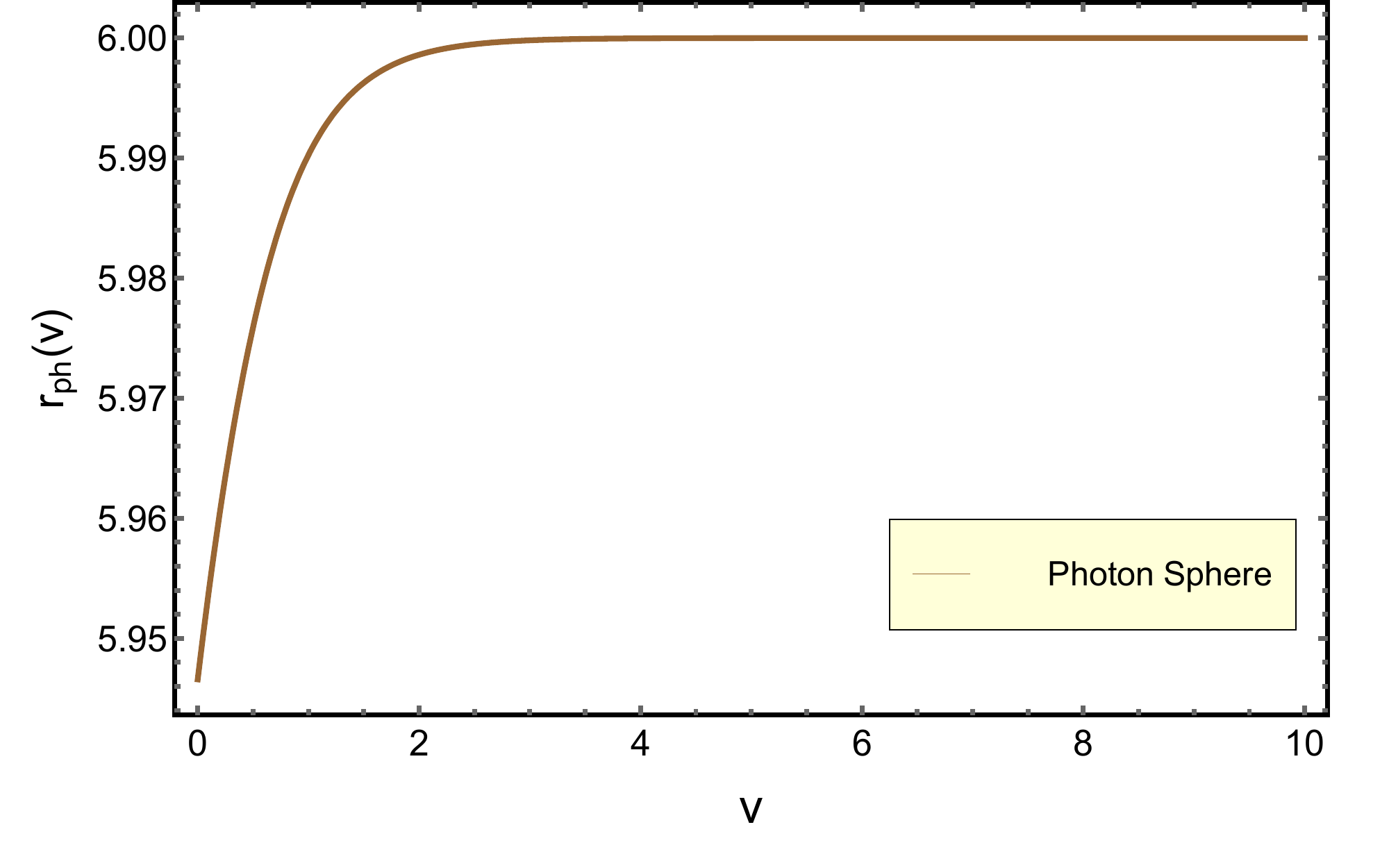} }}%
    \quad
	\subfloat{{\includegraphics[scale=0.375]{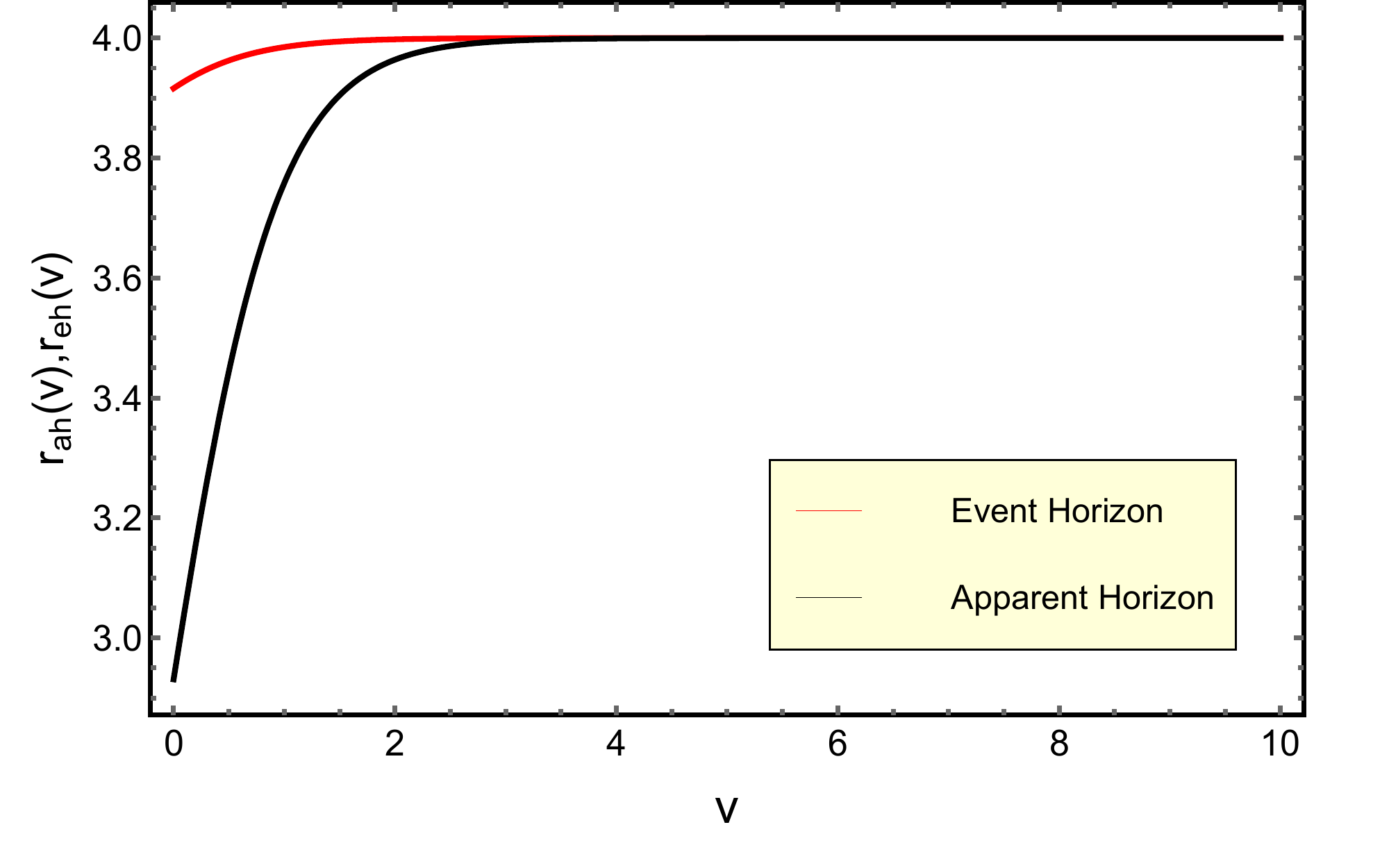} }}
   
    \caption{The top left panel shows the mass and charge function. The top right panel shows the evolution of the radius of \PS. 
In the bottom left panel, we've plotted the evolution of \EH\ and \AH. The bottom right panel shows the evolution of \PS, \AH, \EH\, and \CS\ together for the choice of mass and charge profile in \ref{m_q_1}} 
\label{choice_1}
\end{figure}

\begin{figure}[h]
	\centering
    \subfloat{{\includegraphics[scale=0.375]{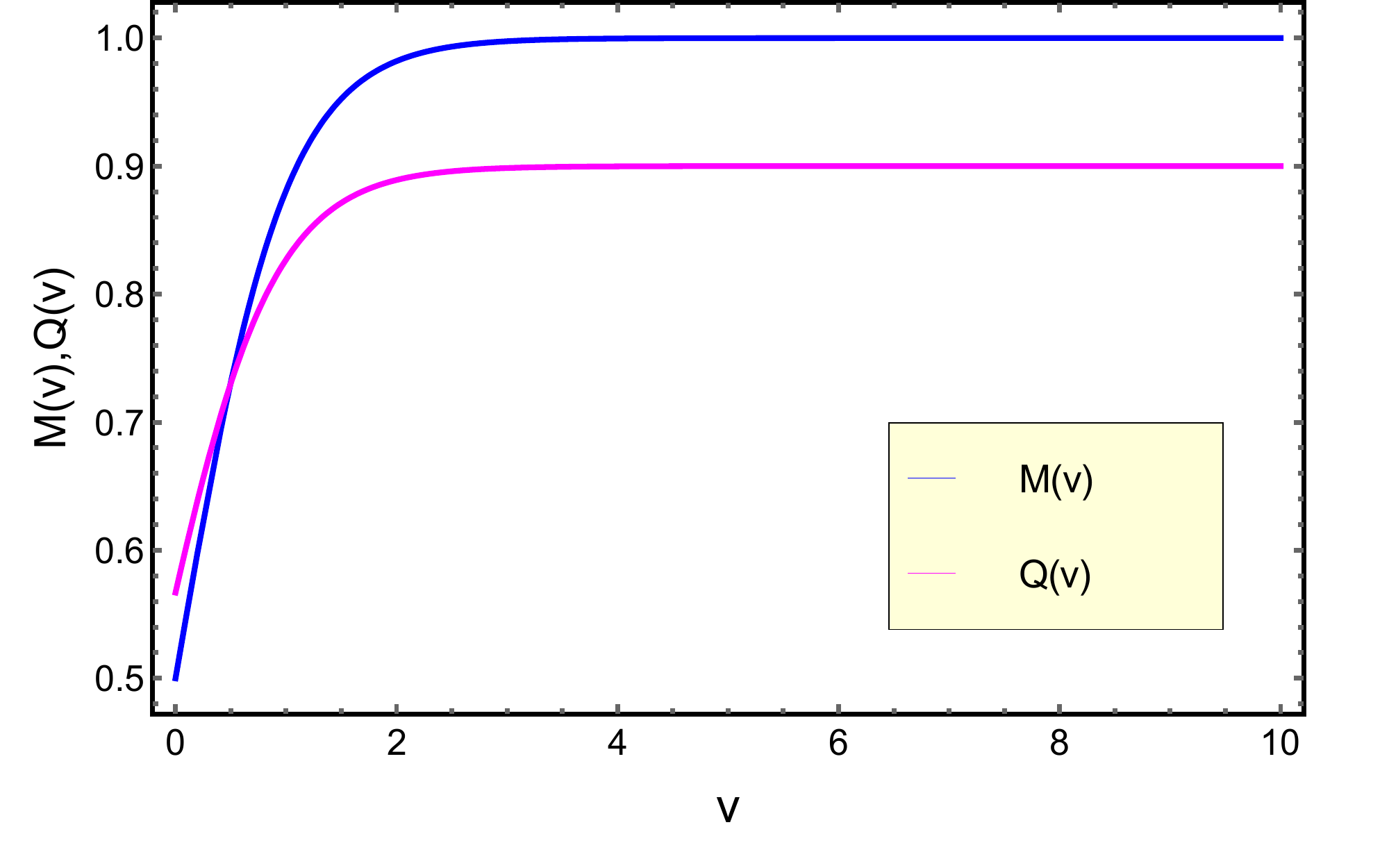} }}
    \qquad
    \subfloat{{\includegraphics[scale=0.375]{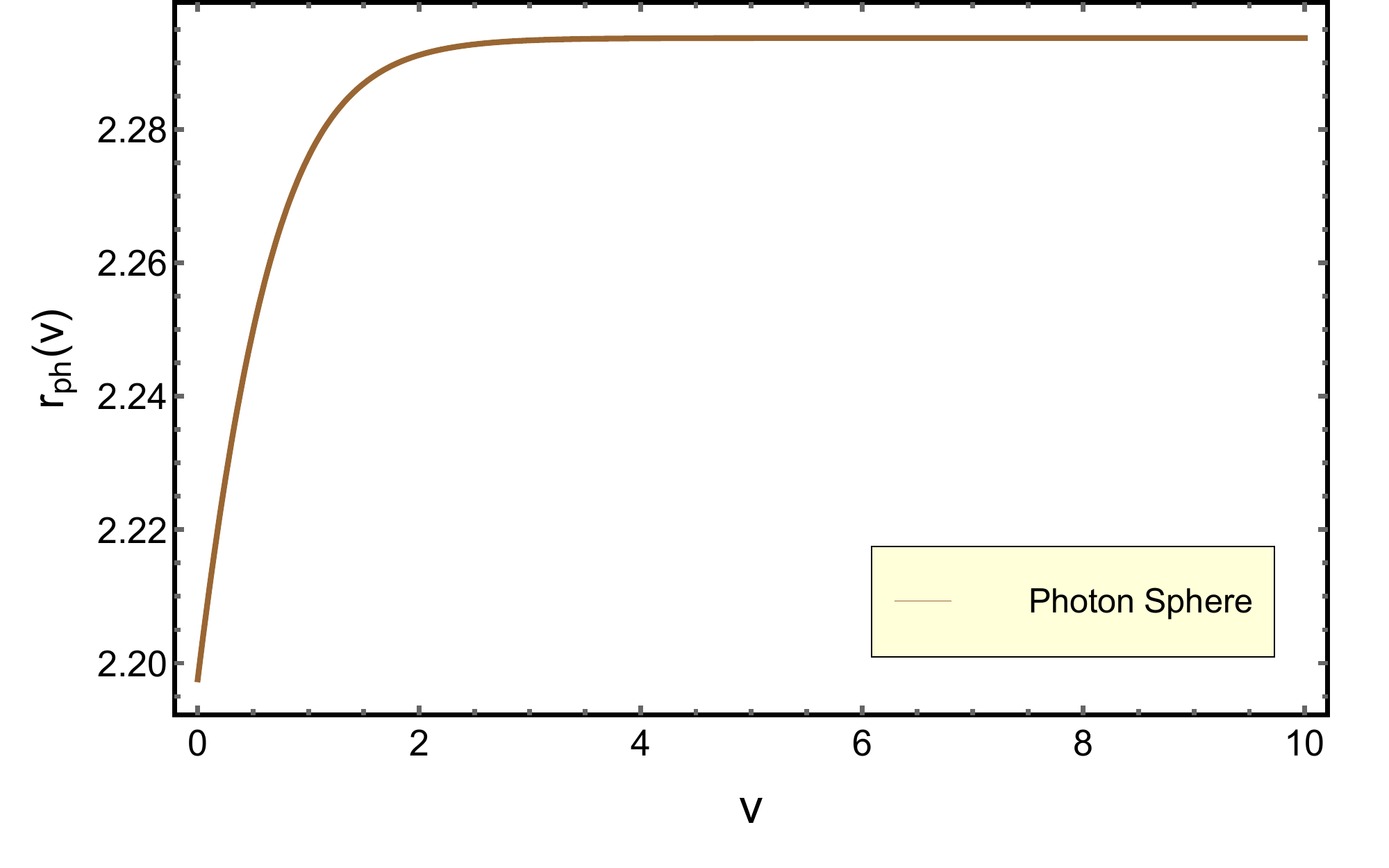} }}%
    \quad
	\subfloat{{\includegraphics[scale=0.375]{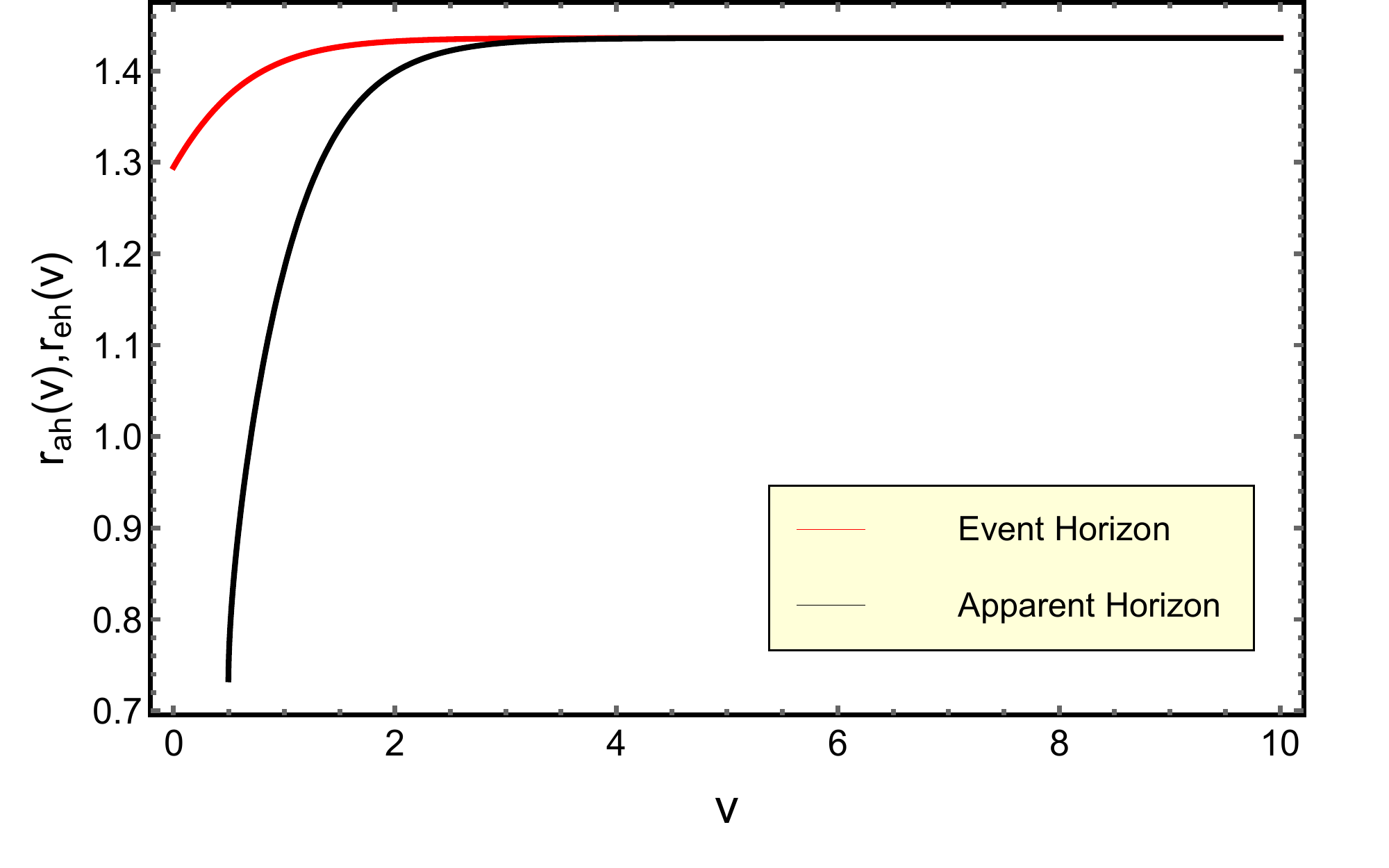} }}
    \qquad
    \subfloat{{\includegraphics[scale=0.375]{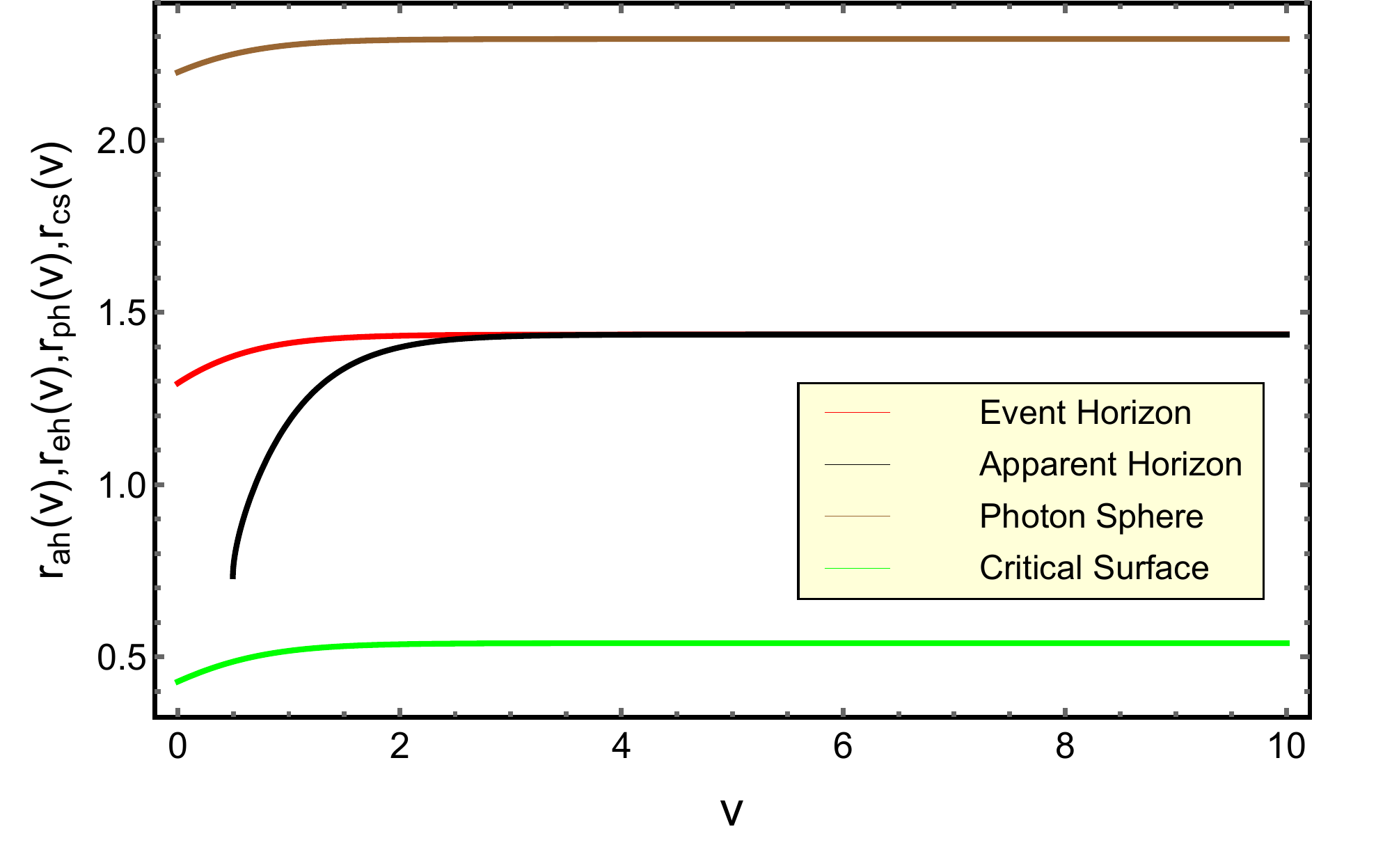} }}%
 \caption{The top left panel shows the mass and charge function. The top right panel shows the evolution of the radius of \PS. 
In the bottom left panel, we've plotted the evolution of \EH\ and \AH. The bottom right panel shows the evolution of \PS, \AH, \EH\, and \CS\ together for the choice of mass and charge profile in \ref{m_q_2}}\label{choice_2}
\end{figure}

\begin{figure}[h]
    \centering
    \subfloat{{\includegraphics[scale=0.375]{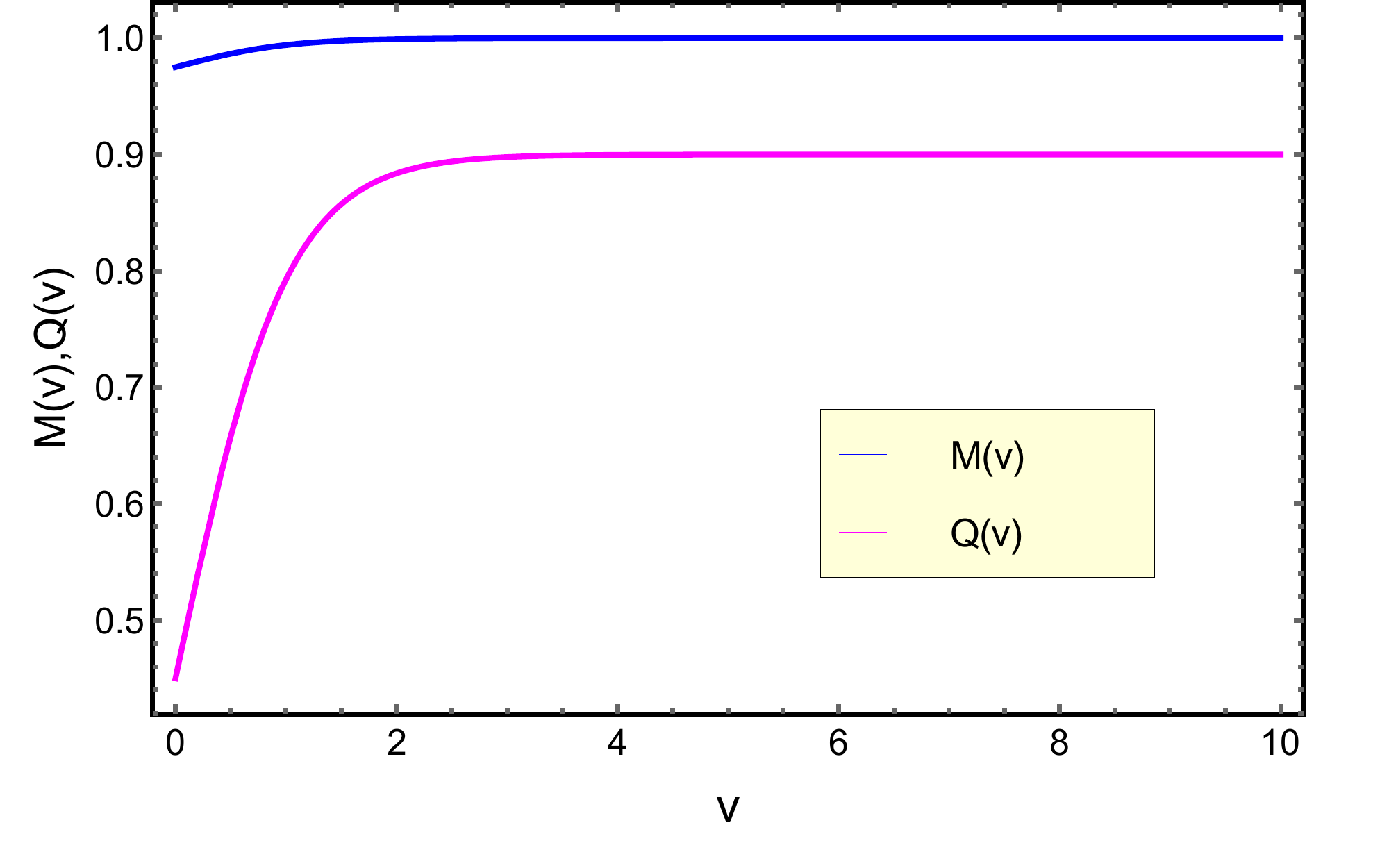} }}
    \qquad
    \subfloat{{\includegraphics[scale=0.375]{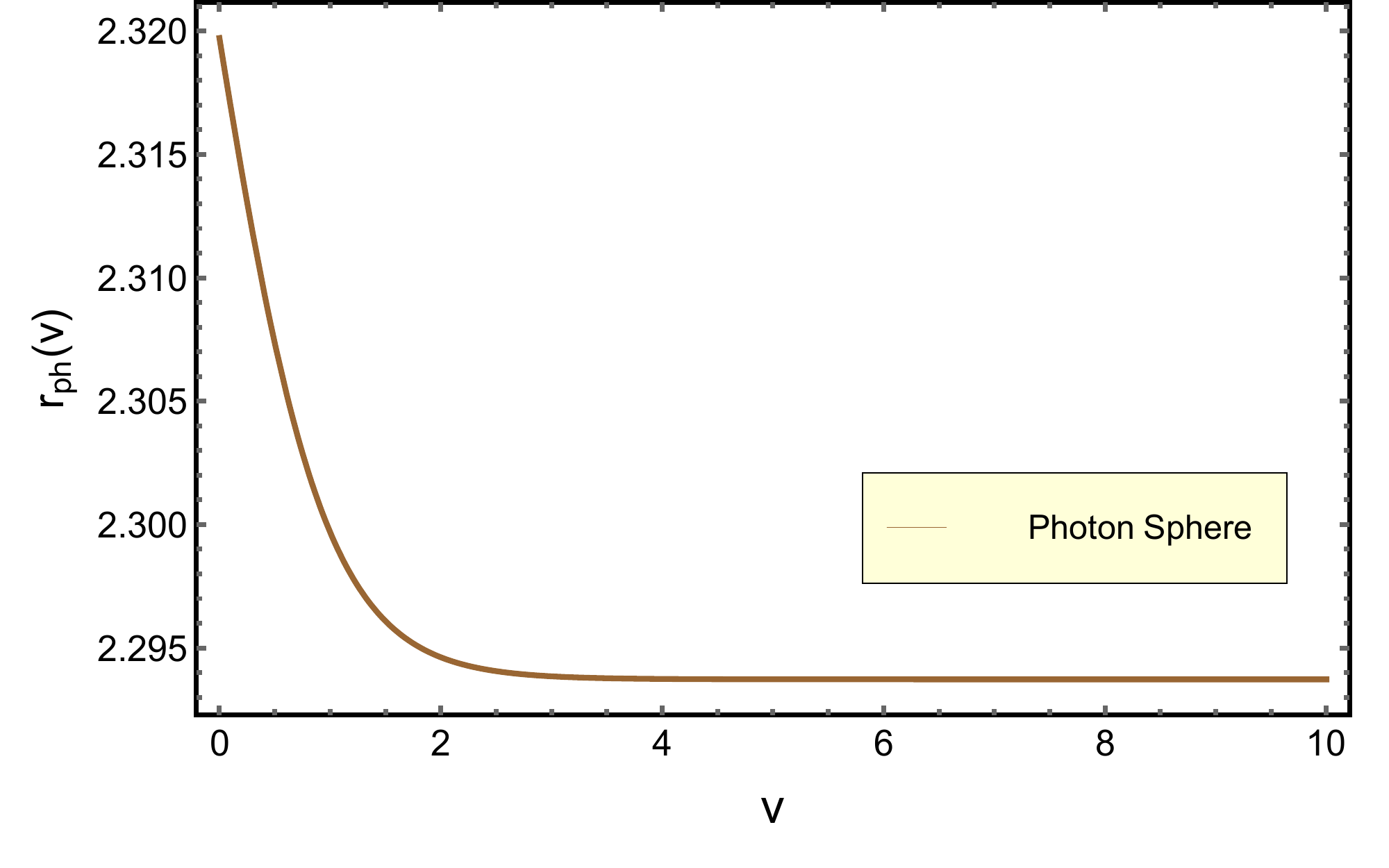} }}%
    \quad
    \subfloat{{\includegraphics[scale=0.375]{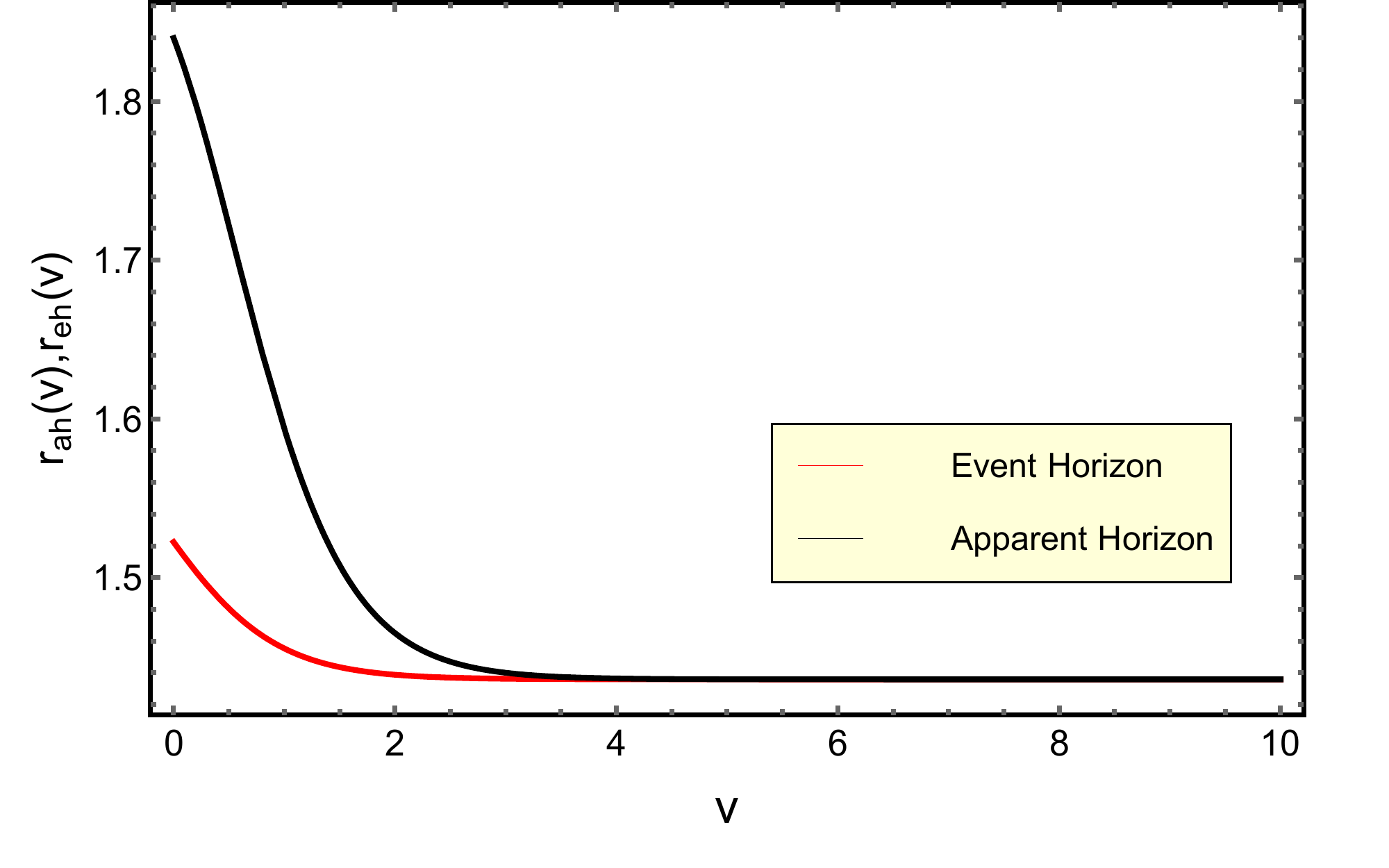} }}
    \qquad
    \subfloat{{\includegraphics[scale=0.375]{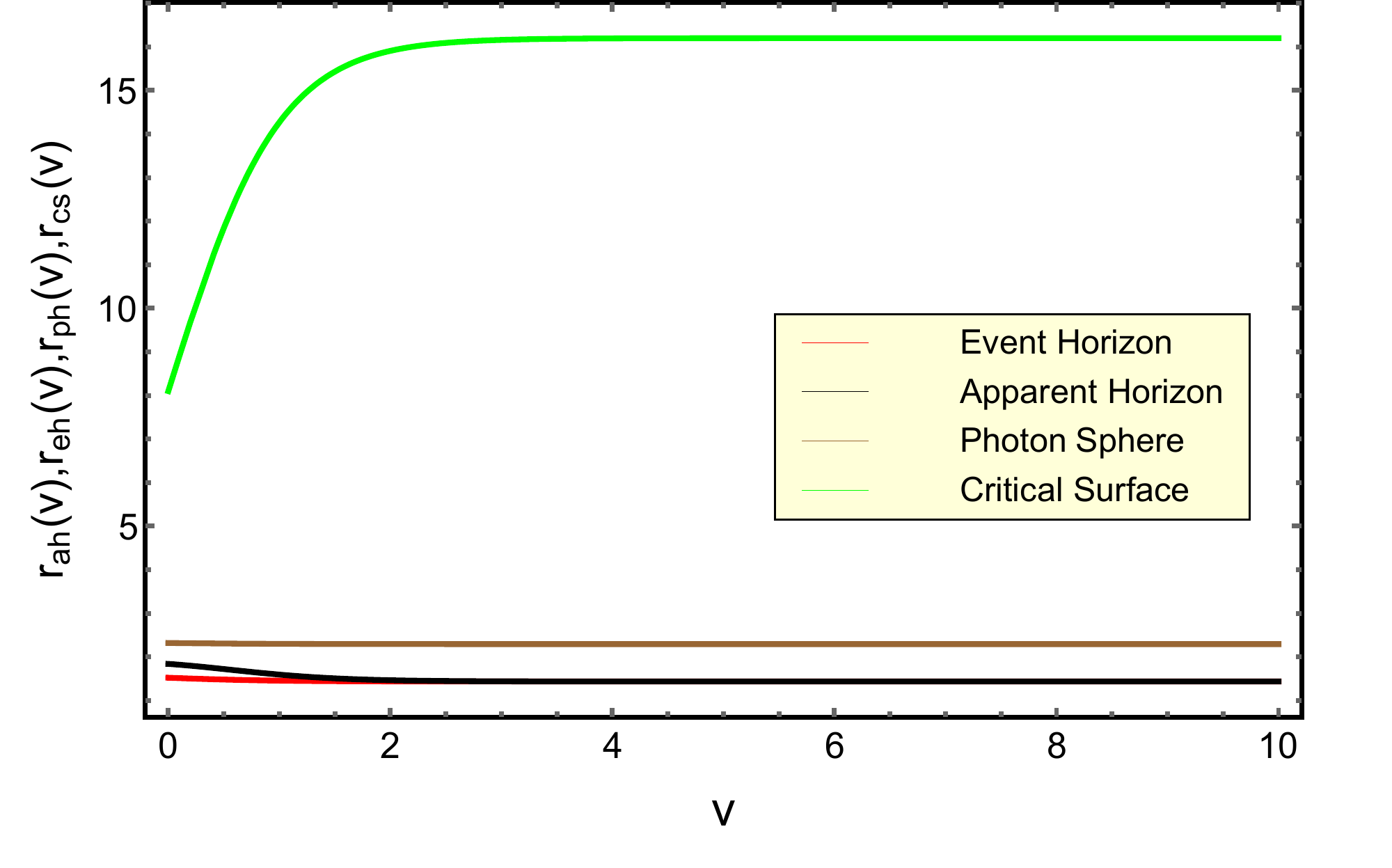} }}%
    \caption{This figure demonstrates the various of photon sphere and related quantities for certain mass and charge functions. The top left panel shows the mass and charge function themselves, taken to be $M(v)=0.95+(0.05/2)\{1+\tanh(v)\}$ and $Q(v)=(0.9/2)\{1+\tanh(v)\}$. The top right panel presents the evolution of the radius of the photon sphere, while the bottom left panel shows the evolution of both the \EH\ and the \AH. Finally, the bottom right panel shows the evolution of the photon sphere, apparent horizon, event horizon, and critical surface together for the above choice of mass and charge functions.}\label{choice_4}
\end{figure}

\begin{figure}[h]
    \centering
    \subfloat{{\includegraphics[scale=0.375]{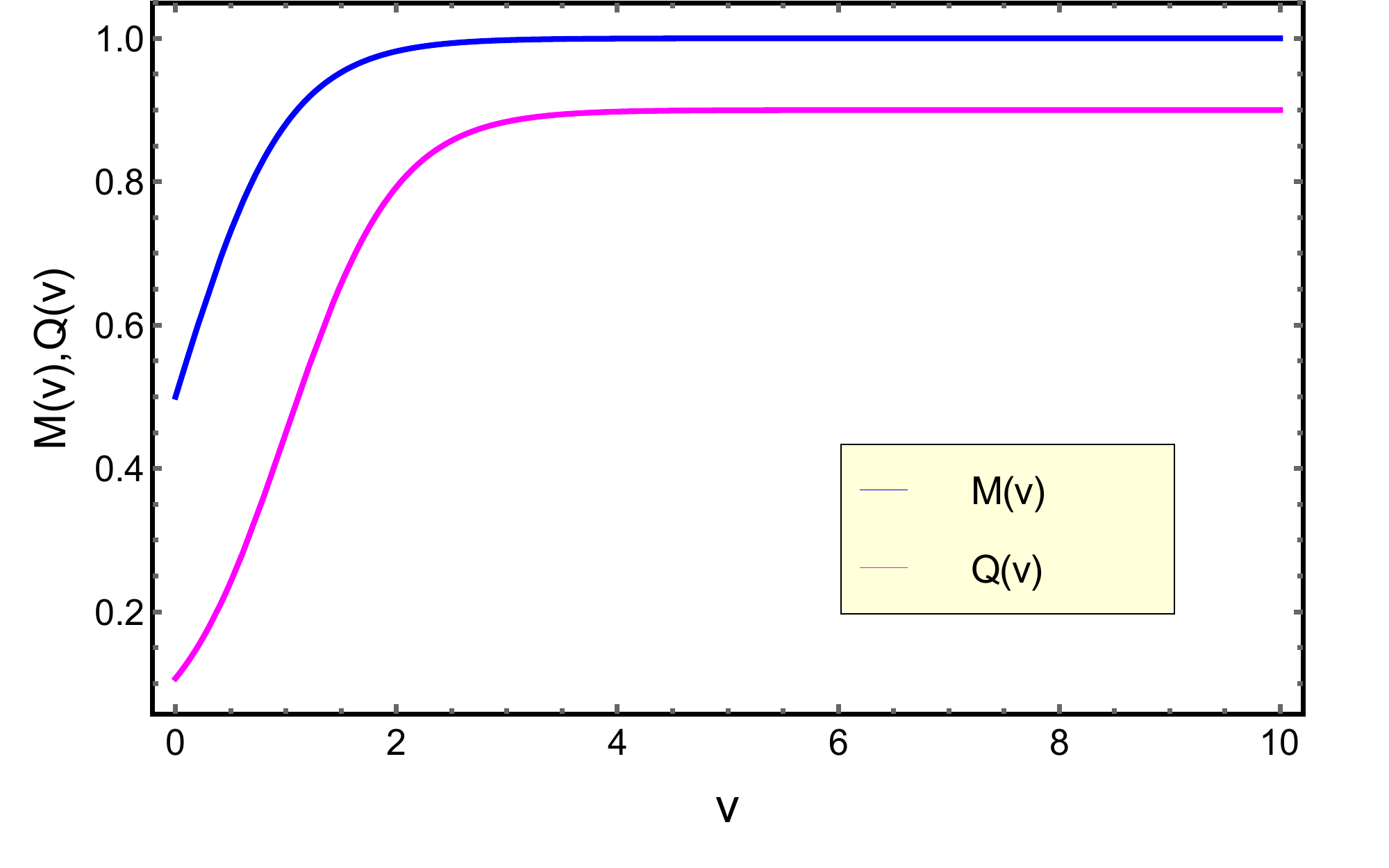} }}
    \qquad
    \subfloat{{\includegraphics[scale=0.375]{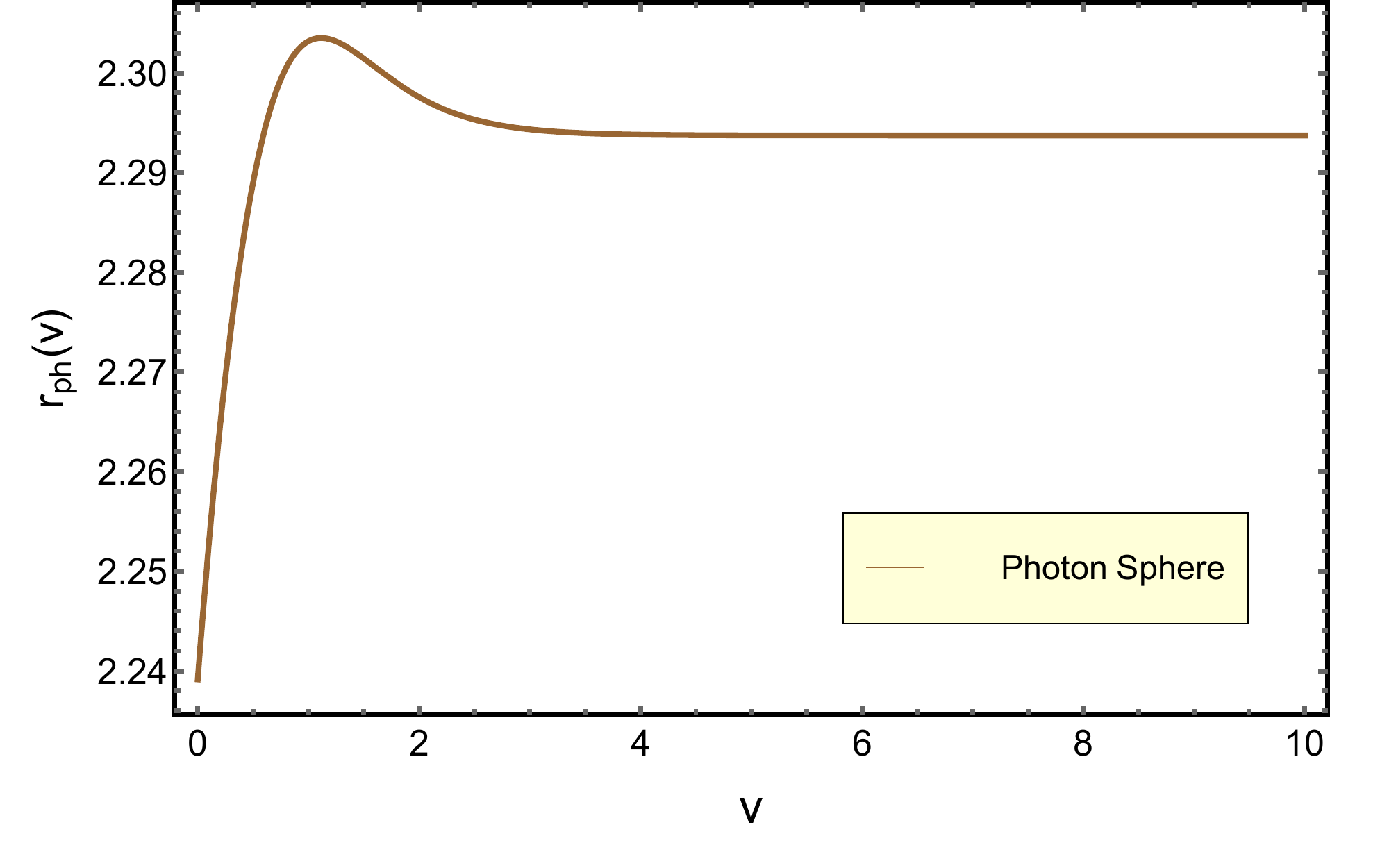} }}%
    \quad
    \subfloat{{\includegraphics[scale=0.375]{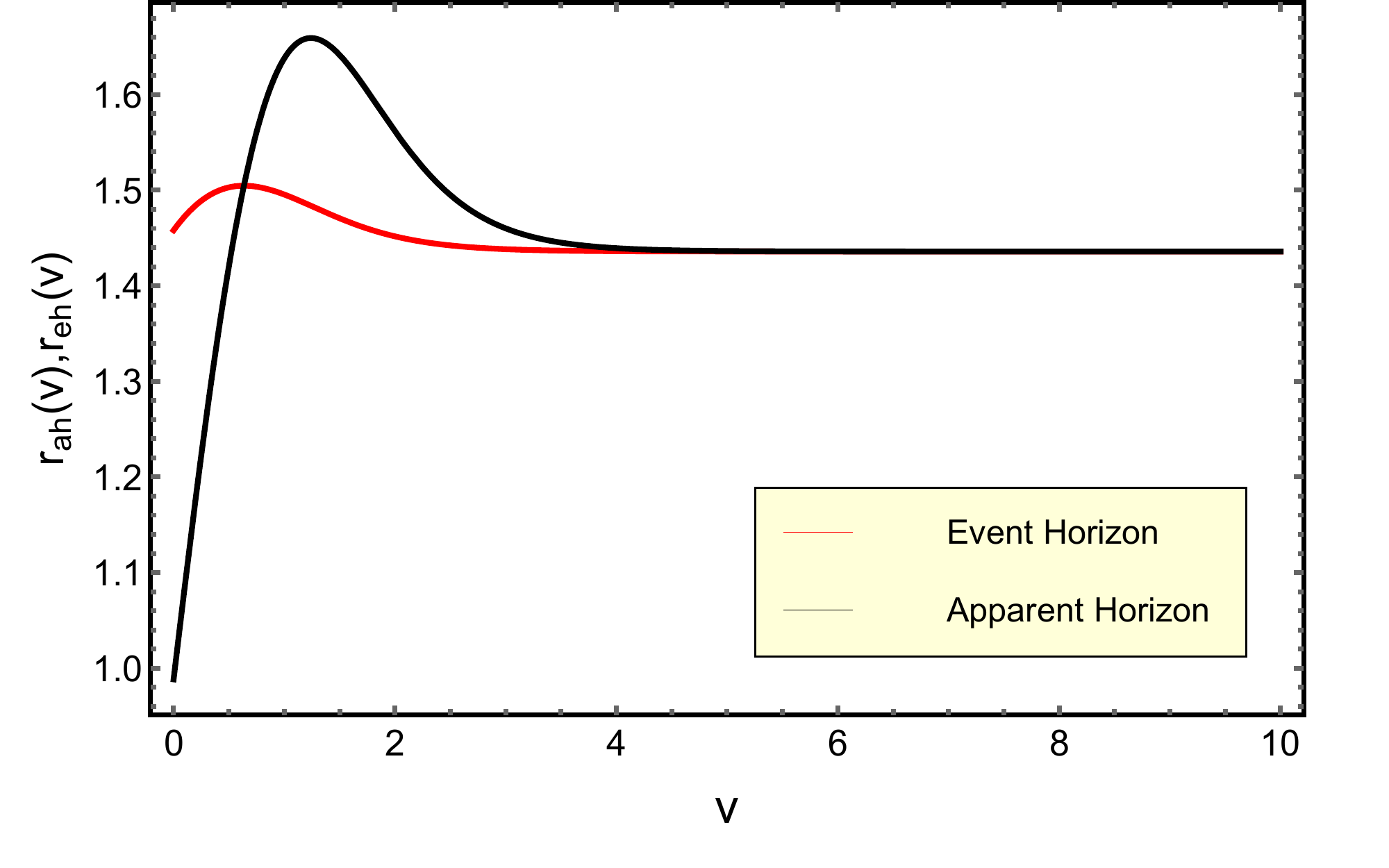} }}
    \qquad
    \subfloat{{\includegraphics[scale=0.375]{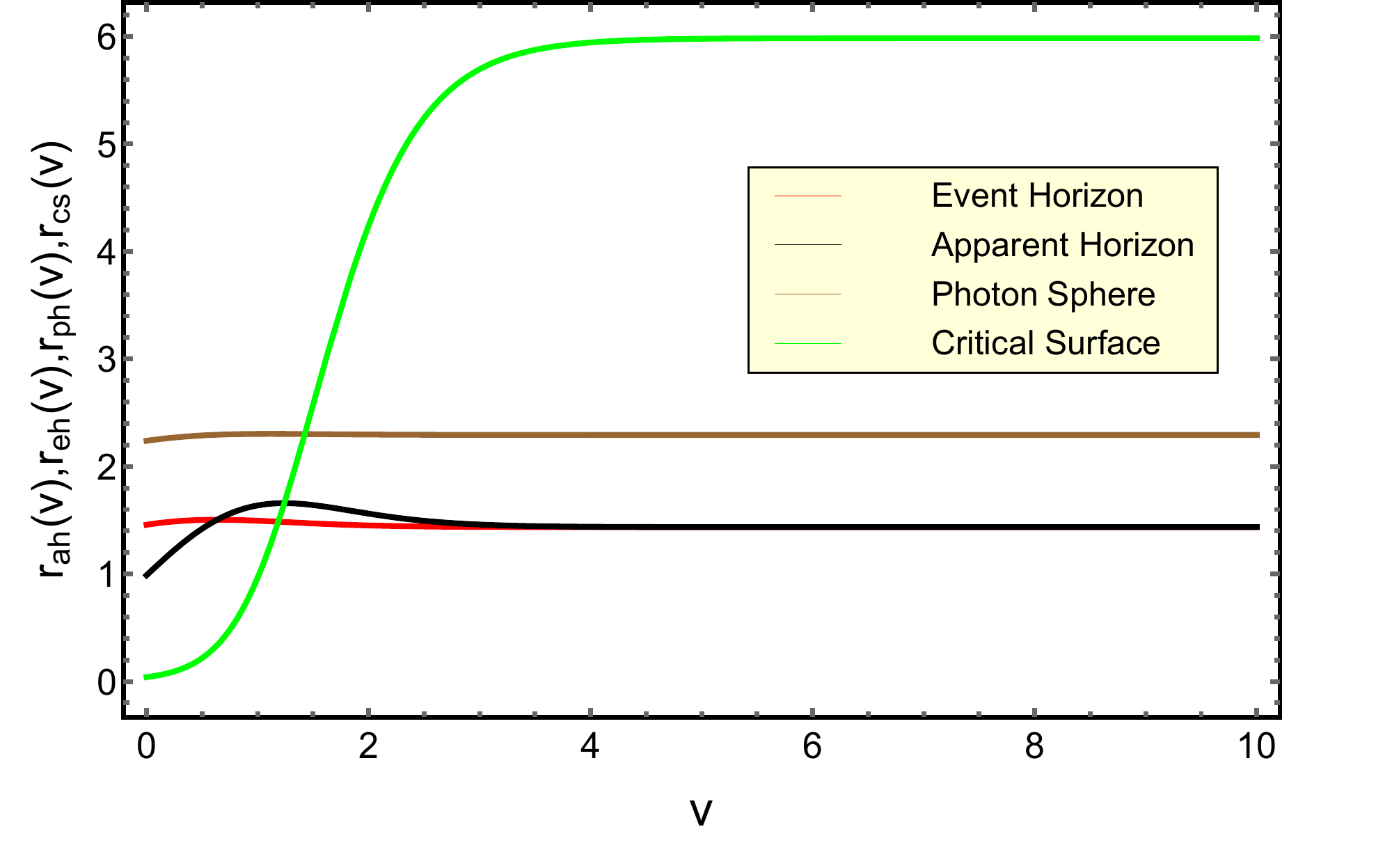} }}%
    \caption{The top left panel demonstrates the mass and charge function for which the critical surface crosses the event horizon and the photon sphere. The exact mass and charge functions are $M(v) =0.5\{1+\tanh(v)\}$ and $Q(v) = 0.45\{1-\tanh(1-v)\}$. The top right panel shows the evolution of the radius of photonthe  sphere, which initially increases and then starts decreasing as null energy condition gets violated. In the bottom left panel we've plotted the evolution of both the event horizon and the apparent horizon, while a collective behaviour of the evolution of photon sphere, apparent horizon, event horizon and critical surface has been presented in the bottom right panel.}\label{choice_6}
\end{figure}

Having described the basic structure, let us now illustrate the evolution of the photon sphere by studying various mass and charge profile and the location of their respective critical surface. As argued earlier, we restrict our attention to smoothly varying mass and charge functions. This is important since to provide a future boundary condition one need to know the entire evolution of the spacetime. Our choice of mass and charge functions are particularly motivated from \cite{Caceres:2013dma}, however, in addition, we have also studied some different mass and charge functions as well. One such mass and charge function can be written down as,
\begin{equation}
M(v) = \frac{M_0}{2}\left[1+\frac{1}{2}\left\{1+\tanh(v)\right\}\right]\quad \text{and}\quad Q(v) = Q_0\,\left\{1-\tanh(v)\right\}\label{m_q_1}
\end{equation}
One can immediately substitute the mass and charge functions introduced above in \ref{NECRNV}, which will ensure that for arbitrary choices of $M_0$ and $Q_0$ the null energy condition will hold and hence the critical surface does not exist. This, in turn, implies that the photon sphere along with the apparent and the event horizon smoothly increases. The differential equation satisfied by the photon sphere requires future boundary conditions to solve for, e.g., $r_{\rm ph}(v\rightarrow \infty)=(1/2)(3\,M_0 + \sqrt{9\,M_0^2-8\, Q_0^2}$ along with $\dot{r}_{\rm ph}(v\rightarrow \infty)=0$. The evolution of the photon sphere associated with the above mass and charge functions along with boundary conditions have been presented in \ref{choice_1}.
\FloatBarrier
For completeness, let us consider another situation in which both the mass and charge function, namely $M(v)$ and $Q(v)$, are such that there exists a critical surface but lies within the event horizon. Since to an outside observer, the region within the event horizon is a black box; it is not possible to probe possible violation of null energy condition. This can be achieved by the following choice of mass and charge profile,
\begin{equation}
M(v) = \frac{M_0}{2}[1+\tanh(v)]\quad \text{and}\quad Q(v) = Q_0\, M(v)^{2/3}\label{m_q_2}
\end{equation}
Again, we solve the differential equation presented in \ref{Photon_Sp_Eqn_in} using the future boundary conditions to obtain the evolution of the photon sphere, and the result is illustrated in \ref{choice_2}. The presence of a critical surface is evident from \ref{choice_2}; however, it remains within the event horizon. In this case, the photon sphere along with the event and apparent horizon grow, ultimately it asymptotes to the static values at late times.
\FloatBarrier
Given the previous scenario, it is straightforward to come up with a different choice of the mass and charge function for which the \CS\ actually lies completely outside the photon sphere. This makes it prone to outside observers. An immediate corollary of the above feature being both the \PS\ and the \EH\ decrease as they evolve. This is because both of them lies in a region where the \NEC\ is violated. This can be realized by considering both $M(v)$ and $Q(v)$ to be proportional to $1+\tanh(v)$, and the result for certain choices of the mass and charge parameters have been presented in \ref{choice_4}. The fact that the photon sphere along with event and apparent horizon decreases with the advanced null coordinate $v$ for certain choices of the mass and charge function is also evident from \ref{choice_4}.
\FloatBarrier
The previous two examples harbor critical surfaces, such that the event horizon and the photon sphere are either completely inside or outside the critical surface. It is certainly possible to come up with a certain mass and charge functions $M(v)$ and $Q(v)$, such that there exists a critical surface, which initially starts being within the event horizon and eventually crosses both \EH\ and photon sphere. Therefore one should expect the \EH\ first to grow (since \NEC\ is satisfied for some time) and eventually starts decreasing. One should also expect to observe the teleological nature of event horizon in this situation\cite{Nielsen:2010gm}, i.e., to see the event horizon starts growing even before it crosses the critical surface. These theoretical expectations are borne out by the plot presented in \ref{choice_6}. The same result can also be derived for the photon sphere as well, i.e., first, it grows for some time and then starts decreasing. Such a situation also comes out of the numerical solution of the differential equation governing the evolution of the photon sphere, illustrated in \ref{choice_6}.
\newpage

\begin{figure}[h]
    \centering
    \subfloat{{\includegraphics[scale=0.375]{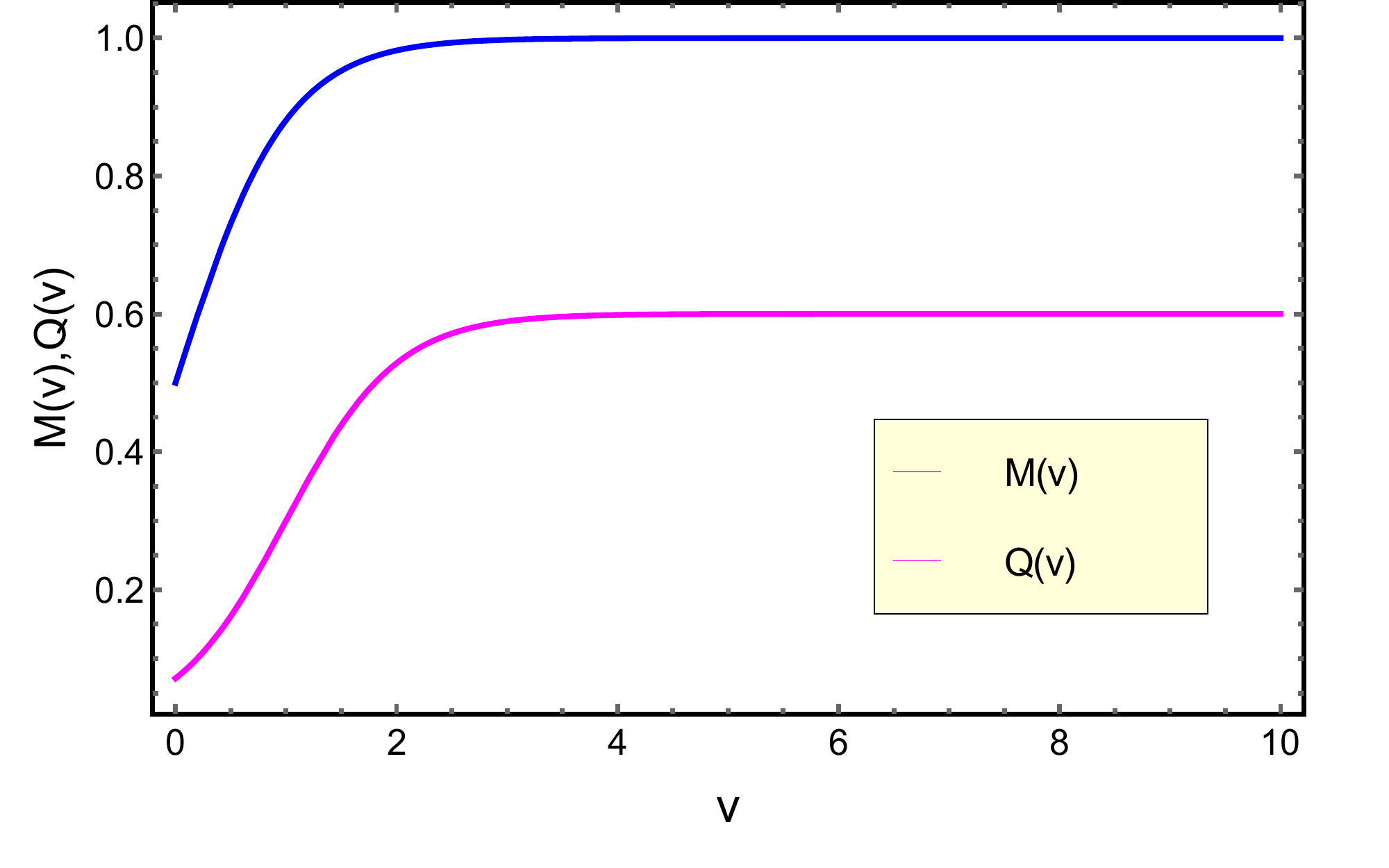} }}
    \qquad
    \subfloat{{\includegraphics[scale=0.375]{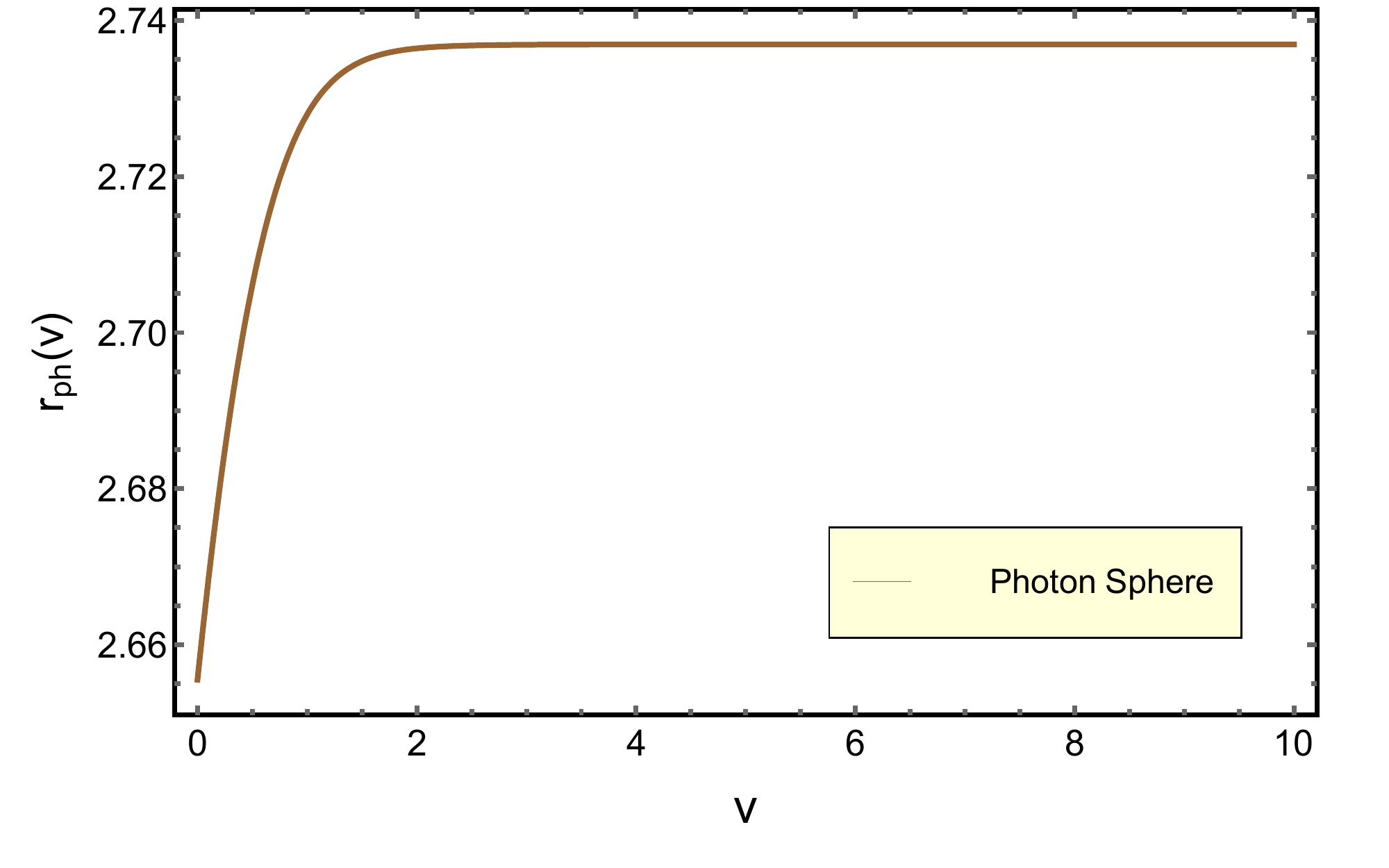} }}%
    \quad
    \subfloat{{\includegraphics[scale=0.375]{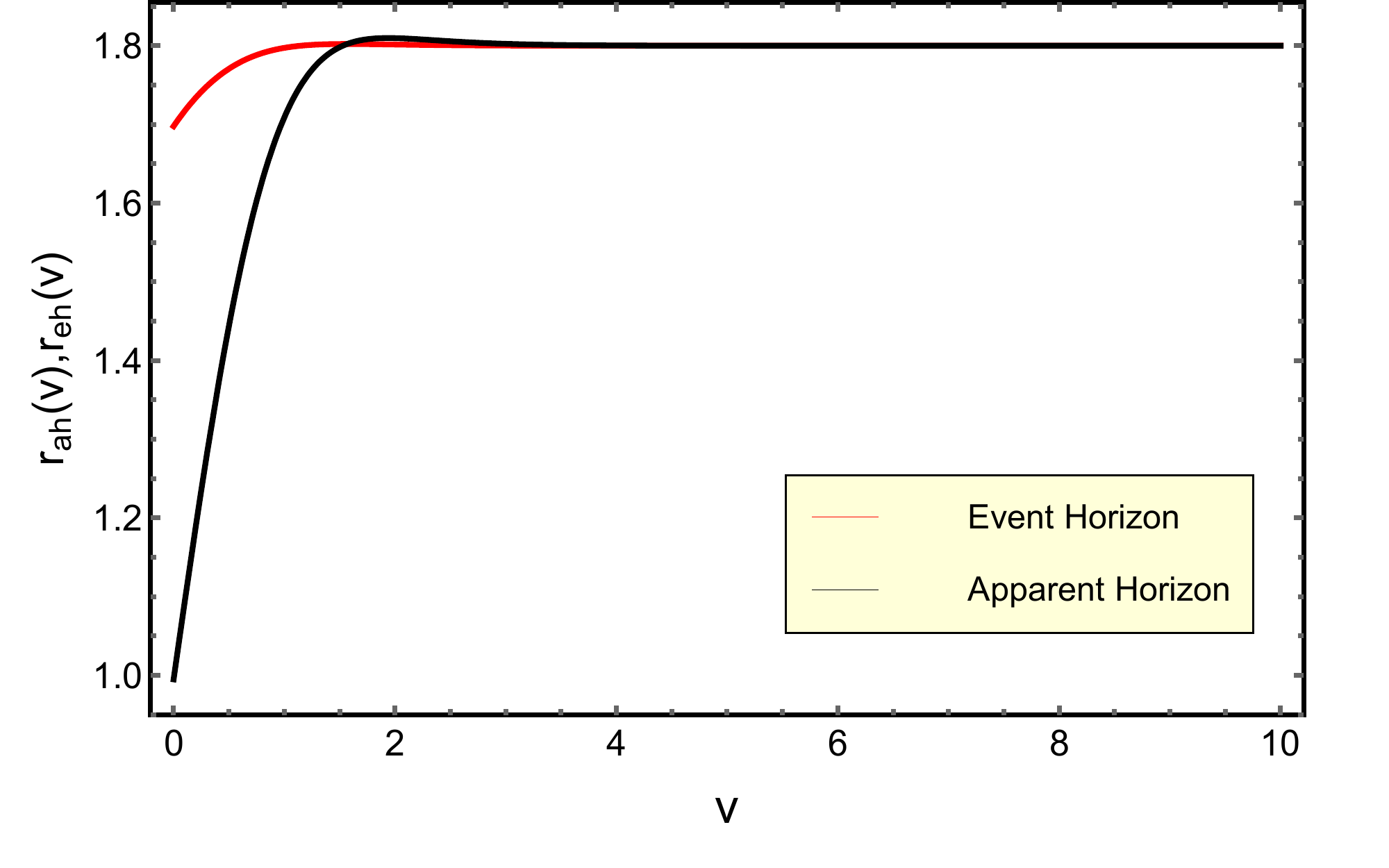} }}
    \qquad
    \subfloat{{\includegraphics[scale=0.375]{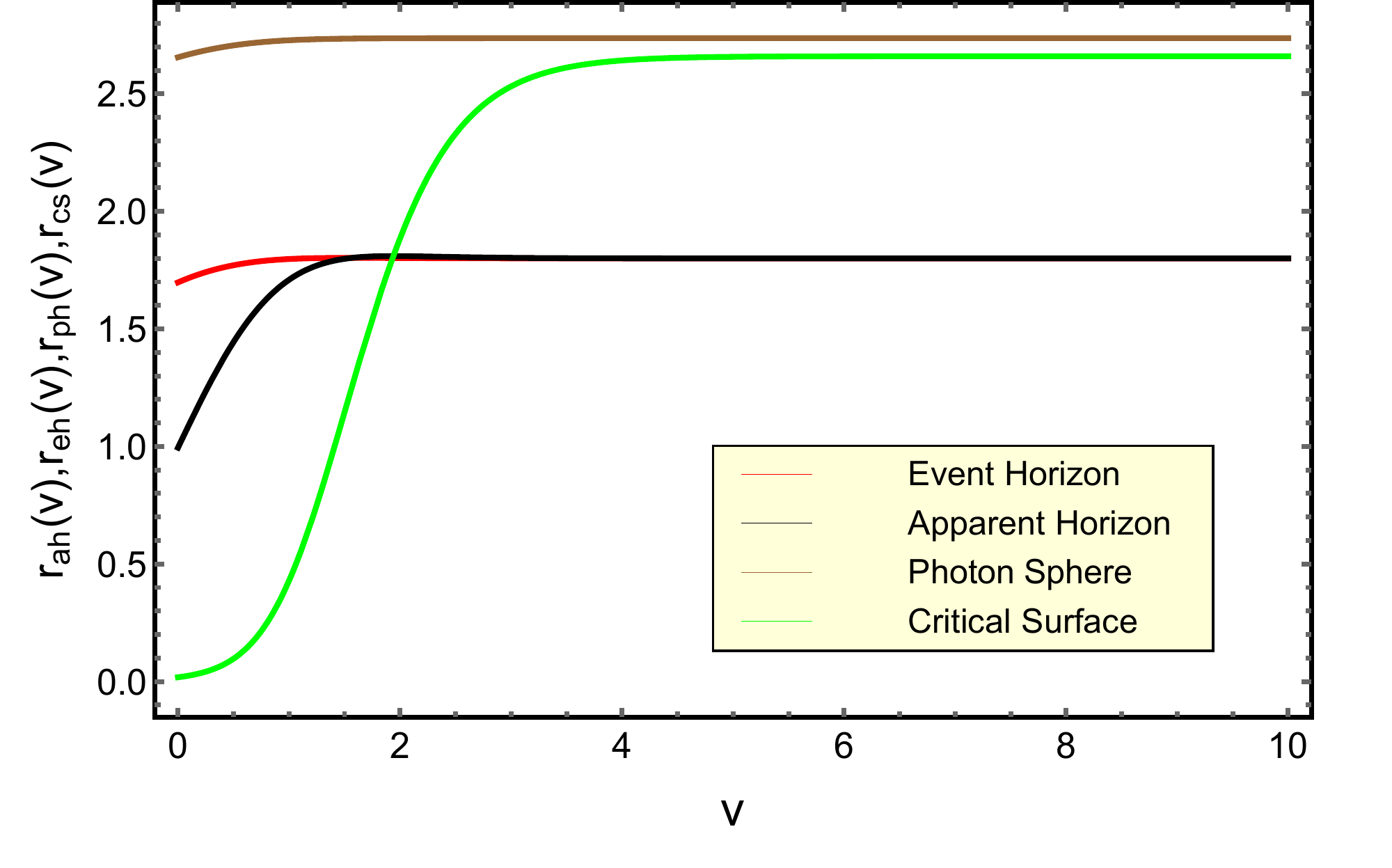} }}%
    \caption{The top left panel shows the variation of the mass and charge function $M(v) =0.5\{1+\tanh(v)\}$ as well as $Q(v) = 0.3\{1-\tanh(1-v)\}$ with the advanced null coordinate $v$. The top right panel, on the other hand, shows the evolution of the radius of photon sphere. In the bottom left panel we have plotted the evolution of event horizon and apparent horizon, while the bottom right panel shows the evolution of photon sphere, apparent horizon, event horizon and critical surface together for the choice of mass and charge profile presented above.}\label{choice_5}
\end{figure}

As a final illustration of the relation between \NEC\ of external matter and evolution of photon sphere for \RNV black holes, consider another choice of $M$ and $Q$ for which the \CS\ crosses the event horizon, but not the photon sphere. Such mass and charge functions take the form, $M(v) =0.5\{1+\tanh(v)\}$ as well as $Q(v) = 0.3\{1-\tanh(1-v)\}$ respectively. For this particular choice, the critical surface starts within the event horizon and eventually cross it, and hence the \EH\ grows for some time and then starts decreasing. However, the \CS\ for this case doesn't cross the \PS\ at any point in the future, and hence the \PS\ always lies in a region where the \NEC\ is satisfied. Therefore the \PS\ can not probe this violation and always increases. This result is illustrated in \ref{choice_5}.
\FloatBarrier

\subsection{\SD Spacetime}

      As a final example of our method developed in \ref{Section_2}, in this section, we shall determine the evolution of the \PS\ surrounding a black hole in the presence of a positive cosmological constant, with its mass being a function of time (or, in-going null coordinate $v$ for an accreting black hole). The metric structure is identical to \ref{Dynamical_in}, with $f(r,v)=1-\{2M(v)/r\}+(\Lambda/3)r^{2}$. Since for the static case with constant mass, the \PS\ does not depend on the presence of the cosmological constant\cite{Stuchlik:1999qk}, it is interesting to look for any effect of the cosmological constant in the dynamical context. To our surprise, it turns out that for a dynamical \SD black hole, i.e., for black hole mass changing with time, the evolution of the \PS\ indeed depends on the value of the cosmological constant $\Lambda$. Again, an analytic solution for the evolution of the photon sphere turned out to be difficult to achieve, however numerically one can indeed solve for the evolution of the \PS. This is what we illustrate next by numerically solving the evolution equation for \PS\ with various choices of the mass function for both accreting and radiating black holes. 

\begin{figure}[h]
    \centering
    \subfloat{{\includegraphics[scale=0.375]{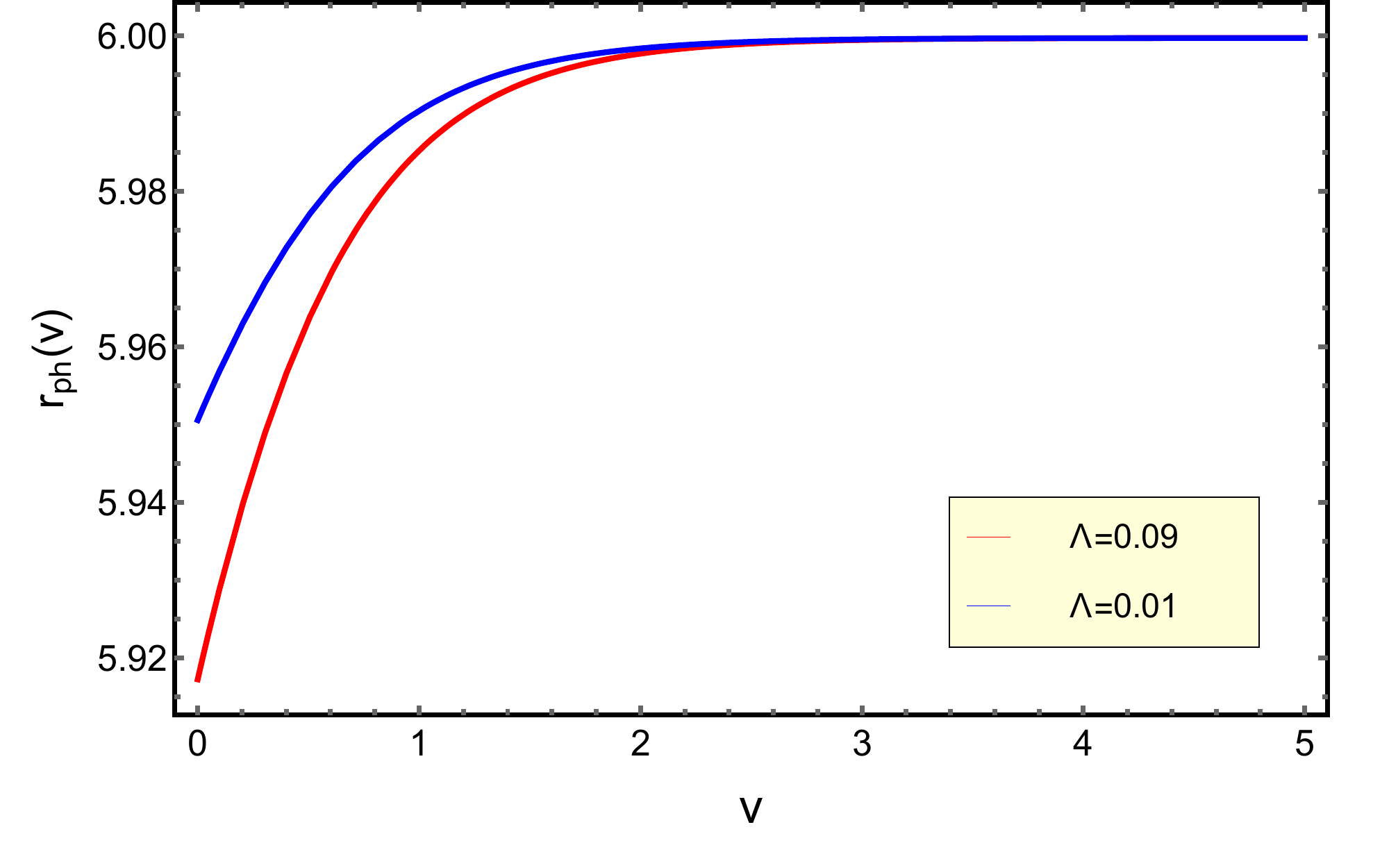} }}
    \qquad
    \subfloat{{\includegraphics[scale=0.375]{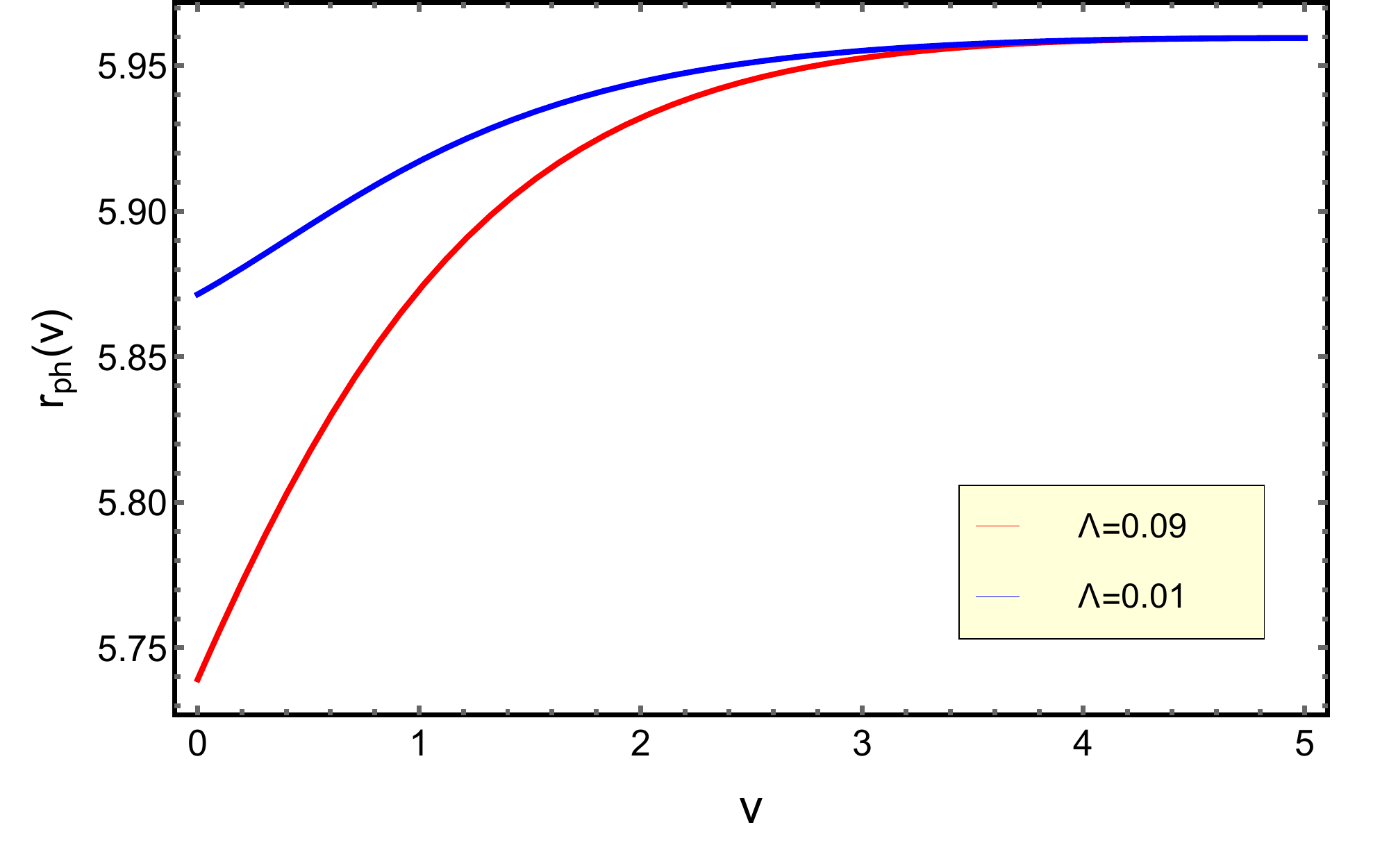} }}
    \caption{Variation of the photon sphere with the advanced time coordinate $v$ has been presented for dynamical Schwarzschild de Sitter spacetime. The left panel shows the evolution of \PS for mass function behaving as $M(v)=M_{0}\{1 + \tanh(v)\}$ and the right panel represents the evolution of \PS\ for mass function $M(v)= M_{0}\{2 - \sech(v)\}$. We have chosen $M_{0}=1$ and for two possible choices of the cosmological constant, taken to be 0.01 and 0.09.}\label{Sch_Des PS}
\end{figure}

As an example of the Schwarzschild de Sitter black hole accreting matter, we consider the mass to be a smoothly increasing function of the in-going time $v$ and using which we solve \ref{Photon_Sp_Eqn_in} to obtain the evolution of the photon sphere for different values of the cosmological constant. To that end we start with the mass functions $M(v)=(M_0/2)\{1+\tanh(v)\}$ and $M(v)=(M_0/2)\{2-\sech(v)\}$ and solve the evolution equation using the future boundary conditions, $r_{\rm ph}(v\rightarrow \infty)=3M_0$ and  $\dot{r}_{\rm ph}(v\rightarrow \infty)=0$. The result of such an evolution is clearly illustrated in \ref{Sch_Des PS} for different choices of the cosmological constant $\Lambda$. 
\FloatBarrier
The above result is interesting in its own right since the photon sphere in the dynamical context depends on the cosmological constant, unlike the static scenario. Moreover, black holes are never in perfect equilibrium, and thus a dynamical study of the photon sphere can be considered as one of the effective tools to explore the value of the cosmological constant of the universe. Here we emphasize that such illustration can also be realized for \RN-\DS black hole having a non-zero time-dependent charge as well. Also, note that the event horizon and cosmological horizon of a static \SD black hole always lies inside and outside the photon sphere respectively. Not surprisingly, this result in the dynamical context holds during the entire course of evolution and have been presented in \ref{Sch_Des PS}. 
\section{Shadow of a Spherically Symmetric Dynamical Black Hole}\label{Section_4}

In the previous sections, we have determined the location of the circular photon orbit on the equatorial plane, which can be used along with spherical symmetry to determine the \PS. However, as far as observational implications are considered, the \PS\ leads to a certain patch of the sky around the black hole to be unobservable. Loosely speaking, the \PS\ casts a shadow, which corresponds to a set of directions in the observer's sky from which light from distant sources doesn't reach the observer. This is what is meant by shadow of a black hole\cite{Vazquez:2003zm,Shaikh:2018lcc,Hou:2018bar,Cunha:2018gql,Tsukamoto:2017fxq,Repin:2018anv,Abdujabbarov:2016hnw,Kumar:2018ple,Cunha:2018cof,Ayzenberg:2018jip,Rahman:2018fgy,Atamurotov:2013sca,Grenzebach:2015oea,Atamurotov:2015xfa,Mars:2017jkk}, and the evolution of photon sphere would be ultimately reflected in the evolution of shadow, which one should expect to be observed  by the Event Horizon Telescope. In this section, taking a cue from our earlier discussion, we will illustrate the evolution of the black hole shadow as the mass and/or charge of the black hole changes with time. We will first work with a general spherically symmetric dynamical spacetime in the in-going null coordinate $v$ with an arbitrary choice of $f(r,v)$. This can also be generalized in a simple manner to the case of out-going null coordinate $u$. Also, in the subsequent sections, we present a generalization of this result to rotating black holes. The geometry of the spacetime we are interested in is given by \ref{Dynamical_in} in which the motion of a test particle is described by the following Lagrangian,
\begin{equation}
\mathcal{L}=\frac{1}{2}\left\{-f(r,v)\left(\frac{dv}{d\lambda}\right)^{2}
+2\left(\frac{dv}{d\lambda}\right)\left(\frac{dr}{d\lambda}\right)
+r^{2}\left(\frac{d\theta}{d\lambda}\right)^{2} 
+r^{2}\sin^2\theta \left(\frac{d\phi}{d\lambda}\right)^2\right\}
\label{lagrangian}
\end{equation}
At this stage, we would like to remind the reader that, a `prime' and `dot' over any quantity represents derivative with respect to the radial coordinate $r$ and the in-going time $v$ respectively. As evident from the Lagrangian presented above, it is independent of the azimuthal coordinate $\phi$ but depends on the in-going null coordinate $v$. Thus the angular momentum is conserved, but the energy is not. However, we can still define a quantity $E(r,v)$, such that,
\begin{equation}\label{conserve quantity}
E(r,v) = -f(r,v)\left(\frac{dv}{d\lambda}\right)+\left(\frac{dr}{d\lambda}\right) \quad \textrm{and} \quad L = r^2 \sin^2\theta \left(\frac{d\phi}{d\lambda}\right)
\end{equation}
Given the above equations one can determine $dv/d\lambda$ and $d\phi/d\lambda$ respectively, in terms of $E(r,v)$ and $L$. This leads to the following expressions,
\begin{equation}
\left(\frac{dv}{d\lambda}\right)=\frac{1}{f(r,v)}\left\{\left(\frac{dr}{d\lambda}\right)- E(r,v)\right\} \quad \textrm{and} \quad \left(\frac{d\phi}{d\lambda}\right) = \frac{L}{r^2 \sin^2\theta} \label{v_phi_dot}
\end{equation}
Since we are interested in the trajectory of the photons, we have to work with the condition $ds^{2}=0$, which implies vanishing of the Lagrangian `$\mathcal{L}$' in \ref{lagrangian}. Finally, use of \ref{v_phi_dot} results into the following differential equation involving both $(dr/d\lambda)$ and $(d\theta/d\lambda)$, such that,
\begin{equation}
r^2 \left(\frac{dr}{d\lambda}\right)^2 - r^2 E^2(r,v) + r^4 \left(\frac{d\theta}{d\lambda}\right)^{2} f(r,v) +\frac{L^2 f(r,v)}{\sin^2\theta} = 0
\end{equation}
The radial and the angular part of the above equation separates out naturally, which is basically achieved by introducing the Carter Constant $K$\cite{Carter:1968rr}, such that the evolution equations for $\theta$ and $r$ becomes,
\begin{align}\label{Cartar_const}
r^4 \left(\frac{d\theta}{d\lambda}\right)^2 = K -\cot^2\theta L^2;
\qquad 
r^{2}\left(\frac{dr}{d\lambda}\right)^2 = E^2 r^2 - \left(K+L^2\right)f(r,v)~.
\end{align}
At this point, we would like to emphasize that for static case one has $r=\textrm{constant}=r_{\rm ph}$ and hence $dr_{\rm ph}/d\lambda=0$. This would give rise to the shadow around a static spherically symmetric black hole. But since we are dealing with dynamical situations, we have $r = r_{\rm ph}(v)$ and hence $(dr_{\rm ph}(v)/d\lambda) = \dot{r}_{\rm ph}(v)\{dv/d\lambda\}$. Then using \ref{v_phi_dot} we obtain,
\begin{equation}
K + L^2 = \frac{E(r_{\rm ph}(v),v)^2r_{\rm ph}(v)^2}{f(r_{\rm ph}(v),v)}\left[ 1- \left\{\frac{\dot{r}_{\rm ph}(v)}{f(r_{\rm ph}(v),v)-\dot{r}_{\rm ph}(v)}\right\}^2\right]
\end{equation}
It is instructive to define the following two quantities $\eta(v)\equiv K/E(r_{\rm ph}(v),v)^2$ and $\xi(v)\equiv L/E(r_{\rm ph}(v),v)$ and the above equation reduces to,
\begin{equation}
\eta(v) + \xi(v)^2 = \alpha(v)^2 +\beta(v)^2 = \frac{r_{\rm ph}(v)^2}{f(r_{\rm ph}(v),v)}\left[ 1- \left\{\frac{\dot{r}_{\rm ph}(v)}{f(r_{\rm ph}(v),v)-\dot{r}_{\rm ph}(v)}\right\}^2\right]\label{spherical shadow}
\end{equation}


\begin{figure}[h]
    \centering
    \subfloat{{\includegraphics[scale=0.35]{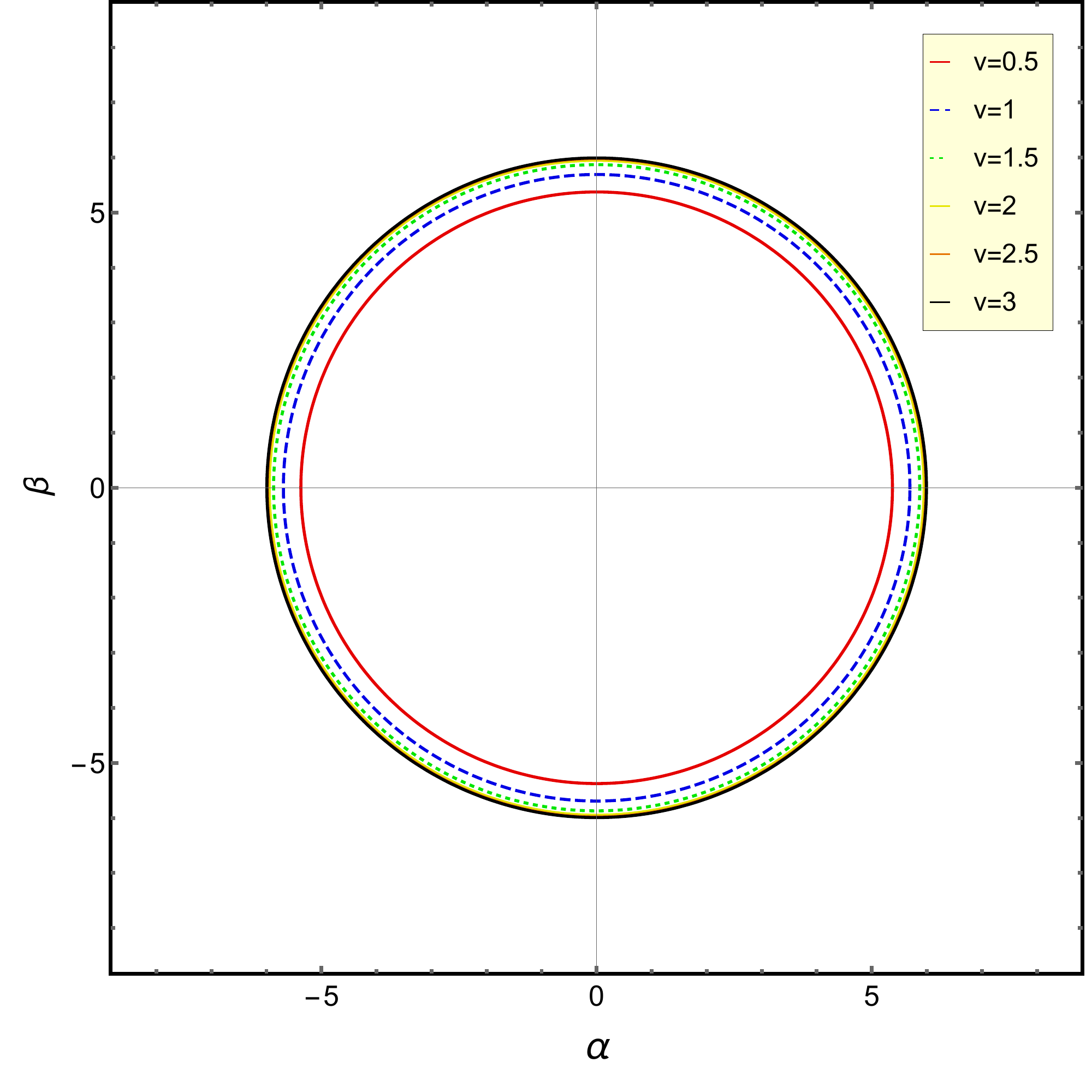} }}
    \qquad
    \subfloat{{\includegraphics[scale=0.35]{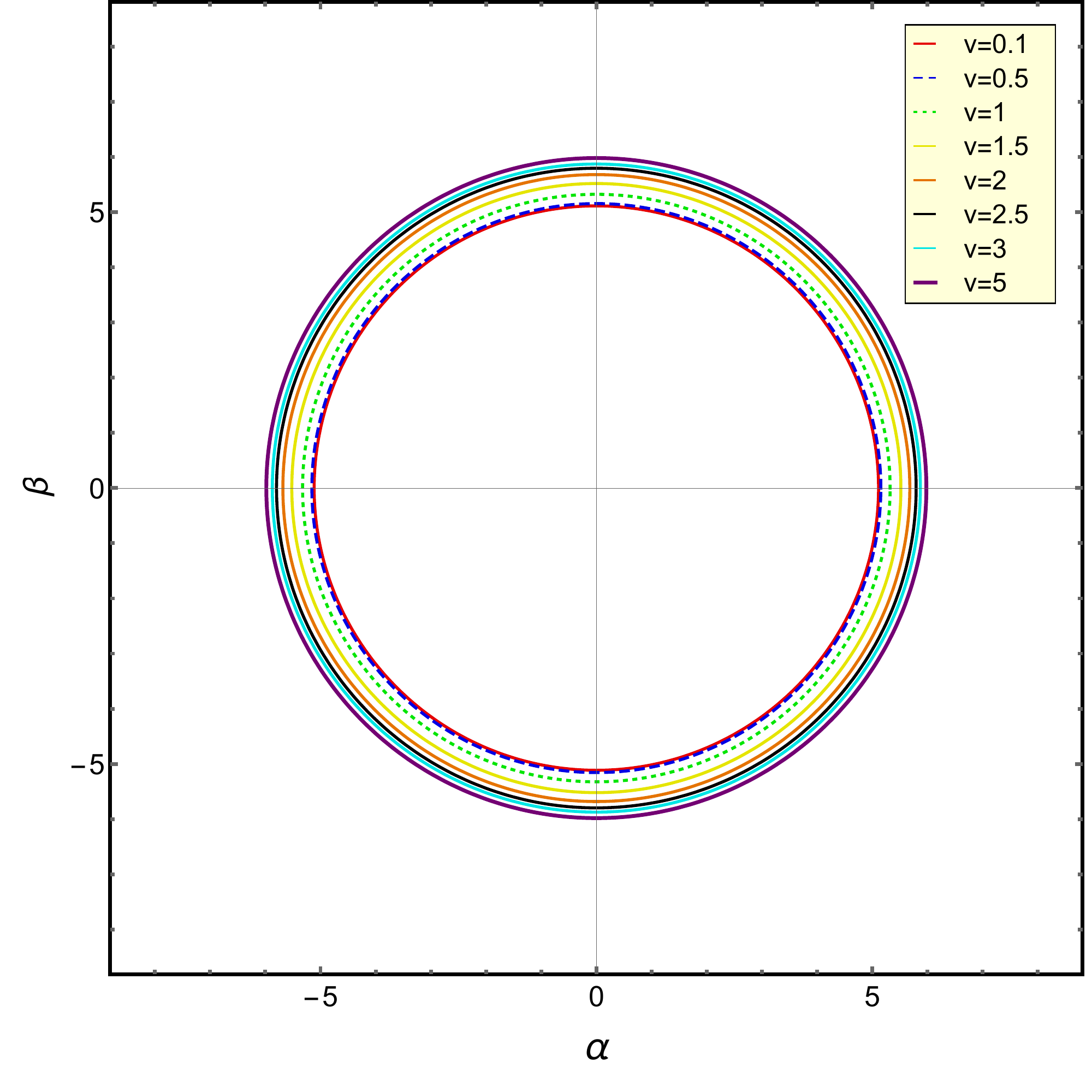} }}%
    
    \subfloat{{\includegraphics[scale=0.35]{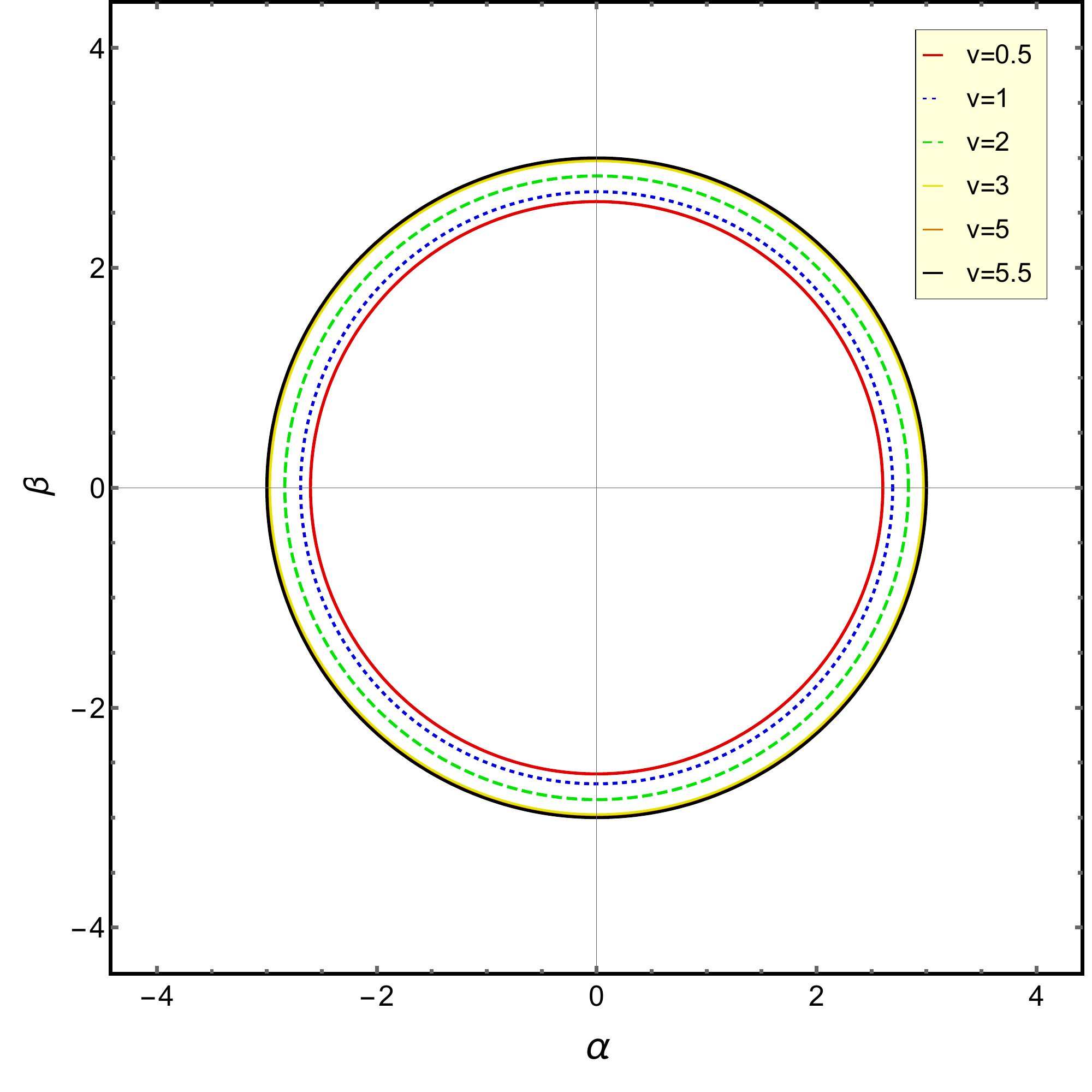} }}
    \qquad
    \subfloat{{\includegraphics[scale=0.35]{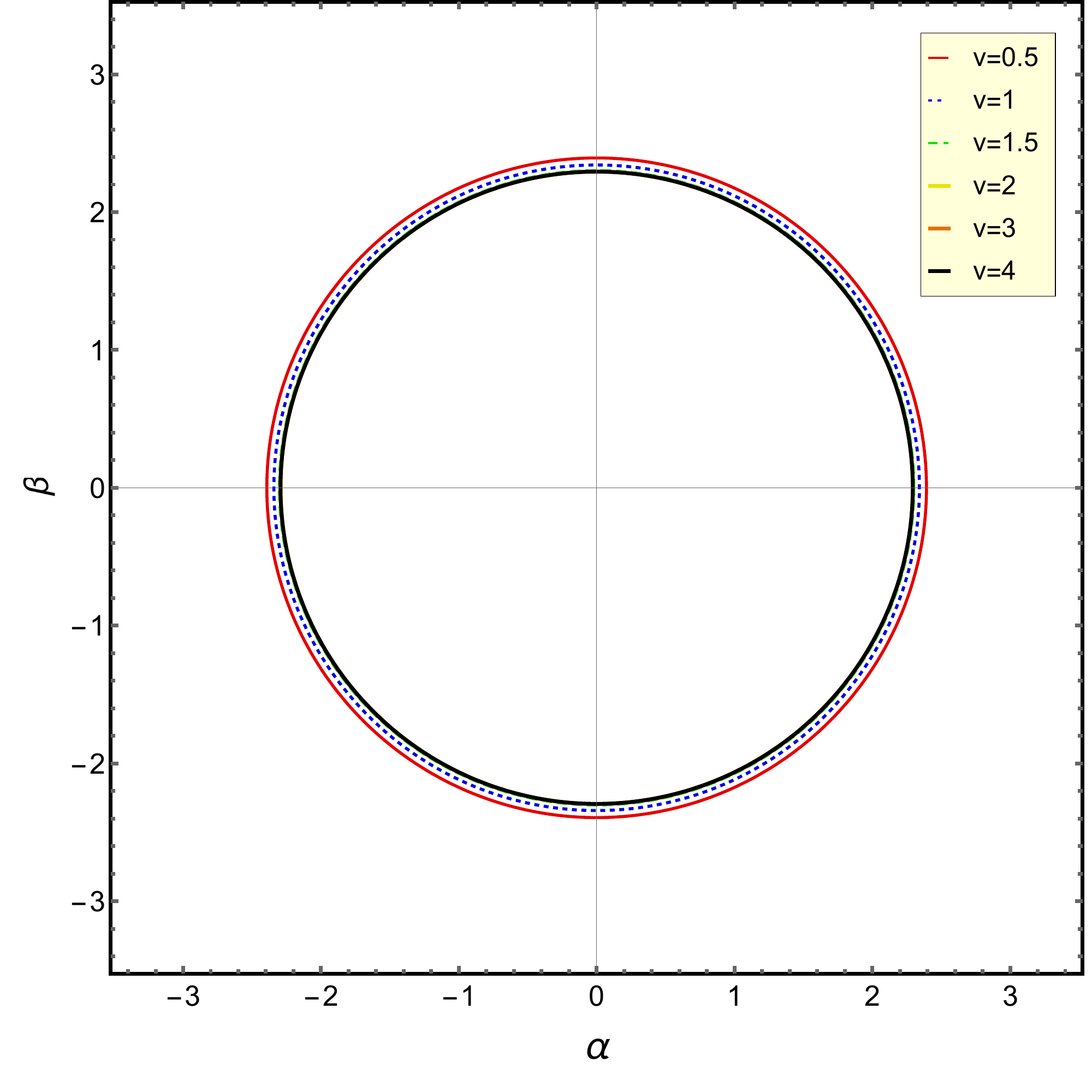} }}%
    \caption{This figure represents the time evolution of shadow casted by a spherically symmetric black hole w.r.t various mass and charge functions. The top left and right panel shows the evolution of shadow for mass functions $1+\tanh(v)$,$2 - \sech(v)$. In the bottom left panel we've plotted the evolution of shadow for the choice of $M(v) = 0.5[1+\tanh(v)], Q(v) = \frac{0.9}{2}[\tanh(v)-\tanh(v-1)]$. The bottom right panel shows the evolution of shadow for $M(v) =0.95+\frac{0.05}{2}[1+\tanh(v)], Q(v) = \frac{0.9}{2}[1+\tanh(v)]$.}\label{shadow_plots}
\end{figure}
\FloatBarrier

Where $\alpha$ and $\beta$ are the celestial coordinates that span the two-dimensional plane(known as the celestial plane) perpendicular to the line of sight w.r.t the observer and defined at spatial infinity\cite{Vazquez:2003zm}. More precisely, each ray of light that reaches the observer from a distant source corresponds to a point $(\alpha_0,\beta_0)$ in the celestial plane. 
The complement of these set of points in the $(\alpha,\beta)$ plane defines the shadow. For consistency, one might check that in the static limit, i.e., $\dot{r}_{\rm ph} =0$ we recover the expression of shadow around the static black hole. In \ref{shadow_plots}, we illustrate the location of shadow at a different instance of time for various choice of mass and charge profile.

\section{Kerr-Vaidya in the Slow Rotation Limit}\label{Section_7}

In this section, we would like to understand the evolution of the circular photon orbit as well as that of black hole shadow for a dynamical black hole with rotation. The situation we will consider in this work corresponds to the Kerr-Vaidya metric. However, a general computation is difficult in this case, since in the Hamilton-Jacobi formalism, the radial and angular part does not get separated. This is because, the dynamical Kerr-Vaidya spacetime has only one Killing field, related to the angular coordinate $\phi$. This prompts us to consider the slow rotation limit\cite{PhysRevD.1.3220}, where such a separation is achievable and under this assumption, we will discuss the evolution of the circular photon orbit along with the nature of shadow it casts.

\subsection{Evolution of Photon Circular Orbit}\label{Section_7.1}
The main aim of this section is to provide the desired equation governing the evolution of the photon circular orbit on the equatorial plane for Kerr-Vaidya metric. As emphasized earlier, for arbitrary values of the rotation parameter it is very difficult to determine the governing differential equation. Thus we will concentrate on the situation in which the rotation parameter is constant and small, so that terms of $\mathcal{O}(a^{2})$ can be neglected. In this slow rotation limit the Kerr-Vaidya metric on the equatorial plane takes the following form\cite{PhysRevD.1.3220},
\begin{align}\label{Photon_Sphere_S_28}
ds^{2}=-\left(1-\frac{2M(v)}{r}\right)dv^{2}+2dv\,dr-2a\,dr\,d\phi-\frac{4M(v)a}{r}dvd\phi
+r^{2}d\phi ^{2}~.
\end{align}
Since we are interested in the motion of a photon on the equatorial plane, we have substituted $\theta=\pi/2$ in the general metric ansatz to arrive at \ref{Photon_Sphere_S_28}. As a first step towards determining the trajectory of the photon on the equatorial plane one requires to set $ds^{2}=0$. This results into the following differential equation,
\begin{align}\label{Photon_Sphere_S_29}
r^{2}\left(\frac{d\phi}{dv}\right)^{2}
+\left(\frac{d\phi}{d\phi}\right)\left\{-2a\left(\frac{dr}{dv}\right)-\frac{4M(v)a}{r} \right\}+2\left(\frac{dr}{dv}\right)-\left(1-\frac{2M(v)}{r}\right)=0~.
\end{align}
The above result holds true for any null trajectory, geodesic or not. However, we are interested in the null geodesics on the equatorial plane of the Kerr-Vaidya solution, for which it is important to write down the corresponding Lagrangian, which takes the following form,
\begin{align}\label{Photon_Sphere_S_30}
L=-\frac{1}{2}\left\{1-\frac{2M(v)}{r}\right\}\left(\frac{dv}{d\lambda}\right)^{2}+\left(\frac{dv}{d\lambda}\right)\left(\frac{dr}{d\lambda}\right)-a\left(\frac{dr}{d\lambda}\right)\left(\frac{d\phi}{d\lambda}\right)
-\frac{2M(v)a}{r}\left(\frac{dv}{d\lambda}\right)\left(\frac{d\phi}{d\lambda}\right)
+\frac{1}{2}r^{2}\left(\frac{d\phi}{d\lambda}\right)^{2}
\end{align}
Here $\lambda$ is the affine parameter associated with the null geodesics. Given the above Lagrangian one can immediately compute various derivatives of the Lagrangian, resulting into the following geodesic equations,
\begin{align}
\frac{d^{2}v}{d\lambda ^{2}}-a\frac{d^{2}\phi}{d\lambda ^{2}}
&=-\frac{1}{r^{2}}M(v)\left(\frac{dv}{d\lambda}\right)^{2}
+\frac{2aM(v)}{r^{2}}\left(\frac{dv}{d\lambda}\right)\left(\frac{d\phi}{d\lambda}\right)
+r\left(\frac{d\phi}{d\lambda}\right)^{2}
\label{Photon_Sphere_S_33a}
\\
\frac{d^{2}r}{d\lambda ^{2}}-\left(1-\frac{2m(v)}{r}\right)\frac{d^{2}v}{d\lambda ^{2}}&-\frac{2aM(v)}{r}\frac{d^{2}\phi}{d\lambda ^{2}}
=-\frac{1}{r}\frac{dM}{dv}\left(\frac{dv}{d\lambda}\right)^{2}
+\frac{2M(v)}{r^{2}}\left(\frac{dv}{d\lambda}\right)\left(\frac{dr}{d\lambda}\right)
-\frac{2M(v)a}{r^{2}}\left(\frac{dr}{d\lambda}\right)\left(\frac{d\phi}{d\lambda}\right)
\label{Photon_Sphere_S_33b}
\\
r^{2}\frac{d^{2}\phi}{d\lambda ^{2}}-a\frac{d^{2}r}{d\lambda ^{2}}
&-\frac{2M(v)a}{r}\frac{d^{2}v}{d\lambda ^{2}}
=\frac{2a}{r}\frac{dM}{dv}\left(\frac{dv}{d\lambda}\right)^{2}
-\frac{2M(v)a}{r^{2}}\left(\frac{dv}{d\lambda}\right)\left(\frac{dr}{d\lambda}\right)
-2r\left(\frac{dr}{d\lambda}\right)\left(\frac{d\phi}{d\lambda}\right)
\label{Photon_Sphere_S_33c}
\end{align}

At this stage it is important to remind us of our goal, which is to construct an equation involving double derivative of $r$ with respect to the in-going null coordinate $v$, which will not involve terms like $\ddot{\phi}$. Keeping this in mind, we first eliminate the term involving $(d^{2}r/d\lambda ^{2})$ in \ref{Photon_Sphere_S_33b} and \ref{Photon_Sphere_S_33c} to arrive at,
\begin{align}\label{Photon_Sphere_S_34}
-a\frac{d^{2}v}{d\lambda ^{2}}+\left(r^{2}-\frac{2M(v)a^{2}}{r}\right)\frac{d^{2}\phi}{d\lambda ^{2}}-\frac{a}{r}\frac{dM}{dv}\left(\frac{dv}{d\lambda}\right)^{2}+\left(2r+\frac{2Ma^{2}}{r^{2}}\right)\left(\frac{dr}{d\lambda}\right)\left(\frac{d\phi}{d\lambda}\right)=0
\end{align}
Further, eliminating $(d^{2}\phi/d\lambda ^{2})$ between \ref{Photon_Sphere_S_34} and \ref{Photon_Sphere_S_33a}, we obtain the following differential equation for $v(\lambda)$ at the lowest order in the rotation parameter,
\begin{align}\label{Photon_Sphere_S_35}
\frac{d^{2}v}{d\lambda ^{2}}+\frac{M(v)}{r^{2}}\left(\frac{dv}{d\lambda}\right)^{2}
-\frac{2M(v)a}{r^{2}}\left(\frac{dv}{d\lambda}\right)\left(\frac{d\phi}{d\lambda}\right)
-r\left(\frac{d\phi}{d\lambda}\right)^{2}
+\frac{2a}{r}\left(\frac{dr}{d\lambda}\right)\left(\frac{d\phi}{d\lambda}\right)=0
\end{align}
From the above equation it is straightforward to read off the expression for $(d^{2}v/d\lambda ^{2})$, which as substituted in \ref{Photon_Sphere_S_33a}, yields,
\begin{align}\label{Photon_Sphere_S_36}
a\frac{d^{2}\phi}{d\lambda ^{2}}=-\frac{2a}{r}\left(\frac{dr}{d\lambda}\right)\left(\frac{d\phi}{d\lambda}\right)
\end{align}
Finally, we can use both the expressions for $(d^{2}v/d\lambda ^{2})$ along with $(d^{2}\phi/d\lambda ^{2})$ from \ref{Photon_Sphere_S_35} and \ref{Photon_Sphere_S_36} in order to rewrite \ref{Photon_Sphere_S_33b} in the following form,
\begin{align}\label{Photon_Sphere_S_37}
\frac{d^{2}r}{d\lambda ^{2}}&=\left\{-\frac{M(v)}{r^{2}}\left(1-\frac{2M(v)}{r}\right)-\frac{1}{r}\frac{dM}{dv} \right\}\left(\frac{dv}{d\lambda}\right)^{2}+\frac{2M}{r^{2}}\left(\frac{dr}{d\lambda}\right)\left(\frac{dv}{d\lambda}\right)-\frac{2M(v)a}{r^{2}}\left(\frac{dr}{d\lambda}\right)\left(\frac{d\phi}{d\lambda}\right)
\nonumber
\\
&+\frac{2M(v)a}{r^{2}}\left(1-\frac{2M(v)}{r}\right)\left(\frac{dv}{d\lambda}\right)\left(\frac{d\phi}{d\lambda}\right)+\left(r-2M\right)\left(\frac{d\phi}{d\lambda}\right)^{2}-\frac{2a}{r}\left(\frac{dr}{d\lambda}\right)\left(\frac{d\phi}{d\lambda}\right)
\end{align}
Note that as desired, the above equation does not involve any factors of $(d^{2}v/d\lambda ^{2})$ or $(d^{2}\phi/d\lambda ^{2})$. Thus one can easily change the variable of differentiation from the affine parameter $\lambda$ to the in-going null coordinate $v$, such that $(d^{2}r/d\lambda ^{2})=(dv/d\lambda)^{2}(d^{2}r/dv^{2})+(dr/dv)(d^{2}v/d\lambda ^{2})$. Performing this transformation of variable we finally arrive at the following differential equation for the photon circular orbit on the equatorial plane $r_{\rm ph}=r_{\rm ph}(v)$,
\begin{align}\label{Photon_Sphere_S_38}
\frac{d^{2}r_{\rm ph}(v)}{dv^{2}}&+\left\{-\frac{3M}{r_{\rm ph}(v)^{2}}\frac{dr_{\rm ph}(v)}{dv}+\frac{M}{r_{\rm ph}(v)^{2}}\left(1-\frac{2M(v)}{r_{\rm ph}(v)}\right)+\frac{1}{r_{\rm ph}(v)}\frac{dM}{dv}\right\}
+\left\{r_{\rm ph}(v)\left(\frac{dr_{\rm ph}(v)}{dv}\right)-\left[r_{\rm ph}(v)-2M\right] \right\}\left(\frac{d\phi}{dv}\right)^{2}
\nonumber
\\
&+\left\{\frac{4M(v)a}{r_{\rm ph}(v)^{2}}\frac{dr_{\rm ph}(v)}{dv}-\frac{2a}{r_{\rm ph}(v)}\left(\frac{dr_{\rm ph}(v)}{dv}\right)^{2}
+\frac{2a}{r_{\rm ph}(v)}\left(\frac{dr_{\rm ph}(v)}{dv}\right)-\frac{2Ma}{r_{\rm ph}(v)^{2}}\left(1-\frac{2M(v)}{r_{\rm ph}(v)}\right)\right\}\frac{d\phi}{dv}=0
\end{align}
However, the above equation is not sufficient to determine the evolution of the photon circular orbit, as the differential equation depends on the $(d\phi/dv)$ term as well. Thus we need to determine $(d\phi/dv)$ in terms of $r$ and $(dr/dv)$, which can be derived separately from \ref{Photon_Sphere_S_29}. This yields,
\begin{align}\label{Photon_Sphere_S_39}
\frac{d\phi}{dv}=\frac{a}{r_{\rm ph}(v)^{2}}\frac{dr_{\rm ph}(v)}{dv}+\frac{2Ma}{r_{\rm ph}(v)^{3}}\pm \frac{1}{r_{\rm ph}(v)}\sqrt{1-\frac{2M}{r_{\rm ph}(v)}-2\left(\frac{dr_{\rm ph}(v)}{dv}\right)}
\end{align}
Thus one need to solve both \ref{Photon_Sphere_S_39} and \ref{Photon_Sphere_S_38} simultaneously in order to determine the evolution of the photon circular orbit $r_{\rm ph}=r_{\rm ph}(v)$. Combining both of these equations and keeping terms up to linear order in the rotation parameter, i.e., neglecting terms depending on $\mathcal{O}(a^{2})$, we obtain the following final form for the evolution equation of the photon circular orbit in the slow rotation limit,
\begin{align}\label{Kerr_PS_Eqn}
\ddot{r}_{\rm ph}(v)+& \frac{3\dot{r}_{\rm ph(v)}}{r_{\rm ph}(v)} + \frac{\dot{M}(v)}{r_{\rm ph}(v)} -\frac{9 M(v)\dot{r}_{\rm ph}(v)}{r_{\rm ph}(v)^2} - \frac{2 \dot{r}_{\rm ph}(v)^2}{r_{\rm ph}(v)}-\frac{1}{r_{\rm ph}(v)} + \frac{5 M(v)}{r_{\rm ph}(v)^2} - \frac{6 M(v)^2}{r_{\rm ph}(v)^3} 
\nonumber
\\
&\pm \frac{6a}{r_{\rm ph}(v)^3}\left[ \left(2M(v)^2 - M(v)r_{\rm ph}(v)+2M(v)r_{\rm ph}(v)\dot{r}_{\rm ph}(v)\right) \sqrt{\frac{r_{\rm ph}(v)-2M(v)-2r_{\rm ph}(v) \dot{r}_{\rm ph}(v)}{r_{\rm ph}(v)^3}}\right]+ \mathcal{O}(a^2) = 0
\end{align}
This is the differential equation governing evolution of the photon circular orbit on the equatorial plane. Thus in order to determine the location of the photon circular orbit we need to solve the above equation with the boundary condition, that for mass functions $M(v)$, asymptotoing to a finite value, $r_{\rm ph}(v)$ at late times must coincide with the photon circular orbits of Kerr black hole. This must hold for both the retrograde orbit and the prograde orbit. 

Moreover, in the slow rotation limit, one can work out the matter stress tensor responsible for the evolution of the black hole mass with time. For this purpose, we point out the non vanishing components of the Ricci tensor for Kerr-Vaidya metric, which reads, $R_{vv} =(2\dot{m}/r^2)$ and $R_{v\phi}=(3\dot{m}a/r^2)$. Interesting the Ricci scalar associated with the Kerr-Vaidya metric is $\mathcal{O}(a^{2})$ and hence does not contribute in the slow rotation limit. Thus in this context the matter stress tensor has only $(v,v)$ and $(v,\phi)$ as the non vanishing components. It turns out that in the slow-rotation limit there exists a vector, $v_i = (1,0,0,(3/2)a\sin^2\theta)$ such that $v_i v^i\sim \mathcal{O}(a^2)\sim 0$ and the stress-energy tensor can be written as\cite{PhysRevD.1.3220}, 
\begin{equation}
T_{ij} = \frac{\dot{M}}{4\pi r^2}v_i v_j
\end{equation} 
Thus up to linear order in the rotation parameter, the Kerr-Vaidya model indeed represents a rotating black hole accreting null fluid and hence just like the \RNV spacetime the evolution of photon sphere in this slowly rotating Kerr-Vaidya geometry is intimately connected to the null energy condition. 

\begin{figure}[h]
    \centering
    \subfloat{{\includegraphics[scale=0.31]{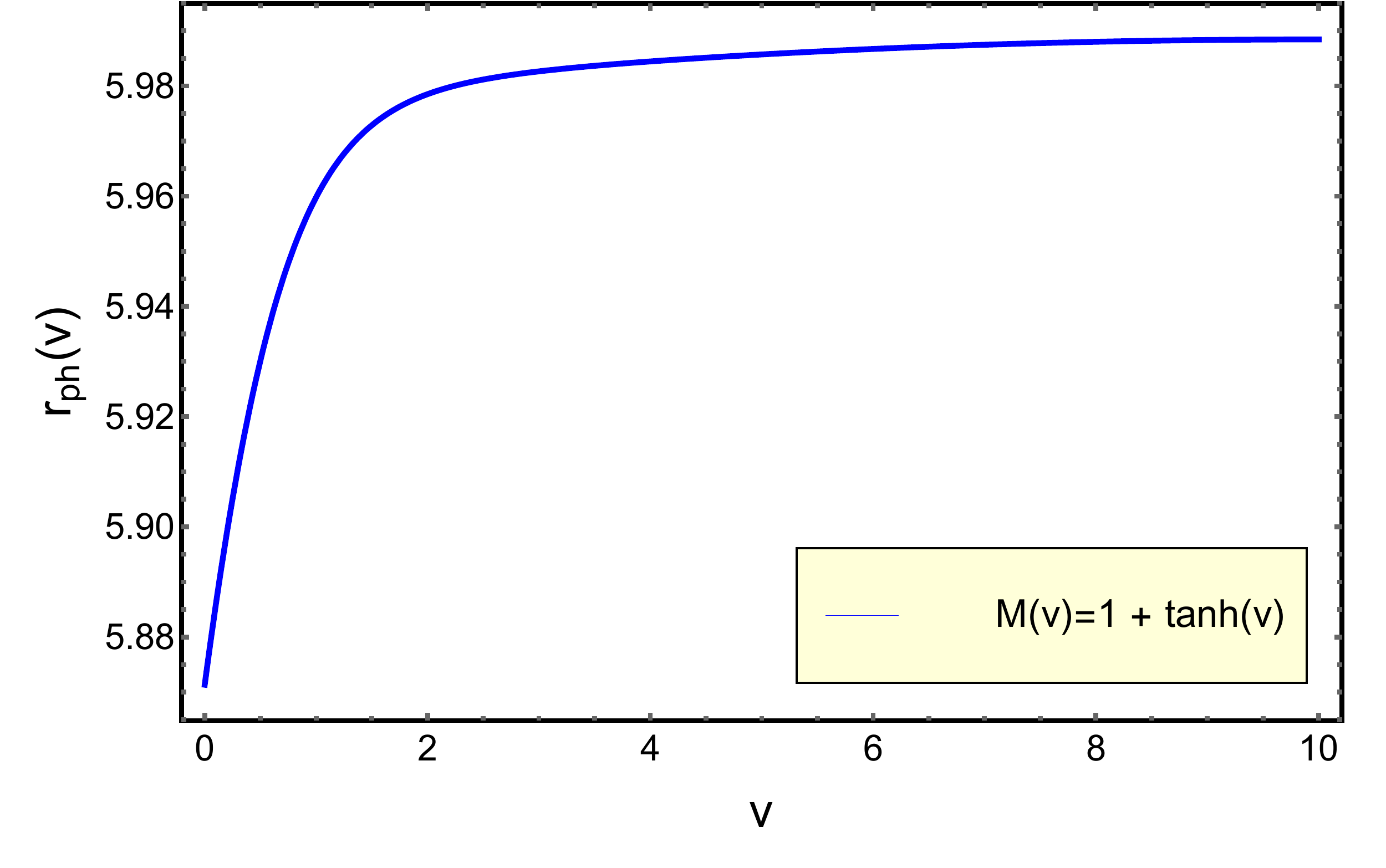} }}
    \qquad
    \subfloat{{\includegraphics[scale=0.31]{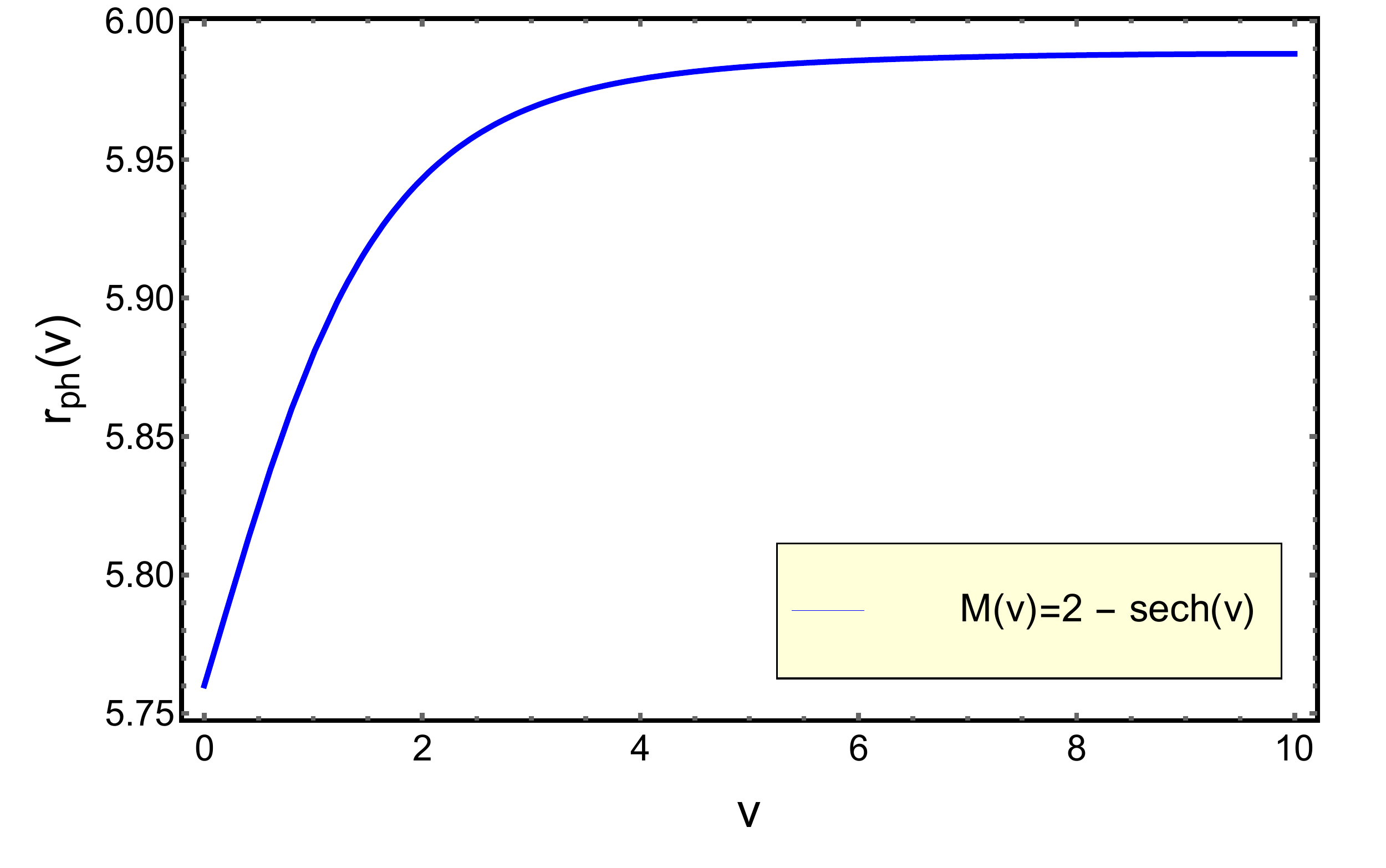} }}%
    	
    \caption{This figure illustrates the evolution of prograde photon orbits for the choice of mass functions $M(v)=1+\tanh(v)$[left panel] and $M(v)=2-\sech(v)$[right panel]. The rotation parameter $`a$' has been chosen to be $0.01$. }.\label{prograde_orbit_plot}
\end{figure}

To illustrate our result by solving \ref{Kerr_PS_Eqn} numerically we need to impose appropriate boundary conditions. Such boundary conditions may be determined for various choices of smoothly increasing mass functions, which asymptotes to a constant value for the mass parameter. In this case for prograde photon orbit one may set the boundary conditions to be $r_{\rm ph}(v\rightarrow \infty)= r_{\rm -}$ and $\dot{r}_{\rm ph}(v\rightarrow \infty)=0$, where $r_{\mp}$ correspond to the location of the photon circular orbit for prograde (or, retrograde) motion in stationary context\cite{Bardeen:1972fi,Misner1973Gravitation}. With these boundary conditions, the growth behavior of the photon sphere has been obtained for mass functions satisfying the above criteria, which has been presented in \ref{prograde_orbit_plot}.
\FloatBarrier
\subsection{Shadow Casted by Slowly-Rotating Kerr-Vaidya Black Hole}\label{Section_7.2}

\begin{figure}[h]
    \centering
    \subfloat{{\includegraphics[scale=0.4]{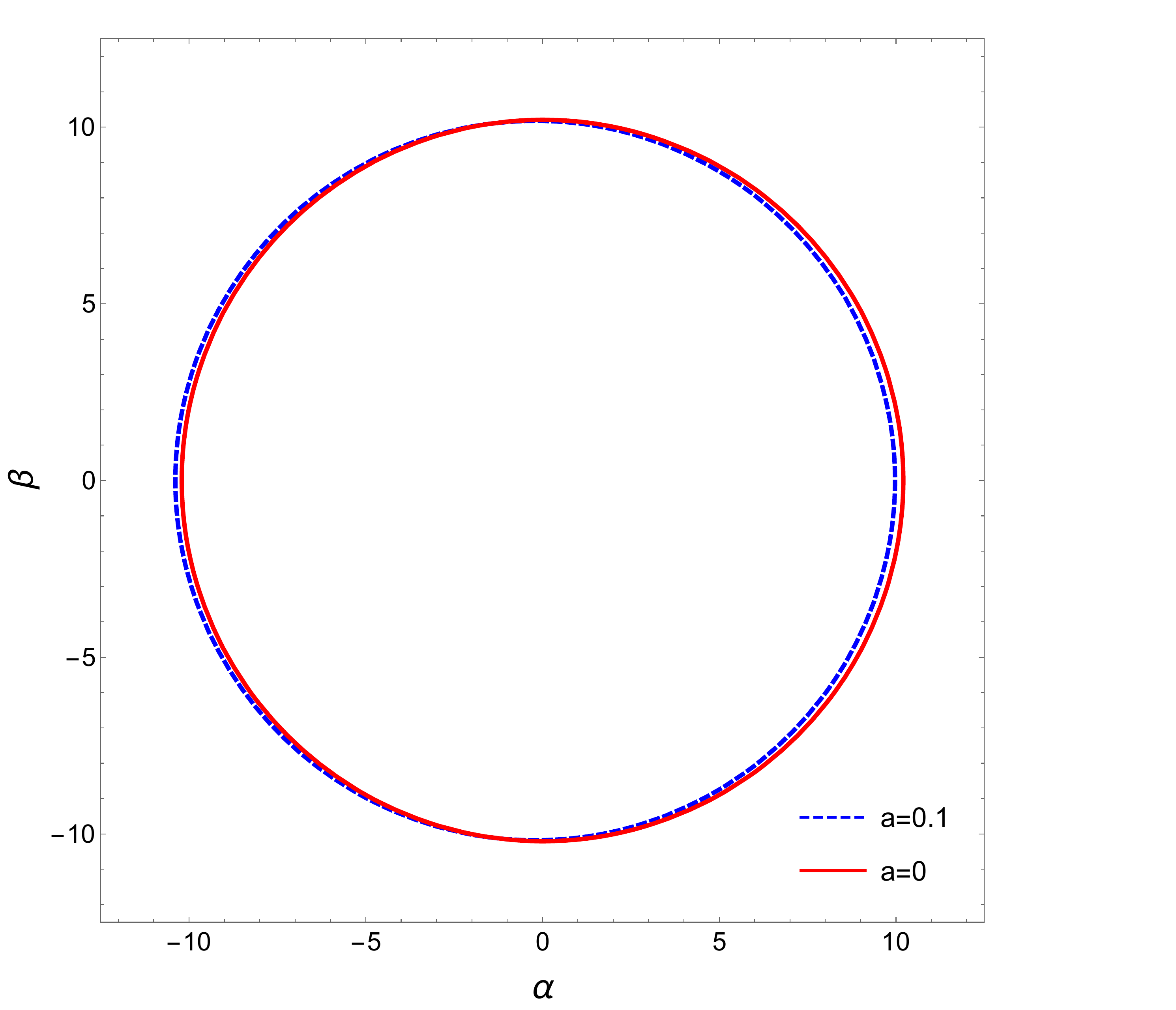} }}

    \caption{This figure presents the comparison between shadow casted by a Kerr-Vaidya black hole with rotation parameter $a=0.1$ to that of a Vaidya black hole$(a=0)$ for the choice of mass function $M(v) = 1+\tanh(v)$ at the in-going time $v=2$. Note that, the shadow of slowly rotating Kerr black hole is still spherical, but with centre shifted}\label{kerr_vs_sch_shadow}
\end{figure}

In the previous section, we had determined the differential equation governing the evolution of the photon orbits on the equatorial plane. Following which, in this section, we will obtain an expression for the shadow region of slowly rotating dynamical black hole, described by the metric ansatz written down in \ref{Photon_Sphere_S_28}. As in the case of spherically symmetric spacetime, here also we start with the Lagrangian of a particle moving in a Kerr background, which up to $\mathcal{O}(a)$ can be written as,
\begin{align}
\mathcal{L}=\frac{1}{2}\Bigg\{-f(r,v)\,\left(\frac{dv}{d\lambda}\right)^2 
&+2\,\left(\frac{dv}{d\lambda}\right)\left(\frac{dr}{d\lambda}\right)  
-2\, a \sin^2\theta\, \left(\frac{dr}{d\lambda}\right)\left(\frac{d\phi}{d\lambda}\right) 
\nonumber
\\
&-\frac{4Ma}{r}\sin^2\theta\,\left(\frac{dv}{d\lambda}\right)\left(\frac{d\phi}{d\lambda}\right) 
+ r^2 \left(\frac{d\theta}{d\lambda}\right)^2 
+ r^2 \sin^2\theta\, \left(\frac{d\phi}{d\lambda}\right)^2\Bigg\}
\end{align}
As evident, the above Lagrangian is independent of the angular coordinate $\phi$, resulting into a conserved angular momentum $L$. However, in the dynamical context, there is no conserved energy. Still we can introduce a quantity $E$, which in this context is dependent on both $r$ and $v$, such that 
\begin{align}
E(r,v)&=\frac{\partial\mathcal{L}}{\partial(dv/d\lambda)} = -f(r,v)\left(\frac{dv}{d\lambda}\right) 
+\left(\frac{dr}{d\lambda}\right) 
-\frac{2 M a}{r} \sin^2\theta \left(\frac{d\phi}{d\lambda}\right) \label{energy}
\\
L&=\frac{\partial\mathcal{L}}{\partial(d\phi/d\lambda)}=-a \sin^2\theta 
-\frac{2 M a}{r} \sin^2\theta\left(\frac{dv}{d\lambda}\right)
+r^2 \sin^2\theta \left(\frac{d\phi}{d\lambda}\right)\label{angular_momentum}
\end{align}
Note that the variation of $E(r,v)$ is determined by the equation $(dE/d\lambda)=(\partial \mathcal{L}/\partial v)$. In the static situation the Lagrangian is independent of $v$ and hence $E(r,v)$ is a constant of motion. From the above two equations, i.e., from \ref{energy} and \ref{angular_momentum} we can immediately solve for $(dv/d\lambda)$ and $(d\phi/d\lambda)$ leading to the following expressions,
\begin{align}
\frac{dv}{d\lambda}&= \frac{(dr/d\lambda)-E(r,v)}{f(r,v)} - \frac{2 M a L}{f(r,v) \, r^3}\label{v_dot}
\\
\frac{d\phi}{d\lambda}&= \frac{L}{r^2 \sin^2\theta}+\frac{a}{r^2}\left(\frac{dr}{d\lambda}\right) 
+\frac{2M a\left[(dr/d\lambda)-E(r,v)\right]}{f(r,v) r^3}\label{phi_dot}
\end{align}
Since we are interested in the trajectory of photons, we can set the line element to be vanishing, which in turn results into $\mathcal{L}$ being zero. Then in that expression we can use \ref{v_dot} and \ref{phi_dot} to express everything in terms of $(dr/d\lambda)$, $(d\theta/d\lambda)$, $E(r,v)$ and $L$. Hence up to $\mathcal{O}(a)$, we obtain, 
\begin{equation}
r^2 \left(\frac{dr}{d\lambda}\right)^2 
-r^2 E(r,v)\left[E(r,v) + \frac{4 M a L}{r^3}\right] + r^4 f(r,v)\left(\frac{d\theta}{d\lambda}\right)^{2} + \frac{f(r,v) L^2}{\sin^2\theta} = 0 \label{r_theta_mixed}
\end{equation}
The above equation can be easily separated into the radial and angular part by introducing the Carter Constant $K$\cite{Carter:1968rr}, such that the angular part takes the following form, 
\begin{equation}
r^4\left(\frac{d\theta}{d\lambda}\right)^{2}=K-L^2\cot^2\theta 
\end{equation}
Substituting this in \ref{r_theta_mixed} and re-arranging terms, we obtain the radial equation to be,
\begin{equation}
\left(\frac{dr}{d\lambda}\right)^2 = E(r,v)\left[E(r,v)+\frac{4M a L}{r^3}\right] - \frac{f(r,v)}{r^2}(K + L^2)\label{r_dot_square}
\end{equation}
In the static case, for circular orbit is at a fixed radius and one would set $r=r_{\rm ph}$ and $\dot{r}=0$. For the dynamical case, the radius of circular orbit would be dependent on the ingoing null coordinate $v$, we have $r = r_{\rm ph}(v)$ and hence $(dr/d\lambda)=\dot{r}_{\rm ph}(v)(dv/d\lambda)$. Since we want to use it in \ref{r_dot_square}, we need an expression for $(dv/d\lambda)^{2}$, which can be obtained from \ref{v_dot} 
\begin{equation}
\left(\frac{dr}{d\lambda}\right)^2= \frac{E(r_{\rm ph},v)^2}{[f(r_{\rm ph},v)-\dot{r}_{\rm ph}(v)]^2} + \frac{4MaL\,E(r_{\rm ph},v)}{r_{\rm ph}(v)^3[f(r_{\rm ph},v) - \dot{r}_{\rm ph}(v)]^2} \label{v_dot_square}
\end{equation}
where we have kept terms up to $\mathcal{O}(a)$. Finally substituting this result in \ref{r_dot_square} we obtain,
\begin{equation}
\frac{f(r_p,v)}{r_p(v)^2}\Big\{\eta+\xi^2\Big\} = 1 + \frac{4 M a}{r_p(v)^3}\xi\left[ 1- \frac{\dot{r}_p(v)^2}{[f(r_p,v)-\dot{r}(v)]^2}\right] - \frac{\dot{r}_p(v)^2}{[f(r_p,v)-\dot{r}_p(v)]^2}
\end{equation}
Here following the spherically symmetric scenario, we have introduced two new quantities, namely, $\eta = K/E(r,v)^2$ and $\xi = L/E(r,v)$. These quantities can be written down trivially in terms of the celestial coordinates $\alpha$ and $\beta$ introduced in the previous section, such that the last equation reduces to,
\begin{equation}
\alpha^2 + \beta^2 = \frac{r_p(v)^2}{f(r,v)}\left[1 - \frac{4 M a}{r_p(v)^3}\alpha 
\left( 1- \frac{\dot{r}_p(v)^2}{[f(r,v)-\dot{r}_p(v)]^2}\right) - \frac{\dot{r}_p(v)^2}{[f(r,v)-\dot{r}_(v)]^2}\right]
\label{Dyn_shadow_slow_rotn}
\end{equation}
The above equation represents the shadow of a Kerr-Vaidya black hole in the slow rotation limit. Note that, the shape of the shadow in this limit is still circular, as that of the spherically symmetric case. This is because of the fact that, in the slow rotation limit, the line element posses spherical symmetry, i.e., $r={\rm const}$ and $v={\rm const}$ surfaces are still sphere. However, because of the small but non-zero value of the rotation parameter $a$, the shadow of a slowly rotating Kerr-Vaidya and Vaidya black hole are indeed different, which is clearly depicted in \ref{kerr_vs_sch_shadow} for a reasonable choice of the mass function.
\FloatBarrier
It is easy to verify the consistency of the above expression, depicting black hole shadow in the slow rotation limit, to other situations discussed in the paper or in the literature. The first test of consistency corresponds to the non-rotating case, i.e., with $a=0$, where \ref{Dyn_shadow_slow_rotn} becomes,
\begin{equation}
\alpha^2 + \beta^2 = \frac{r_p(v)^2[f(r_p,v)-2\dot{r}_p(v)]}{[f(r_p,v) - \dot{r}_p(v)]^2}
\end{equation}
which exactly matches with \ref{spherical shadow}, the expression for the non-rotating dynamical black hole. Secondly for stationary slowly rotating case, we have $\dot{r}_{\rm ph}(v)=0$ and we have the following equation governing the photon sphere,
\begin{equation}
\alpha(v)^2 + \beta(v)^2 = \frac{r_p^2}{f(r_p)}\left(1 - \frac{4 M a}{r_p^3}\alpha \right)
\end{equation}
As one can immediately verify this exactly matches with the shadow of a stationary black hole in the slow rotation limit \cite{Tsukamoto:2017fxq}. Finally, it is possible to study the case for static non-rotating black hole, where both $\dot{r}_{\rm ph}(v)=0$ and $a=0$. In this limit also we recover the earlier result, i.e., $\alpha^2 + \beta^2 =(r_{\rm ph}^2/f(r_{\rm ph}))$.
\section{Effective Graviton Metric in Gauss-Bonnet Gravity}\label{Section_6}

Our discussion so far has been at the level of General Relativity, i.e., we've considered the time evolution of the photon sphere and shadow around dynamical black holes that are solutions of Einstein's equation. Now we would like to extend our analysis to theories beyond General Relativity. Particularly, we would be interested in the \LL correction, which is unique generalization over the Einstein-Hilbert action in dimensions higher than four, with the field equation containing at most second derivative of the metric\cite{Lovelock:1971yv,Zwiebach:1985uq,Boulware:1985wk,Zumino:1985dp}. \LL theories are interesting in many respect and possess some unique properties that are not present in GR. One such feature is the existence of superluminal propagating modes, and hence the issue of causality is far from obvious in \LL theories. 

It is well known that, in a theory of gravity where higher order curvature terms are considered, the gravitational degree of freedoms propagates at a different speed than that of the background ones. Here we shall refer the background metric as the photon metric, and it's correction due to the higher curvature terms as the effective graviton metric. Such results have been studied extensively by numerous authors in the static case\cite{Aragone:1987jm,ChoquetBruhat:1988dw,Izumi:2014loa,Reall:2014pwa,Papallo:2015rna,Reall:2014sla,Brustein:2017iet,Andrade:2016yzc}, by explicitly obtaining effective graviton metric. This is ultimately a consequence of the fact that, the causal structure of a system of PDE is determined by the characteristics hyper-surface, which turns out to be non-null for \LL theories\cite{Izumi:2014loa,Reall:2014sla}. As a result, graviton and photon attain different speed of propagation and consequently different radius of circular null orbit and shadow. In this section, we would like to understand how these results generalize to the dynamical context, which will enable us to compare the evolution of photon and graviton sphere. We start by presenting an explicit calculation of the graviton effective metric for five-dimensional Einstein-Gauss-Bonnet gravity(a Lovelock theory), which admits black hole solutions\cite{Boulware:1985wk,Wiltshire:1988uq}. For computational ease, without giving up any physical insights, we study the limiting case of small Gauss-Bonnet coupling constant and distinctly obtain the evolution of graviton and photon sphere. This analysis also enables us to understand the shadow cast by graviton and photon and their respective evolution. Let us start with the \LL Lagrangian,
\begin{equation}
\mathcal{L} = \sum_{k=0}^{k_{max}} \lambda_k \mathcal{L}_k
\end{equation}
where,
\begin{equation}
\mathcal{L}_k = \frac{1}{2^k} \delta^{aba_1b_1...a_kb_k}_{cdc_1d_1...c_kd_k} R_{ab}\,^{cd} R_{a_1b_1}\,^{c_1d_1}....R_{a_kb_k}\,^{c_kd_k}
\end{equation}

For such higher curvature corrections, the form of effective metric has been obtained(Eqn 2.24 of \cite{Brustein:2017iet}) for arbitrary order Lovelock terms. The strategy developed in \cite{Brustein:2017iet}, is to start with a background metric and study its tensor perturbation. The effective metric is then identified by looking for the coefficient of the second-order derivative of the transverse-traceless perturbation $h_{ab}$ in the linearized theory, which represents the gravitational degrees of freedom. 
For Gauss-Bonnet correction of Einstein-Hilbert Lagrangian, the effective metric takes the form,

\begin{equation}
\left[G^b\,_d\right] \nabla_b \nabla^d h_q^p = \left[(\delta^{pab}_{qcd} - \delta^p_q \delta^{ab}_{cd}) - \lambda_2(\delta^{paba_1b_1}_{qcdc_1d_1} R_{a_1b_1}\,^{c_1d_1} - \delta^p_q \delta^{aba_1b_1}_{cdc_1 d_1} R_{a_1b_1}\,^{c_1d_1})\right]\nabla_b \nabla^d h_a^c \label{effective metric_form}
\end{equation}
Note that, the first term in the right-hand side is the background metric $g^a_b$ and the second term, i.e., the coefficient of $\lambda_2$ corresponds to the Gauss-Bonnet correction.
In Ref. \cite{Brustein:2017iet}, the above form of the effective metric for graviton degree of freedom was obtained by assuming the background metric to be static. Here in our analysis, since we are interested in obtaining the effective graviton metric in the dynamical case, we start with the following background metric ansatz,

\begin{equation}
ds^2 = -f(r,v) dv^2 +2dv dr +r^2 d\Omega_{D-2}^2
\end{equation}
The non-vanishing component of the Riemann tensor for this line element are given by,

\begin{equation}
R_{vr}\,^{vr} = -\frac{f''(r,v)}{2}\label{Riemann_1}
\end{equation}

\begin{equation}
R_{ij}\,^{kl} = \frac{1-f(r,v)}{r^2}\delta_{ij}^{kl}\label{Riemann_2}
\end{equation}

\begin{equation}
R_{\alpha i}\,^{\alpha j} = -\frac{f'(r,v)}{2 r}\delta_{i}^{j}\label{Riemann_3}
\end{equation}

\begin{equation}
R_{vi}\,^{rj} = -\frac{\dot{f}(r,v)}{2r} \delta_i^j\label{dyn_extra}
\end{equation}
The indices $i,j,k,l, \rm{etc} = 1,2,......(D-2)$ denotes the angular coordinates, while $\alpha=(v,r)$. \ref{dyn_extra} represents the additional contribution to the Riemann tensor due to the time dependence of the metric. Now by following an identical line of calculation in\cite{Brustein:2017iet},  we can  obtain various components of the effective graviton metric\footnote{For a complete derivation refer to \ref{Appendix_B}.}, i.e.,
\begin{equation}
G^v_v = 1 - 2\lambda_2\left[(D-4) \left(\frac{f'(r,v)}{r}\right) -\, (D-4)(D-5) \left(\frac{1-f(r,v)}{r^2}\right)\right]\label{eff_vup_vdn}
\end{equation}

\begin{equation}
G^r_v = 2\lambda_2 (D-4) \frac{\dot{f}(r,v)}{r}\label{eff_rv}
\end{equation}
And, the other components of the effective metric remains the same as that of the static case derived in \cite{Brustein:2017iet}. Note that, $G_{vv} = G^v_v g_{vv} + G^r_v g_{rv}$ and $G_{rv} = G^v_v $. Therefore, the effective graviton metric finally takes the form,

\begin{equation}
ds^2_{eff} = G_{vv} dv^2 + 2 G_{rv} dv dr + G_{ij} dx^i dx^j
\end{equation}
In our analysis, we are interested in the graviton circular null orbit, for which case we have to work with the condition $ds^2_{eff} = 0$. This further allows us to divide the line element by $G_{rv}$ and write the effective graviton metric in a somewhat simplified and more intuitive form, i.e.,

\begin{equation}
ds^2_{eff} = \frac{G_{vv}}{G_{rv}} dv^2 + 2  dv dr + \frac{G_{ij}}{G_{rv}} dx^i dx^j
\end{equation}
Using \ref{eff_vup_vdn} and \ref{eff_rv}, and defining $\alpha = \lambda_2 (D-4)(D-3)$, this finally reduces to,

\begin{equation}
ds^2_{eff} = -\left(f(r,v) - \frac{2 \alpha r \dot{f}(r,v)}{(D-3)r^2 + 2\alpha\left[ (1-f(r,v))(D-5) - r f'(r,v) \right]} \right) dv^2 + 2\,  dv\, dr + g(r,v)\, d\Omega_{D-2}^2\label{final_eff_metric}
\end{equation}
where $g(r,v) = G_{ij}/G^v_v$. In five-dimensions, the effective metric takes the following form,

\begin{equation}
ds^2_{eff} = -\left[f(r,v)- \frac{\alpha\dot{f}(r,v)}{r-\alpha\, f'(r,v)} \right] dv^2 + 2  dv dr + \left(\frac{1-\alpha f''(r,v)}{1-\frac{\alpha f'(r,v)}{r}}\right)\, d\Omega_{3}^2\label{final_eff_metric_5D}
\end{equation}
The above expression of the effective graviton metric is analogous to \ref{Dynamical_in} and allow us to obtain the evolution of the radius of graviton circular null orbit by proceeding in a similar approach developed in \ref{Section_2} as that of the photon. Again for consistency, one might check that, in the static limit we have $\dot{f}(r,v) = 0$ and the photon and graviton event horizon coincides but they possess different radius of the circular null orbit, which is in agreement with all earlier results\cite{Reall:2014pwa,Papallo:2015rna,Brustein:2017iet}. With \ref{final_eff_metric_5D} as the effective graviton metric, we are now set to obtain the evolution of graviton sphere for the various choice of mass functions.

\subsection{Photon Vs. Graviton Sphere}\label{Section_8.1}
\begin{figure}[h]
    \centering
    \subfloat{{\includegraphics[scale=0.34]{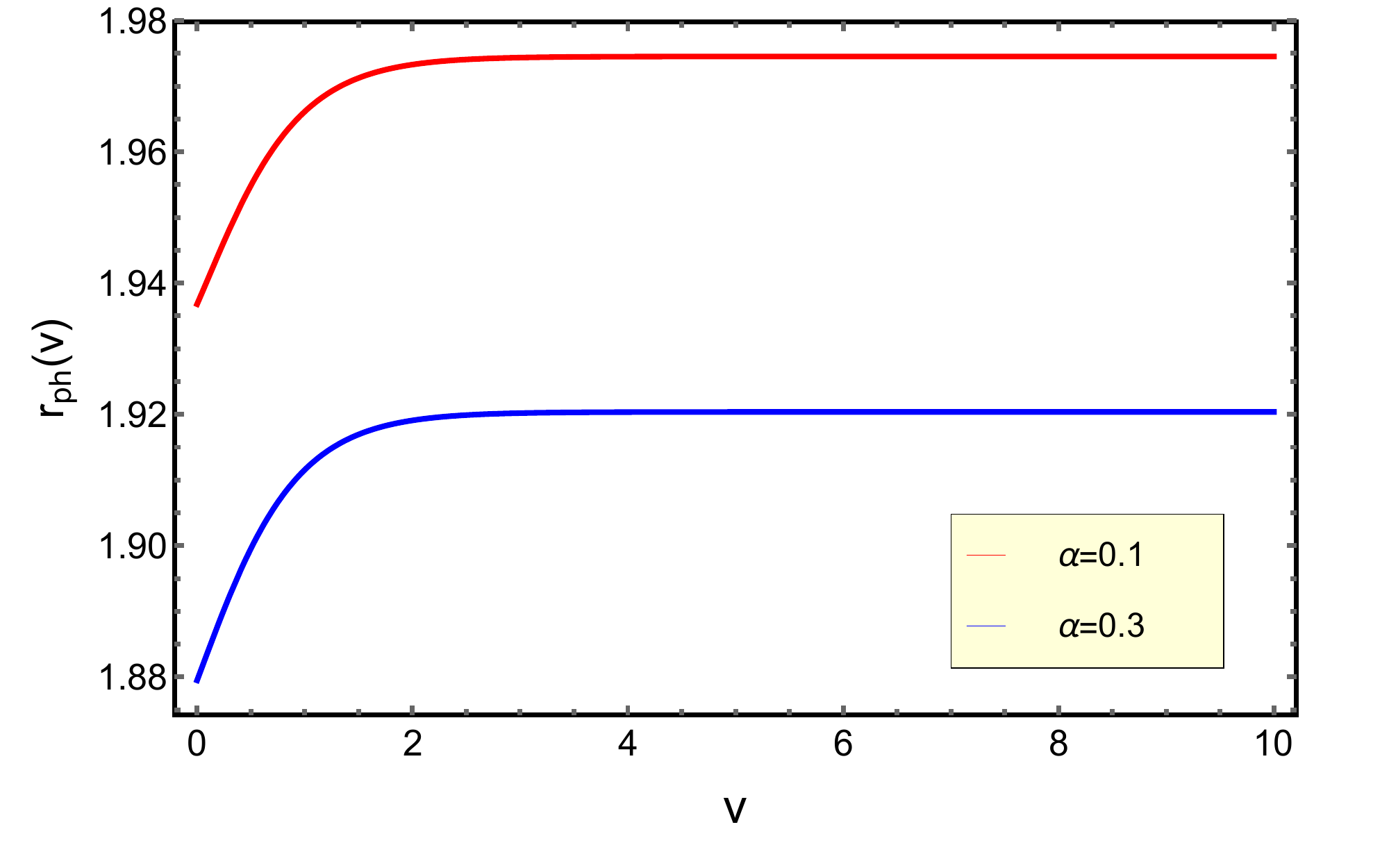} }}
    \qquad
    \subfloat{{\includegraphics[scale=0.34]{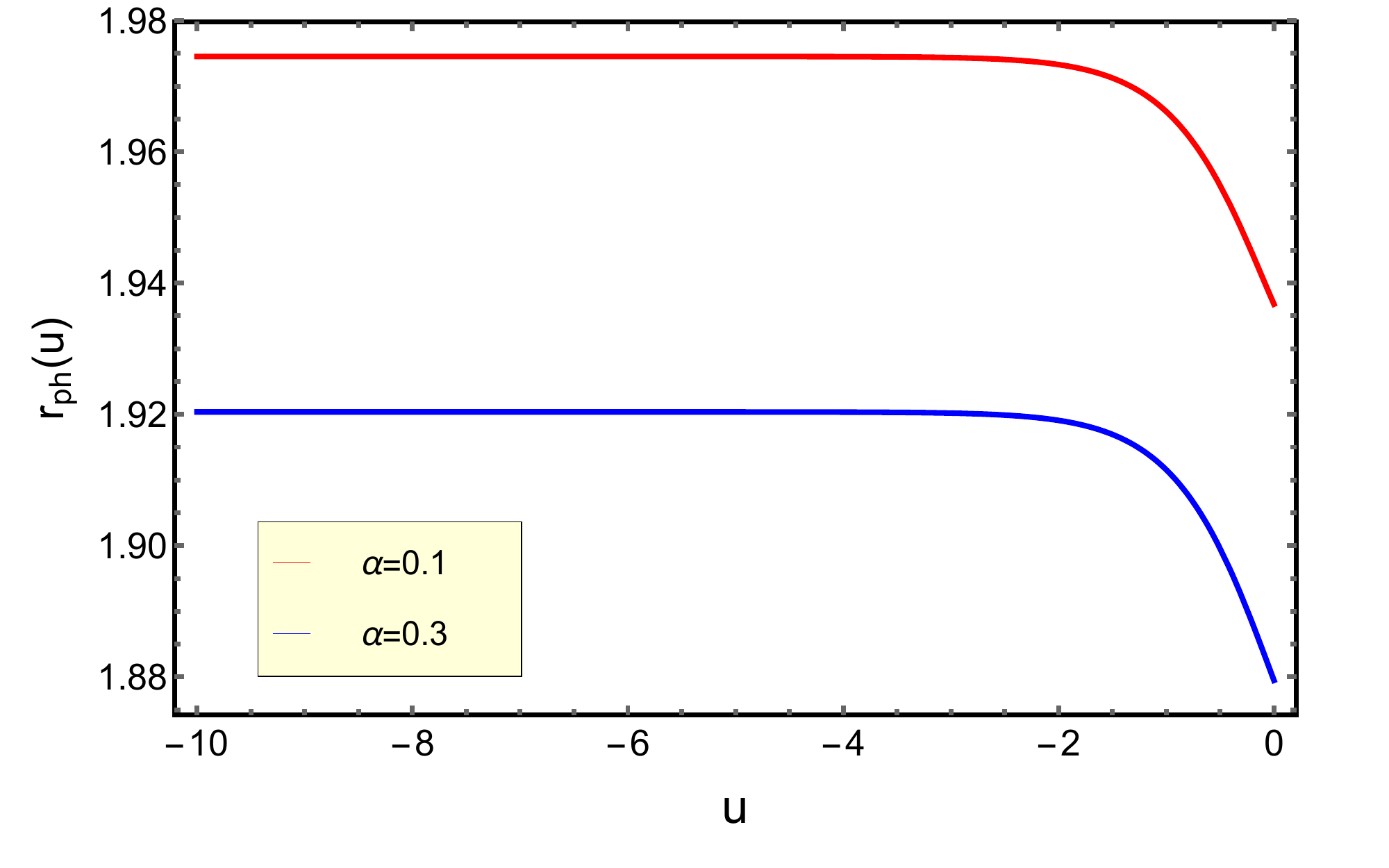} }}%
    \quad
    \subfloat{{\includegraphics[scale=0.31]{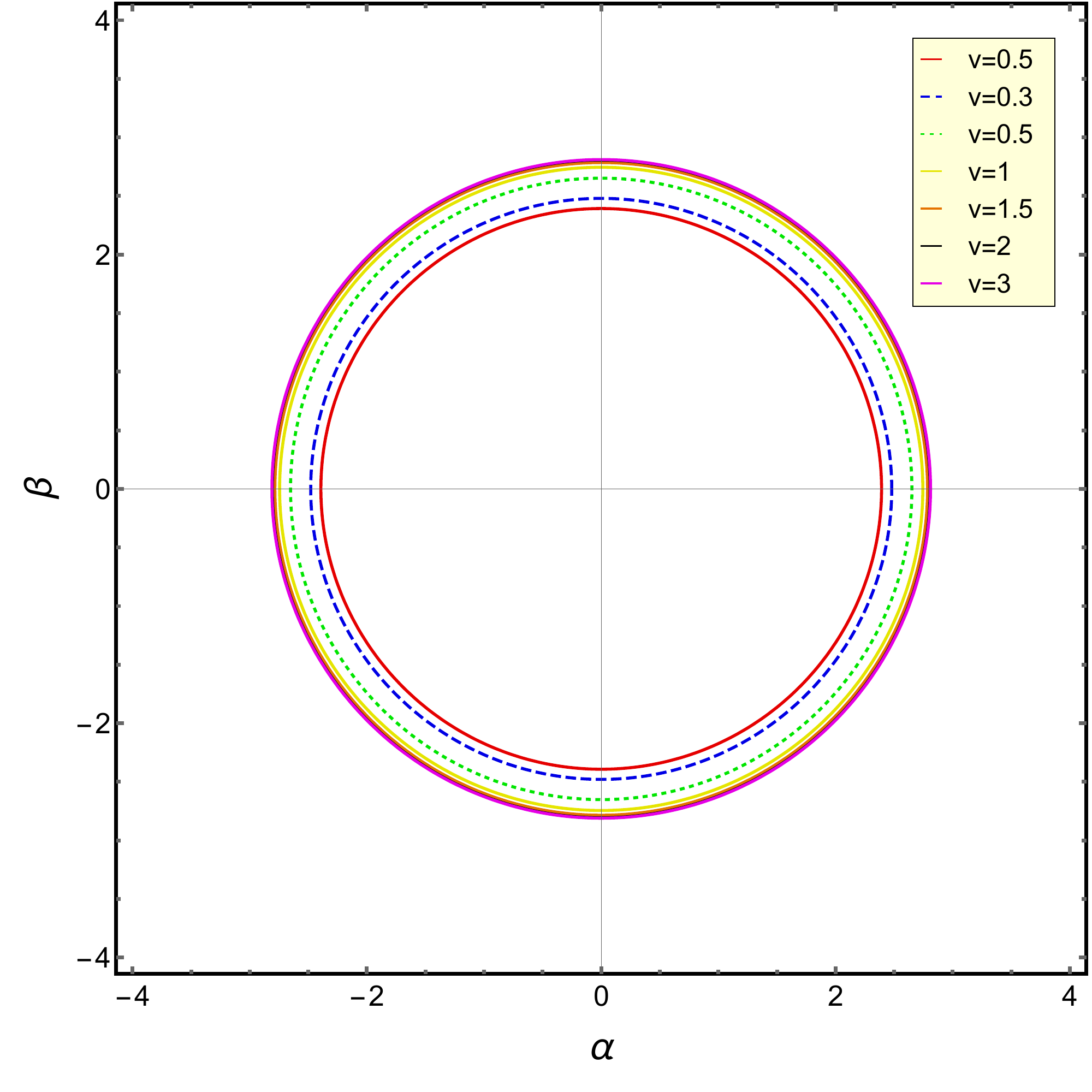} }}
    \qquad
    \subfloat{{\includegraphics[scale=0.31]{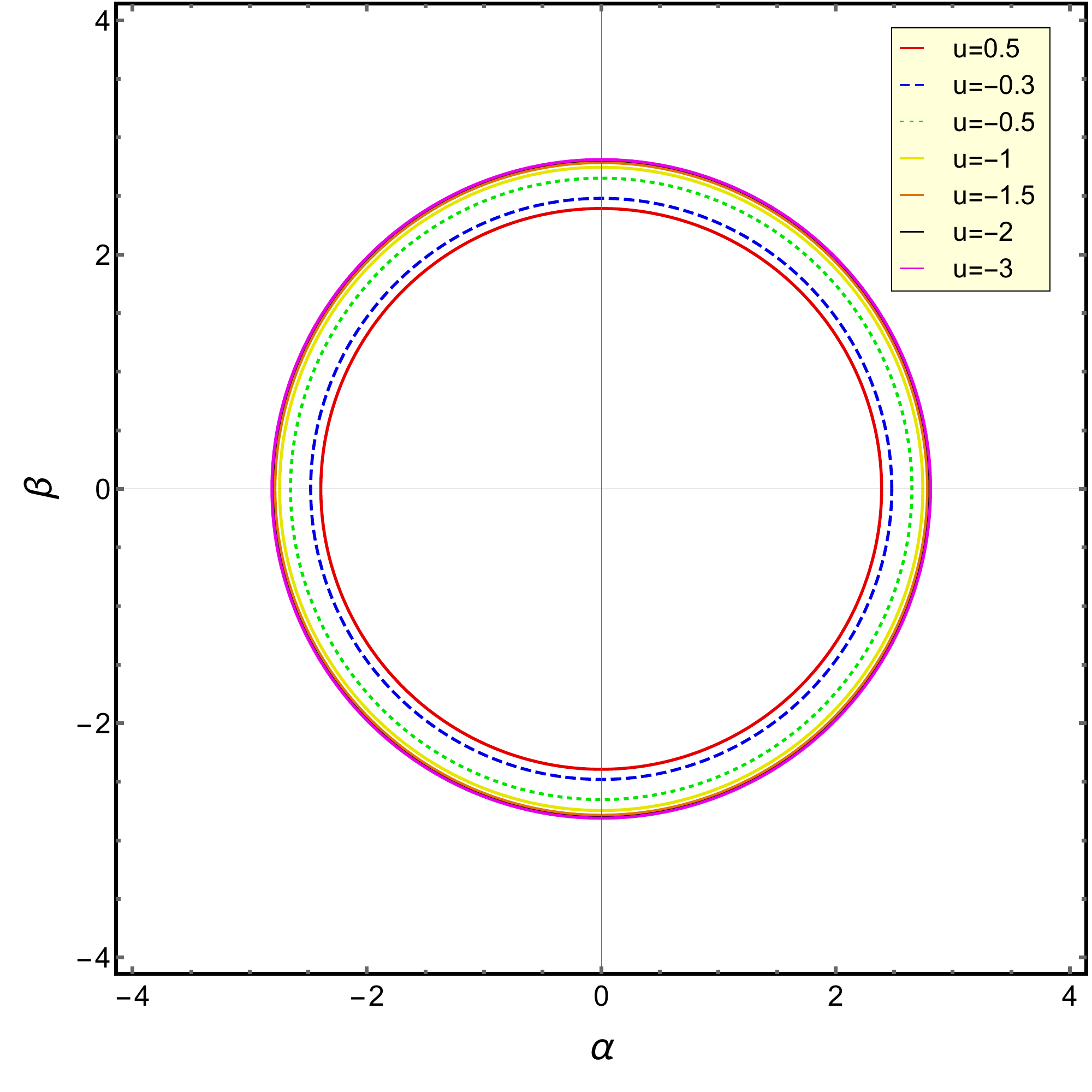} }}%
    \caption{This figure illustrates the evolution of the photon sphere and shadow around a five-dimensional Gauss-Bonnet black hole for the various choice of coupling constant $\alpha$. The top left and right panel shows the evolution of photon sphere for $m(v) = 1+\tanh(v)$ and $m(v)=1-\tanh(u)$ respectively. In the bottom left and right panel we have plotted the corresponding evolution of shadow. }\label{GB_PS}
\end{figure}

Before addressing the more complicated case of graviton sphere, which requires some special care, first, we would like to study the time evolution of the photon sphere in five-dimensional Einstein-Gauss-Bonnet theory. Gauss-Bonnet term represents the quadratic order \LL correction to general relativity. Such a theory admits spherically symmetric black hole solution, which in terms of the in-going coordinate has the form\cite{Boulware:1985wk,PhysRevD.38.2434,Wheeler:1985nh},

\begin{equation}
ds^2 = -f(r,v) dv^2 + 2 dv dr + r^2 d\Omega_3^2
\end{equation}
where,
\begin{equation}
f(r,v) = 1+ \frac{r^2}{2\alpha}\left( 1-\sqrt{1+\frac{4\alpha M(v)}{r^4}}\right)\label{f_EGB}
\end{equation}
Here $d\Omega_3^2 = d\theta^2 +\sin^2\theta \,d\phi^2 + \,\sin^2\theta\,\sin^2\phi\, d\psi^2$, represents the volume of the three-dimensional sphere spanned by the angular coordinates ($\theta,\phi,\psi$) and $\alpha$ is the Gauss-Bonnet coupling constant. 
Note that, by virtue of the spherical symmetry, one can always set $\theta=\psi =\pi/2$ and restrict only to the equatorial orbits for which the evolution equation is of the same form as \ref{Photon_Sp_Eqn_in}, with $f(r,v)$ replaced by \ref{f_EGB}. For acreeting matter, i.e., increasing mass, we we solve this differential equation w.r.t the future boundary conditions $r_{\rm ph}(v\rightarrow\infty) =\sqrt{2}(M^2 - M\,\alpha)^{1/4}$ and $\dot{r}_{\rm ph}(v\rightarrow\infty)=0$ to obtain the time evolution of the radius of the photon sphere around a Einstein-Gauss-Bonnet black hole. Similarly, for the outgoing coordinate we solve \ref{Photon Sp Eqn_out} w.r.t the boundary conditions $r_{\rm ph}(u\rightarrow -\infty) =\sqrt{2}(M^2 - M\,\alpha)^{1/4}$ and $\dot{r}_{\rm ph}(u\rightarrow -\infty)=0$. For the evolution of shadow we use \ref{spherical shadow} with the choice of $f(r,v)$ in \ref{f_EGB}. The results are illustrated in \ref{GB_PS}.

\begin{figure}[h]
    \centering
    \subfloat{{\includegraphics[scale=0.375]{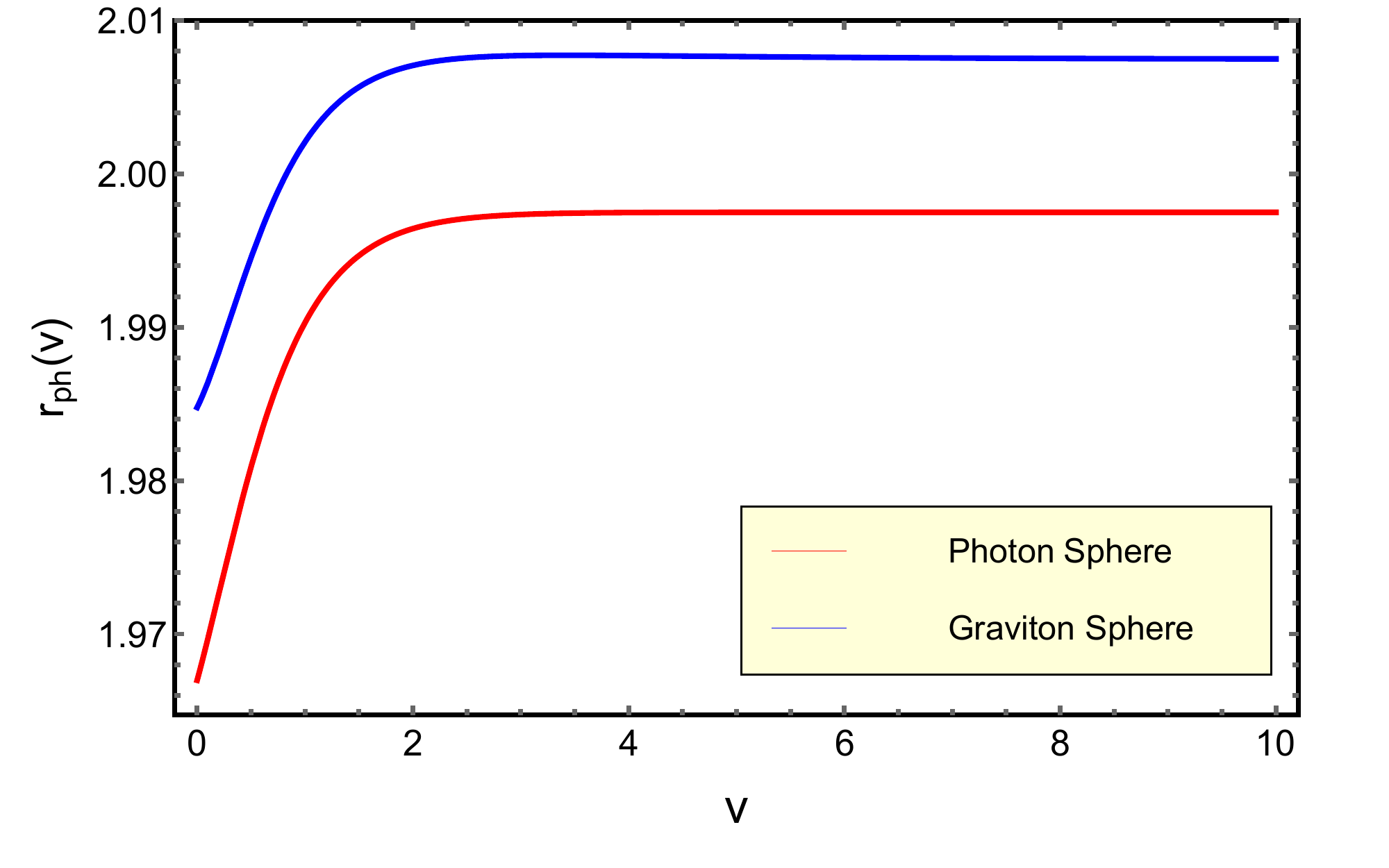} }}
    \qquad
    \subfloat{{\includegraphics[scale=0.375]{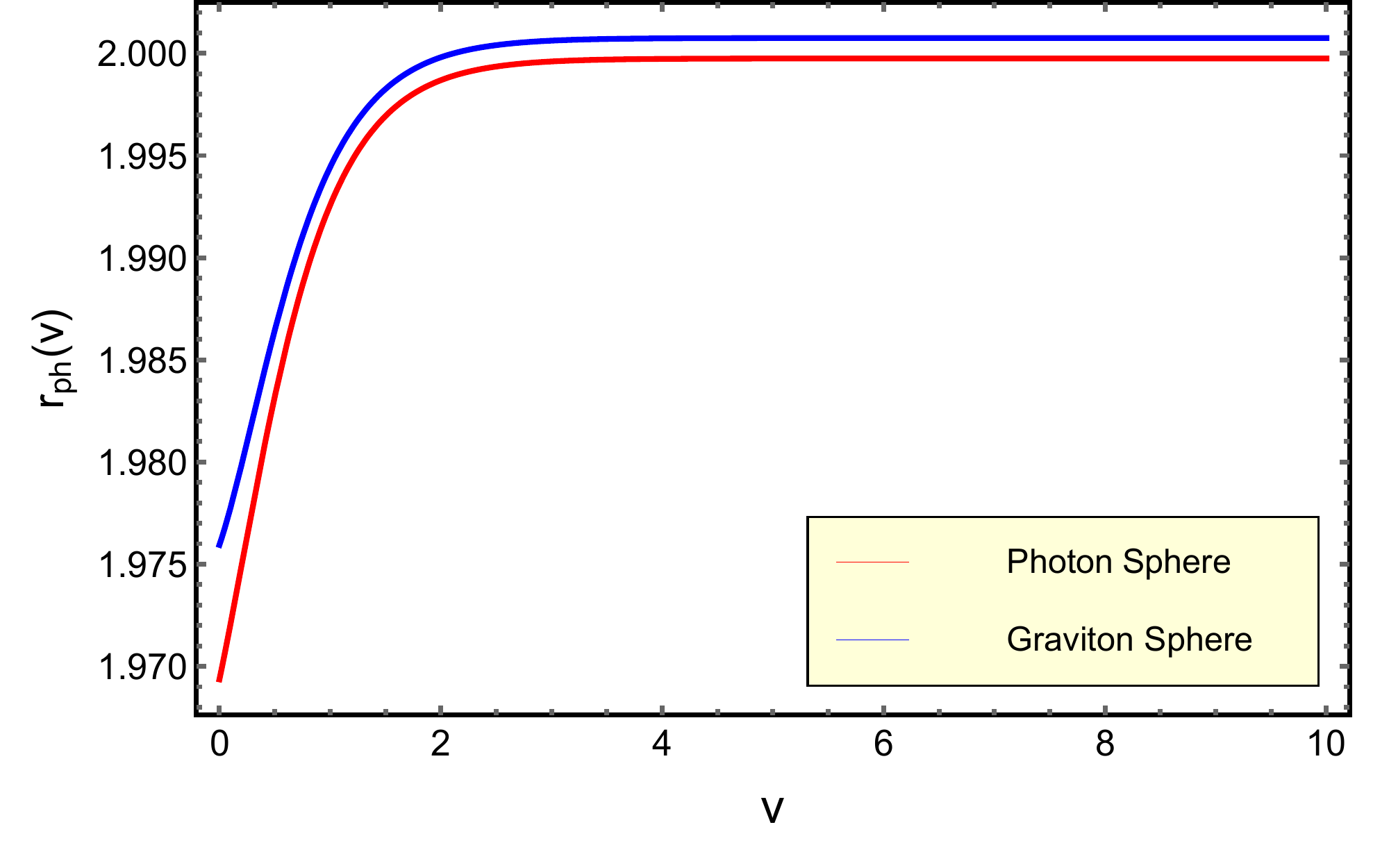} }}%
    \quad
	\subfloat{{\includegraphics[scale=0.375]{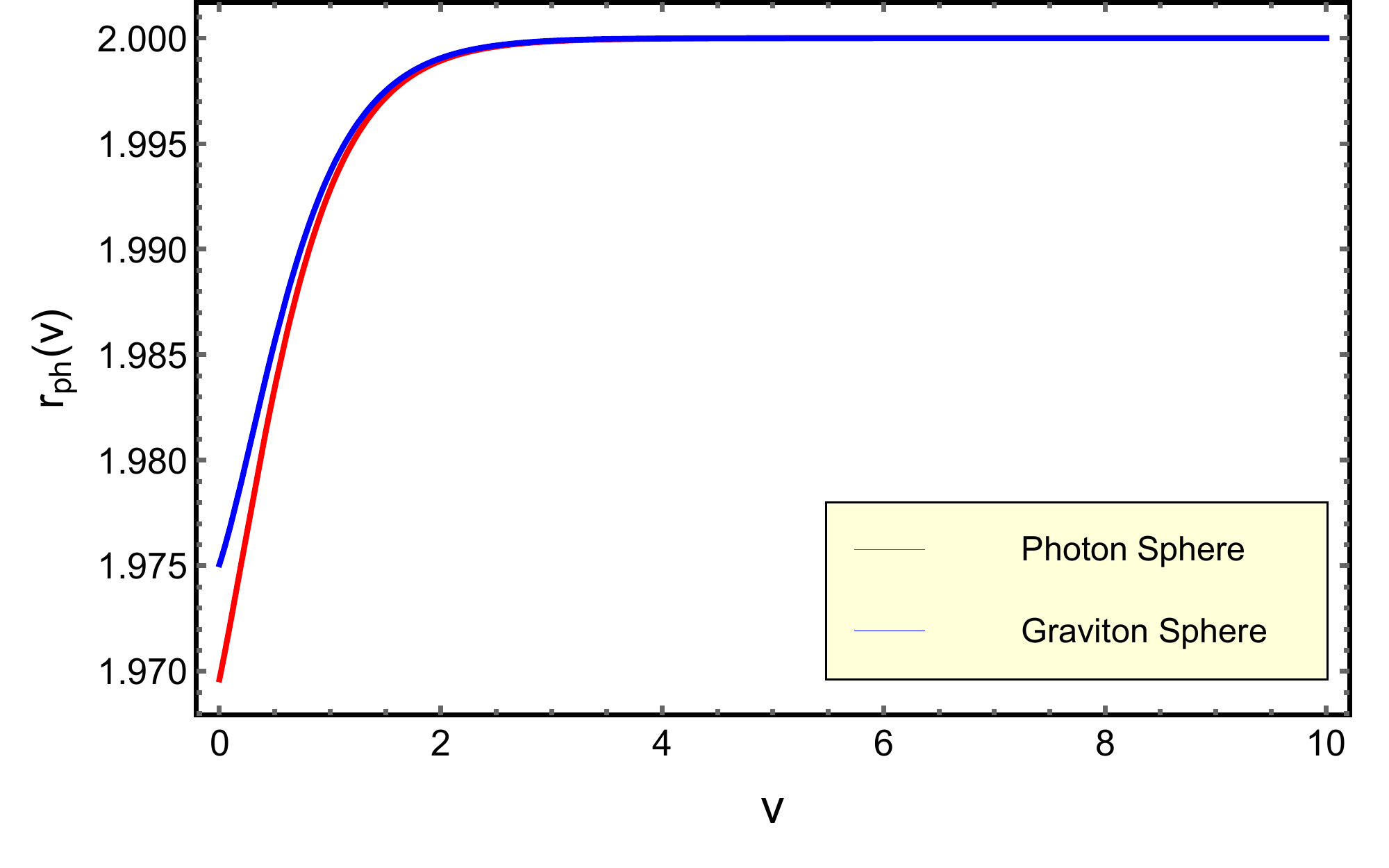} }}
   
    \caption{In this figure we compare the evolution of photon and graviton sphere around a five-dimensional Gauss-Bonnet black hole for the choice of mass function $M(v)=1+\tanh(v)$ and the coupling constant $\alpha =0.01(\text{top left}),0.001(\text{top right}),0.00001(\text{bottom})$ respectvelly.}\label{PSvsGS}
\end{figure}

\begin{figure}[h]
\centering
\includegraphics[scale=0.5]{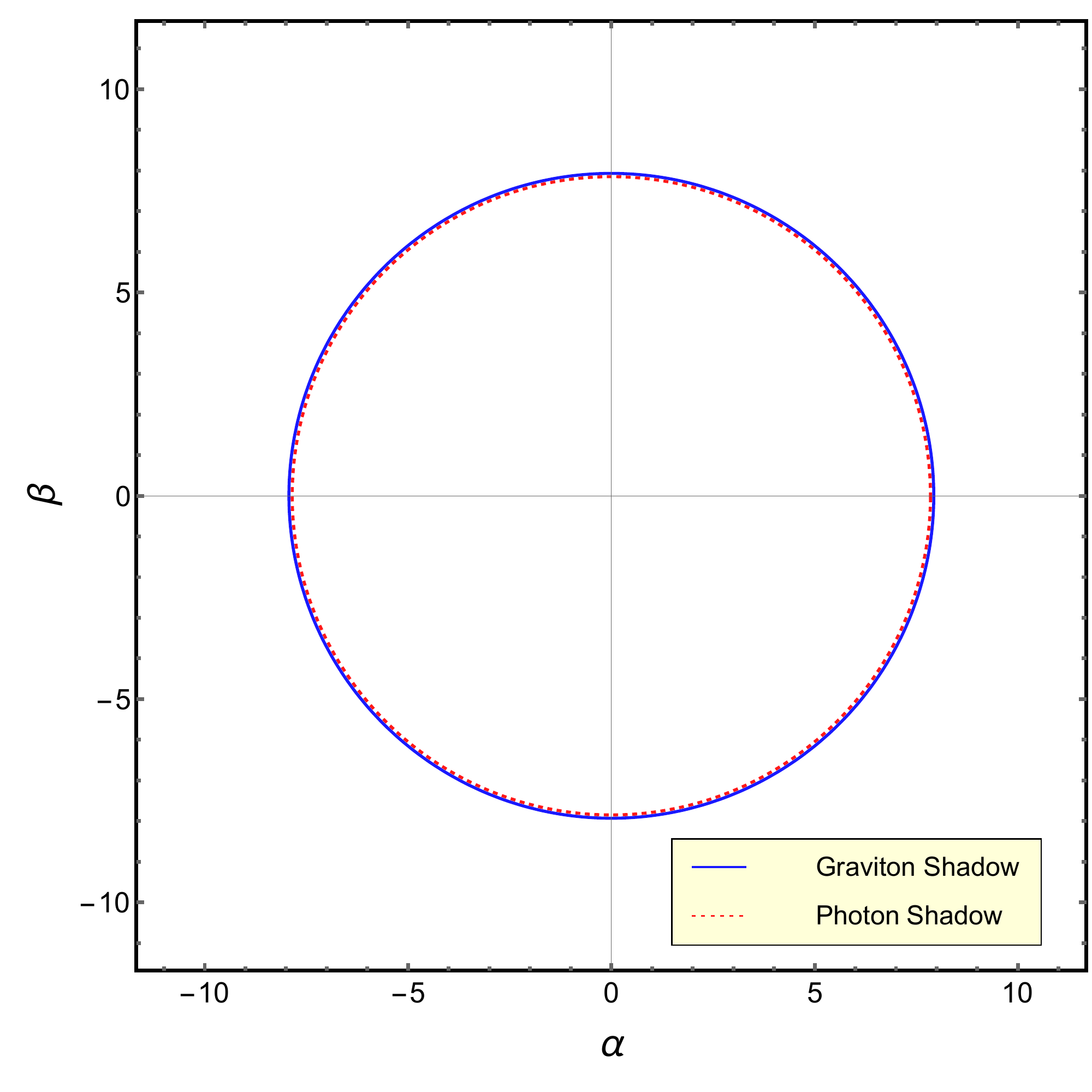}
\caption{This figure represents a snapshot of the evolution of graviton and photon shadow at in-going time $v=2$. The mass function corresponding to this evolution is $M(v)=1+\tanh(v)$ and the coupling constant $\alpha$ is chosen to be $0.01$.}\label{PSvsGS_shadow}
\end{figure}

\FloatBarrier
Having discussed the evolution of the photon sphere in the context of five-dimensional Gauss-Bonnet gravity, now let us move to the more interesting case of graviton circular null orbit. To that end, we start with the effective graviton metric given in \ref{final_eff_metric_5D} which is of the form,
\begin{equation}
ds^2_{\rm eff} = -f_{\rm eff}(r,v) dv^2 + 2  dv dr + g(r,v), d\Omega_{D-2}^2\
\end{equation}
Following an identical line of calculation as in \ref{Section_2}, we can obtain the following second order differential equation which governs the evolution of graviton sphere,
\begin{align}
\ddot{r}_{\rm gr}(v) + \frac{1}{2}[\dot{r}_{\rm gr}(v) g'(r,v) - \dot{g}(r,v) &-f_{\rm eff}(r,v) g'(r,v)] \left( \frac{f_{\rm eff}(r,v)-2\dot{r}_{\rm gr}}{g(r,v)}\right) -\frac{3}{2}\dot{r}_{\rm gr}(v)f_{\rm eff}'(r,v) 
\nonumber
\\
&+\frac{1}{2}[f_{\rm eff}(r,v)f_{\rm eff}'(r,v) -\dot{f}_{\rm eff}(r,v)]=0\label{Graviton Sp Eqn_in}
\end{align}
For consistency, one might check that when $g(r,v) = r^2$, we recover the previously derived \ref{Photon_Sp_Eqn_in}. 
Our aim here is to obtain the time evolution for the graviton sphere and understand how it is different from that of the photon sphere. As emphasized earlier, for computational simplicity, we restrict our attention to the small value of the Gauss-Bonnet coupling constant, $\alpha$. This doesn't ruin any physical insights since the distinction between the photon and graviton sphere would be still significant. Hence in this limit we have,
\begin{equation}
f_{\rm eff}(r,v) = 1-\frac{M(v)}{r^2} + \left(\frac{M(v)^2}{r^6} - \frac{\dot{M}(v)}{r^3} \right)\alpha + \mathcal{O}(\alpha^2)
\end{equation}

\begin{equation}
g(r,v) = r^2 + \frac{8 M(v) a}{r^2} + \mathcal{O}(\alpha^2)
\end{equation}
Now we feed in these $\mathcal{O}(\alpha)$ expressions of $f_{\rm eff}(r,v)$ and $g(r,v)$ in \ref{Graviton Sp Eqn_in} to obtain the evolution of the graviton sphere for a given choice of smoothly increasing mass function. In order to solve \ref{Graviton Sp Eqn_in} numerically, one requires two future boundary conditions. Therefore, first we need to derive an expression for the radius of graviton sphere in the  static limit, i.e., by setting $\dot{f}_{\rm eff}(r,v) = \dot{g}(r,v)=0$ in \ref{Graviton Sp Eqn_in}, which further reduces to,
\begin{equation}
g(r) f'(r)-f(r)g'(r)\Big|_{r=R_{\rm gr}}=0\label{static graviton sp}
\end{equation}
Up to $\mathcal{O}(\alpha)$, this leads to the following algebraic equation, 
\begin{equation}
r^6 - 2r^4 M - 8 r^2 M\, \alpha + 4M^2 \alpha\Big|_{r=R_{\rm gr}} = 0\label{graviton sp_linear eqn}
\end{equation}
Note that, when $\alpha = 0$, we obtain $R_{\rm gr}= \sqrt{2M}$, which is precisely the radius of photon sphere for the five-dimensional Schwarzschild black hole. Therefore, this allows us to expand the solution of the above algebraic equation around $\sqrt{2M}$ as,
\begin{equation}
R_{\rm gr} = \sqrt{2M} + A\, \alpha +\mathcal{O}(\alpha^2)
\end{equation}
with $A$ being some unknown factor to be determined by substituting $R_{\rm gr}$ in \ref{graviton sp_linear eqn} and keeping terms up to $\mathcal{O}(\alpha)$. This leads to,
\begin{equation}
R_{\rm gr} = \sqrt{2M} + \frac{3\, \alpha}{2\sqrt{2M}} + \mathcal{O}(\alpha^2)
\end{equation}
This represents the radius of the graviton sphere around a static spherically symmetric Gauss-Bonnet black hole in the limit when the coupling constant is small.
Now, with the boundary condition $r_{\rm gr}(v\rightarrow\infty) = R_{\rm gr} $ and $\dot{r}_{\rm gr}(v\rightarrow\infty)=0$, we solve \ref{Graviton Sp Eqn_in} to obtain the evolution of graviton sphere. We illustrate this result in \ref{PSvsGS} for the choice of mass function $M(v) = 1+\tanh(v)$. 
\FloatBarrier
From \ref{PSvsGS} , we see a distinction between the evolution of photon and graviton sphere. Note that, as the Gauss-Bonnet coupling constant $\alpha$ becomes smaller and smaller, both photon and graviton sphere approach each other, which one should expect. This implies, for a five-dimensional Schwarzschild black hole, there is no distinction between photon graviton sphere. It is not surprising since the effective metric contribution comes from the higher curvature correction, which is here is the Gauss-Bonnet correction. 
Similarly one can follow an identical approach developed in \ref{Section_4} to obtain the dynamical evolution of the graviton shadow with respect to the effective graviton metric, which reads,
\begin{equation}
\alpha(v)^2 +\beta(v)^2 = \frac{g(r_g,v)}{f_e(r,v)}\left[ 1- \left(\frac{\dot{r}_g(v)}{f_e(r,v)-\dot{r}_g(v)}\right)^2\right]\label{graviton_shadow}
\end{equation}
Again, for consistency one might set $g(r,v) = r^2$ to recover the evolution equation of photon shadow. The shadow casted by the graviton lensing is clearly different than that of the photon and 
we illustrate this distinction in \ref{PSvsGS_shadow}.
\FloatBarrier

\section{Conclusion}\label{Section_9}

Conclusive evidence towards the existence of black holes is building up with detection of gravitational waves and existence of supermassive compact objects at the center of most of the galaxies. However, the most direct test in this respect will be the detection of black hole shadow, a dark region surrounding the black hole due to the existence of a photon sphere. Since most of the astrophysical black holes are accreting it is legitimate to understand the evolution of the photon sphere in the context of dynamical black holes. In this work, we have achieved this goal, i.e., determining the evolution of the photon sphere and black hole shadow in the context of dynamical black holes. Starting from a spherically symmetric situation, we have derived the differential equation governing the evolution of the photon sphere and have subsequently applied it for Vaidya, \RNV and de Sitter Vaidya spacetime. For Vaidya spacetime, an appropriate mass function results into a well-behaved evolution of the photon sphere. It is clear that as the mass function grows the photon sphere should also grow.
On the other hand, in \RNV spacetime, there can be well-behaved mass and charge functions violating the null energy condition. Then it appears that in those situations besides the event and apparent horizon the photon sphere also starts decreasing in radius. This feature is counter-intuitive, and it is interesting that violation of energy condition is so ingrained in the evolution of the photon sphere, that its nature changes. Furthermore, interestingly, for de Sitter Vaidya spacetime the photon sphere starts depending on the choice of the cosmological constant, unlike the Schwarzschild-de Sitter spacetime, where the photon sphere did not depend on the choice of the cosmological constant. As evident from the above discussion, in the dynamical context, the shadow of the black hole will also get modified, which is also borne out by our computation of black hole shadow as well. The same story continues to hold for rotating spacetimes as well. However, in the dynamical contexts, it turns out that the Hamilton-Jacobi equation is not separable unless slow-rotation limit is assumed. Thus in the slow rotation limit, we have presented the evolution equation of the photon sphere and have demonstrated the same using numerical analysis. Besides, the evolution of the black hole shadow has also been presented. So far we have been considering dynamical solutions within the framework of \gr, which subsequently have been generalized to dynamical solutions for Einstein-Gauss-Bonnet gravity, the first non-trivial correction over and above \gr, which keeps field equations second order. In this context besides the photon sphere, we have also studied the evolution of graviton sphere. This is because, the photon moves in null trajectory, while the graviton does not in theories of gravity beyond \gr. But one can circumvent this problem by arguing that there is some effective metric different from the actual one, where gravitons propagate along null lines. The fact that photon and graviton sphere differs has been demonstrated explicitly, along with their evolution as the black hole accretes matter. 
\section*{Acknowledgement}

AKM would like to thank IACS Kolkata for hospitality, where part of this work was carried out. Research of SC is supported by INSPIRE Faculty Fellowship (Reg. No. DST/INSPIRE/04/2018/000893) from Department of Science and Technology, Government of India. Research of SS is supported by the Department of Science and Technology, Government of India under the SERB Fast Track Scheme for Young Scientists (YSS/2015/001346). We thank Xian O. Camanho for sharing his unpublished notes on the effective graviton metric calculation. We also thank Avirup Ghosh, Yotam Sherf and Ram Brustein for helpful discussions.

\section*{Appendices}
	\appendix 
	\labelformat{section}{Appendix #1} 
	\labelformat{subsection}{Appendix #1} 

\section{Derivation of the Evolution Equation for \PS}\label{Appendix_A}
In this section we provide a complete derivation of the evolution equation of the photon sphere, i.e., \ref{Photon_Sp_Eqn_in}. For in-going case we have $r_{\rm ph}=r_{\rm ph}(v)$. Hence,
\begin{equation}
dr_{\rm ph}(v) = \frac{\partial r_{\rm ph}(v)}{\partial v} dv = \dot{r}_{\rm ph}(v) dv\label{dr}
\end{equation}
For null orbits we now put $ds^2=0$ from \ref{Dynamical_in} but keeping in mind that, now we don't have $dr_{\rm ph}=0$ rather $dr_{\rm ph}$ is given by  \ref{dr}. This leads to the expression,
\begin{equation}
\left(\frac{d\phi}{dv}\right)^2 = \frac{1}{r_{\rm ph}(v)^2} f(r_{\rm ph}(v),v)-\frac{2}{r_{\rm ph}(v)^2} \dot{r}_{\rm ph}(v)\label{dynamical_dphi_dv}
\end{equation}
Now for the metric in \ref{Dynamical_in} we have the following non-vanishing components of the Christoffel connection,
$$\Gamma^v_{vv} = \frac{1}{2}f'(r,v), \quad \Gamma^r_{vv} = \frac{1}{2}[\dot{f}(r,v)+f(r,v)f'(r,v)],\quad \Gamma^v_{rv} = -\frac{1}{2}f'(r,v)$$

$$\Gamma^v_{\theta\theta} = -r, \quad \Gamma^r_{r\theta} = \frac{1}{r},\quad \Gamma^v_{\phi\phi} = -r \sin^2\theta,\quad \Gamma^{\phi}_{r\phi} = \frac{1}{r} \quad \Gamma^{\theta}_{\phi\phi} = -\cos\theta \sin\theta, \quad \Gamma^{\phi}_{\theta\phi} = \cot\theta$$
and the geodesic equations can be written as, 
\begin{equation}
\frac{d^2r}{d\lambda^2}-\frac{\partial f}{\partial r}\left(\frac{dr}{d\lambda}\right)\left(\frac{dv}{d\lambda}\right) + \frac{1}{2} \left(f \frac{\partial f}{\partial r }-\frac{\partial f}{\partial v}\right) \left(\frac{dv}{d\lambda}\right)^2 -rf\left(\frac{d\phi}{d\lambda}\right)^2=0\label{r_eqn}
\end{equation}

\begin{equation}
\frac{d^2v}{d\lambda^2} + \frac{1}{2} \frac{\partial f}{\partial r}\left(\frac{dv}{d\lambda}\right)^2 - r\left(\frac{d\phi}{d\lambda}\right)^2 =0\label{v_eqn}
\end{equation}
Now we set $r=r_{\rm ph}(v)$ for the evolution of the radius of photon sphere. From \ref{dr} we obtain,
$$
\frac{d^2r_{\rm ph}(v)}{d\lambda^2} = \dot{r}_{\rm ph}(v) \frac{d^2v}{d\lambda^2} + \ddot{r}_{\rm ph}(v) \left(\frac{dv}{d\lambda}\right)^2
$$
\begin{equation}
=  \dot{r}_{\rm ph}(v)\left[r_{\rm ph}(v)\left(\frac{d\phi}{d\lambda}\right)^2 - \frac{1}{2} \frac{\partial f}{\partial r}\Big|_{r_{\rm ph}(v),v}\left(\frac{dv}{d\lambda}\right)^2 \right]+ \ddot{r}_{\rm ph}(v) \left(\frac{dv}{d\lambda}\right)^2\label{3}
\end{equation}
Now we plug in \ref{dynamical_dphi_dv} and \ref{3} in \ref{r_eqn} to obtain
\begin{align}\label{abc}
\ddot{r}_{\rm ph}(v) &+ \dot{r}_{\rm ph}(v)\left[\frac{3}{r_{\rm ph}(v)} f(r_{\rm ph},v) 
- \frac{3}{2}\frac{\partial f}{\partial r}\Big|_{r_{\rm ph}(v),v} \right] 
- \frac{2}{r_{\rm ph}(v)} \dot{r}_{\rm ph}(v)^2
\nonumber
\\
&+\frac{1}{2} \left( f \frac{\partial f}{\partial r}\Big|_{r_{\rm ph}(v),v} -
 \frac{\partial f}{\partial v}\Big|_{r_{\rm ph}(v),v}\right)- 
 \frac{1}{r_{\rm ph}(v)} f(r_{\rm ph}(v),v)^2 =0
\end{align}
Similar procedure can be carried out for the evolution equation of photon sphere in terms of out-going coordinate, i.e., \ref{Dynamical_out} to obtain  \ref{Photon Sp Eqn_out}.

\section{Derivation of the  effective graviton metric}\label{Appendix_B}
We shall start with the $G^v_v$ component, for which we set $b=d=v$ in \ref{effective metric_form} and obtain the correction as,
\begin{equation}
\lambda_2(\delta^{pava_1b_1}_{qcvc_1d_1} - \delta^p_q \delta^{ava_1b_1}_{cvc_1 d_1} ) R_{a_1b_1}\,^{c_1d_1}\nabla_v \nabla^v h_a^c  = 
\lambda_2(\delta^{\hat{p}\hat{a}\hat{a_1}\hat{b_1}}_{\hat{q}\hat{c}\hat{c_1}\hat{d_1}} - \delta^p_q \delta^{\hat{a}\hat{a_1}\hat{b_1}}_{\hat{c}\hat{c_1}\hat{d_1}} ) R_{\hat{a_1}\hat{b_1}}\,^{\hat{c_1}\hat{d_1}}\nabla_v \nabla^v h_{\hat{a}}^{\hat{c}} \label{step_1}
\end{equation}
Here $\hat{a},\hat{b}= 1,2,........D-1$, are the spatial indexes, i.e., the radial coordinate $r$ and angular coordinates. We shall denote the angular coordinates as $i,j = 1,2,.......D-2$. In the above result we've used the identity,
\begin{equation}
\delta^{a_1 a_2 v ......a_k}_{b_1 b_2 v ......b_k} = \delta^{\hat{a_1} \hat{a_2}......\hat{a_k}}_{\hat{b_1}\hat{ b_2} ......\hat{b_k}} 
\end{equation}
Hence \ref{step_1} becomes,
$$\lambda_2\left(4\delta^{\hat{p}\hat{a}r\hat{b_1}}_{\hat{q}\hat{c}r\hat{d_1}} R_{r\hat{b_1}}\,^{r\hat{d_1}}  +\delta^{\hat{p}\hat{a}\hat{a_1}\hat{b_1}}_{\hat{q}\hat{c}\hat{c_1}\hat{d_1}} R_{\hat{a_1}\hat{b_1}}\,^{\hat{c_1}\hat{d_1}}- 4\delta^p_q \delta^{\hat{a}r\hat{b_1}}_{\hat{c}r\hat{d_1}} R_{r\hat{b_1}}\,^{r\hat{d_1}} - \delta^p_q \delta^{\hat{a}\hat{a_1}\hat{b_1}}_{\hat{c}\hat{c_1}\hat{d_1}} R_{\hat{a_1}\hat{b_1}}\,^{\hat{c_1}\hat{d_1}}\right) $$

$$=\lambda_2\left(4\delta^{\hat{p}\hat{a}ri}_{\hat{q}\hat{c}rj} R_{ri}\,^{rj} +\delta^{\hat{p}\hat{a}ij}_{\hat{q}\hat{c}kl} R_{ij}\,^{kl}- 4\delta^p_q \delta^{\hat{a}ri}_{\hat{c}rj} R_{ri}\,^{rj} - \delta^p_q \delta^{\hat{a}ij}_{\hat{c}kl} R_{ij}\,^{kl}\right)\nabla_v \nabla^v h_{\hat{a}}^{\hat{c}}   $$
Note that, we are using gauge-invariant transverse and traceless tensor perturbation, i.e., $h^{\mu\nu} = h^{\mu i} = h^i_i = \nabla_i h^{ij} = 0$, where $\mu,\nu = r,v$.
To understand the calculation further, let us concentrate on the first term, i.e., $\delta^{\hat{p}\hat{a}ri}_{\hat{q}\hat{c}rj}$. Note that, because of the antisymmetric properties of $\delta$ tensor, it can not have two same indexes either in the contravariant or in the covariant position. Hence all other indexes in this term apart from $r$ are angular coordinates and so on for other terms. Now we put the component of Riemann tensor from \ref{Riemann_1},\ref{Riemann_2},\ref{Riemann_3} to obtain,
$$\lambda_2\left[4\delta^{\hat{p}\hat{a}ri}_{\hat{q}\hat{c}rj} \left(\frac{-f'}{2r}\right)\delta_i^j +\delta^{\hat{p}\hat{a}ij}_{\hat{q}\hat{c}kl} \left(\frac{1-f}{r^2}\right)\delta_{ij}^{kl}- 4\delta^p_q \delta^{\hat{a}ri}_{\hat{c}rj} \left(\frac{-f'}{2r}\right)\delta_i^j - \delta^p_q \delta^{\hat{a}ij}_{\hat{c}kl} \left(\frac{1-f}{r^2}\right)\delta_{ij}^{kl}\right]\nabla_v \nabla^v h_{\hat{a}}^{\hat{c}}   $$

$$=\lambda_2\left[4\delta^{\hat{p}\hat{a}i}_{\hat{q}\hat{c}i} \left(\frac{-f'}{2r}\right) +2\, \delta^{\hat{p}\hat{a}ij}_{\hat{q}\hat{c}ij} \left(\frac{1-f}{r^2}\right)- 4\delta^p_q \delta^{\hat{a}i}_{\hat{c}i} \left(\frac{-f'}{2r}\right) - 2\delta^p_q \delta^{\hat{a}ij}_{\hat{c}ij} \left(\frac{1-f}{r^2}\right)\right]\nabla_v \nabla^v h_{\hat{a}}^{\hat{c}}   $$
\small
$$=\lambda_2\left[4\delta^{\hat{p}\hat{a}}_{\hat{q}\hat{c}}(D-4) \left(\frac{-f'}{2r}\right) +2\, \delta^{\hat{p}\hat{a}}_{\hat{q}\hat{c}}(D-4)(D-5) \left(\frac{1-f}{r^2}\right)- 4\delta^p_q \delta^{\hat{a}}_{\hat{c}}(D-3) \left(\frac{-f'}{2r}\right) - 2\delta^p_q \delta^{\hat{a}}_{\hat{c}}(D-3)(D-4) \left(\frac{1-f}{r^2}\right)\right]\nabla_v \nabla^v h_{\hat{a}}^{\hat{c}}   $$
\normalsize
Here it is understood that $f$ is a function of $r$ and $v$, i.e., $f=f(r,v)$, which we've considered for simplicity.
Deriving the above result we've used the following identities,
\begin{eqnarray}
\delta^{ijm}_{klm} = (D-4)\delta^{ij}_{kl}\\
\delta^{ijmn}_{klmn} = (D-4)(D-5) \delta^{ij}_{kl}\\
\delta^{imn}_{jmn} = (D-3)(D-4)\delta^i_j\\
\delta^{im}_{jm} = (D-3)\delta^i_j\\
\delta^{ij mno}_{kl mno} = (D-4)(D-5)(D-6)\delta^{ij}_{kl}
\end{eqnarray}
Note that, the last two terms in the above expression doesn't contribute, because they contain a $\delta^{\hat{a}}_{\hat{c}}$ term, which multiplies with $h_{\hat{a}}^{\hat{c}}$ to give zero from the traceless condition. Hence we are left with only,
$$\lambda_2\left[4\delta^{\hat{p}\hat{a}}_{\hat{q}\hat{c}}(D-4) \left(\frac{-f'}{2r}\right) +2\, \delta^{\hat{p}\hat{a}}_{\hat{q}\hat{c}}(D-4)(D-5) \left(\frac{1-f}{r^2}\right)\right]\nabla_v \nabla^v h_{\hat{a}}^{\hat{c}}$$

$$=\lambda_2\left[4\left(\delta^{\hat{p}}_{\hat{q}}\delta^{\hat{a}}_{\hat{c}}-\delta^{\hat{p}}_{\hat{c}}\delta^{\hat{a}}_{\hat{q}}\right)(D-4) \left(\frac{-f'}{2r}\right) +2\, \left(\delta^{\hat{p}}_{\hat{q}}\delta^{\hat{a}}_{\hat{c}}-\delta^{\hat{p}}_{\hat{c}}\delta^{\hat{a}}_{\hat{q}}\right)(D-4)(D-5) \left(\frac{1-f}{r^2}\right)\right]\nabla_v \nabla^v h_{\hat{a}}^{\hat{c}}$$
Again using the fact that, $\delta^{\hat{a}}_{\hat{c}}$ term doesn't contribute, we further get,
$$\lambda_2\left[2\delta^{\hat{p}}_{\hat{c}}\delta^{\hat{a}}_{\hat{q}}(D-4) \left(\frac{f'}{r}\right) -2\, \delta^{\hat{p}}_{\hat{c}}\delta^{\hat{a}}_{\hat{q}}(D-4)(D-5) \left(\frac{1-f}{r^2}\right)\right]\nabla_v \nabla^v h_{\hat{a}}^{\hat{c}}$$

$$=\lambda_2\left[2(D-4) \left(\frac{f'}{r}\right) -2\, (D-4)(D-5) \left(\frac{1-f}{r^2}\right)\right]\nabla_v \nabla^v h_{\hat{q}}^{\hat{p}}$$
The coefficient of the kinetic term we identify as the correction to the background metric and hence the $(v,v)$ component of the effective metric is given by,
\begin{equation}
G^v_v = 1 - 2\lambda_2\left[(D-4) \left(\frac{f'}{r}\right) -\, (D-4)(D-5) \left(\frac{1-f}{r^2}\right)\right]
\end{equation}
Now let us calculate $G^r_v$ component. For this we set $b =r,d=v$ in \ref{effective metric_form} to obtain,
$$
\lambda_2(\delta^{pava_1b_1}_{qcrc_1d_1} - \delta^p_q \delta^{ava_1b_1}_{crc_1 d_1} ) R_{a_1b_1}\,^{c_1d_1}\nabla_v \nabla^v h_a^c  $$

$$=\lambda_2\left[4\delta^{pavri}_{qcrvj}R_{vi}\,^{rj} + \delta^{pavij}_{qcrkl}R_{ij}\,^{kl} - 4\delta^p_q \delta^{avri}_{crvj}R_{vi}\,^{rj}-\delta^p_q \delta^{avij}_{crkl}R_{ij}\,^{kl}\right]\nabla_v \nabla^v h_{a}^c$$
Note that $\delta^{pavri}_{qcrvj} = -\delta^{pai}_{qcj}$ and $\delta^{pavij}_{qcrkl} =0$ in the above expression. The first identity is because of antisymmetric properties of $\delta$ tensor. The second identity is because all other indexes apart from `$v$' in the contravariant position and index apart from `$r$' in the covariant position are angular index. This gives zero when the determinant is taken.
Using this and by replacing the components of Riemann tensor, the above expression becomes,
$$\lambda_2\left[-4\delta^{pai}_{qcj}\left(\frac{-\dot{f}}{2r}\delta_i^j\right) - 4\delta^p_q \delta^{ai}_{cj}\left(\frac{-\dot{f}}{2r}\delta_i^j\right)\right]\nabla_v \nabla^v h_{a}^c$$

$$=\lambda_2\left[2\delta^{pai}_{qci}\left(\frac{\dot{f}}{r}\right) + 2\delta^p_q \delta^{ai}_{ci}\left(\frac{\dot{f}}{r}\right)\right]\nabla_v \nabla^v h_{a}^c$$

$$=\lambda_2\left[2\delta^{pa}_{qc}(D-4)\left(\frac{\dot{f}}{r}\right) + 2\delta^p_q \delta^{a}_{c}(D-3)\left(\frac{\dot{f}}{r}\right)\right]\nabla_v \nabla^v h_{a}^c$$
The last term again doesn't contribute because of the traceless condition, and finally, we have\cite{Izumi:2014loa},
\begin{equation}
G^r_v = 2\lambda_2 (D-4) \frac{\dot{f}}{r}
\end{equation}

\bibliography{Photon_Graviton_Sphere}

\providecommand{\href}[2]{#2}\begingroup\raggedright\begin{thebibliography}{10}

\bibitem{bardeen1973}
J.~M. Bardeen, B.~Carter, and S.~W. Hawking, ``The four laws of black hole
  mechanics,'' {\em Comm. Math. Phys.} {\bfseries 31} no.~2, (1973) 161--170.
  \url{https://projecteuclid.org:443/euclid.cmp/1103858973}.

\bibitem{hawking1972}
S.~W. Hawking, ``Black holes in general relativity,'' {\em Comm. Math. Phys.}
  {\bfseries 25} no.~2, (1972) 152--166.
  \url{https://projecteuclid.org:443/euclid.cmp/1103857884}.

\bibitem{hawking1975}
S.~W. Hawking, ``Particle creation by black holes,'' {\em Comm. Math. Phys.}
  {\bfseries 43} no.~3, (1975) 199--220.
  \url{https://projecteuclid.org:443/euclid.cmp/1103899181}.

\bibitem{hawking_ellis_1973}
S.~W. Hawking and G.~F.~R. Ellis,
  \href{http://dx.doi.org/10.1017/CBO9780511524646}{{\em The Large Scale
  Structure of Space-Time}}.
\newblock Cambridge Monographs on Mathematical Physics. Cambridge University
  Press, 1973.

\bibitem{Bekenstein:1972tm}
J.~D. Bekenstein, ``{Black holes and the second law},''
\href{http://dx.doi.org/10.1007/BF02757029}{{\em Lett. Nuovo Cim.} {\bfseries
  4} (1972) 737--740}.

\bibitem{Bekenstein:1973ur}
J.~D. Bekenstein, ``{Black holes and entropy},''
\href{http://dx.doi.org/10.1103/PhysRevD.7.2333}{{\em Phys. Rev.} {\bfseries
  D7} (1973) 2333--2346}.

\bibitem{Wald:1995yp}
R.~M. Wald, {\em {Quantum Field Theory in Curved Space-Time and Black Hole
  Thermodynamics}}.
\newblock Chicago Lectures in Physics. University of Chicago Press, Chicago,
  IL,
1995.
\newblock

\bibitem{Abbott:2016blz}
{\bfseries LIGO Scientific, Virgo} Collaboration, B.~P. Abbott {\em et~al.},
  ``{Observation of Gravitational Waves from a Binary Black Hole Merger},''
  \href{http://dx.doi.org/10.1103/PhysRevLett.116.061102}{{\em Phys. Rev.
  Lett.} {\bfseries 116} no.~6, (2016) 061102},
\href{http://arxiv.org/abs/1602.03837}{{\ttfamily arXiv:1602.03837 [gr-qc]}}.

\bibitem{TheLIGOScientific:2016wfe}
{\bfseries LIGO Scientific, Virgo} Collaboration, B.~P. Abbott {\em et~al.},
  ``{Properties of the Binary Black Hole Merger GW150914},''
  \href{http://dx.doi.org/10.1103/PhysRevLett.116.241102}{{\em Phys. Rev.
  Lett.} {\bfseries 116} no.~24, (2016) 241102},
\href{http://arxiv.org/abs/1602.03840}{{\ttfamily arXiv:1602.03840 [gr-qc]}}.

\bibitem{Abbott:2016nmj}
{\bfseries LIGO Scientific, Virgo} Collaboration, B.~P. Abbott {\em et~al.},
  ``{GW151226: Observation of Gravitational Waves from a 22-Solar-Mass Binary
  Black Hole Coalescence},''
  \href{http://dx.doi.org/10.1103/PhysRevLett.116.241103}{{\em Phys. Rev.
  Lett.} {\bfseries 116} no.~24, (2016) 241103},
\href{http://arxiv.org/abs/1606.04855}{{\ttfamily arXiv:1606.04855 [gr-qc]}}.

\bibitem{Abbott:2017vtc}
{\bfseries LIGO Scientific, VIRGO} Collaboration, B.~P. Abbott {\em et~al.},
  ``{GW170104: Observation of a 50-Solar-Mass Binary Black Hole Coalescence at
  Redshift 0.2}'' \href{http://dx.doi.org/10.1103/PhysRevLett.118.221101,
  10.1103/PhysRevLett.121.129901}{{\em Phys. Rev. Lett.} {\bfseries 118}
  no.~22, (2017) 221101}, \href{http://arxiv.org/abs/1706.01812}{{\ttfamily
  arXiv:1706.01812 [gr-qc]}}.
[Erratum: Phys. Rev. Lett.121,no.12,129901(2018)].

\bibitem{Ghez:2008ms}
A.~M. Ghez {\em et~al.}, ``{Measuring Distance and Properties of the Milky
  Way's Central Supermassive Black Hole with Stellar Orbits},''
  \href{http://dx.doi.org/10.1086/592738}{{\em Astrophys. J.} {\bfseries 689}
  (2008) 1044--1062},
\href{http://arxiv.org/abs/0808.2870}{{\ttfamily arXiv:0808.2870 [astro-ph]}}.

\bibitem{Schodel:2002vg}
R.~Schodel {\em et~al.}, ``{A Star in a 15.2 year orbit around the supermassive
  black hole at the center of the Milky Way},''
  \href{http://dx.doi.org/10.1038/nature01121}{{\em Nature} (2002) },
  \href{http://arxiv.org/abs/astro-ph/0210426}{{\ttfamily
  arXiv:astro-ph/0210426 [astro-ph]}}.
[Nature419,694(2002)].

\bibitem{Kormendy:2013dxa}
J.~Kormendy and L.~C. Ho, ``{Coevolution (Or Not) of Supermassive Black Holes
  and Host Galaxies},''
  \href{http://dx.doi.org/10.1146/annurev-astro-082708-101811}{{\em Ann. Rev.
  Astron. Astrophys.} {\bfseries 51} (2013) 511--653},
\href{http://arxiv.org/abs/1304.7762}{{\ttfamily arXiv:1304.7762
  [astro-ph.CO]}}.

\bibitem{Vazquez:2003zm}
S.~E. Vazquez and E.~P. Esteban, ``{Strong field gravitational lensing by a
  Kerr black hole},'' \href{http://dx.doi.org/10.1393/ncb/i2004-10121-y}{{\em
  Nuovo Cim.} {\bfseries B119} (2004) 489--519},
\href{http://arxiv.org/abs/gr-qc/0308023}{{\ttfamily arXiv:gr-qc/0308023
  [gr-qc]}}.

\bibitem{Shaikh:2018lcc}
R.~Shaikh, P.~Kocherlakota, R.~Narayan, and P.~S. Joshi, ``{Shadows of
  spherically symmetric black holes and naked singularities},''
  \href{http://dx.doi.org/10.1093/mnras/sty2624}{{\em Mon. Not. Roy. Astron.
  Soc.} {\bfseries 482} (2019) 52},
\href{http://arxiv.org/abs/1802.08060}{{\ttfamily arXiv:1802.08060
  [astro-ph.HE]}}.

\bibitem{Hou:2018bar}
X.~Hou, Z.~Xu, M.~Zhou, and J.~Wang, ``{Black hole shadow of Sgr A$^{*}$ in
  dark matter halo},''
  \href{http://dx.doi.org/10.1088/1475-7516/2018/07/015}{{\em JCAP} {\bfseries
  1807} no.~07, (2018) 015},
\href{http://arxiv.org/abs/1804.08110}{{\ttfamily arXiv:1804.08110 [gr-qc]}}.

\bibitem{Cunha:2018gql}
P.~V.~P. Cunha, C.~A.~R. Herdeiro, and M.~J. Rodriguez, ``{Does the black hole
  shadow probe the event horizon geometry?},''
  \href{http://dx.doi.org/10.1103/PhysRevD.97.084020}{{\em Phys. Rev.}
  {\bfseries D97} no.~8, (2018) 084020},
\href{http://arxiv.org/abs/1802.02675}{{\ttfamily arXiv:1802.02675 [gr-qc]}}.

\bibitem{Tsukamoto:2017fxq}
N.~Tsukamoto, ``{Black hole shadow in an asymptotically-flat, stationary, and
  axisymmetric spacetime: The Kerr-Newman and rotating regular black holes},''
  \href{http://dx.doi.org/10.1103/PhysRevD.97.064021}{{\em Phys. Rev.}
  {\bfseries D97} no.~6, (2018) 064021},
\href{http://arxiv.org/abs/1708.07427}{{\ttfamily arXiv:1708.07427 [gr-qc]}}.

\bibitem{Repin:2018anv}
S.~V. Repin, D.~A. Kompaneets, I.~D. Novikov, and V.~A. Mityagina, ``{Shadow of
  rotating black holes on a standard background screen},''
\href{http://arxiv.org/abs/1802.04667}{{\ttfamily arXiv:1802.04667 [gr-qc]}}.

\bibitem{Abdujabbarov:2016hnw}
A.~Abdujabbarov, M.~Amir, B.~Ahmedov, and S.~G. Ghosh, ``{Shadow of rotating
  regular black holes},''
  \href{http://dx.doi.org/10.1103/PhysRevD.93.104004}{{\em Phys. Rev.}
  {\bfseries D93} no.~10, (2016) 104004},
\href{http://arxiv.org/abs/1604.03809}{{\ttfamily arXiv:1604.03809 [gr-qc]}}.

\bibitem{Kumar:2018ple}
R.~Kumar and S.~G. Ghosh, ``{Black hole parameters estimation from its
  shadow},''
\href{http://arxiv.org/abs/1811.01260}{{\ttfamily arXiv:1811.01260 [gr-qc]}}.

\bibitem{Cunha:2018cof}
P.~V.~P. Cunha, C.~A.~R. Herdeiro, and M.~J. Rodriguez, ``{Shadows of Exact
  Binary Black Holes},''
  \href{http://dx.doi.org/10.1103/PhysRevD.98.044053}{{\em Phys. Rev.}
  {\bfseries D98} no.~4, (2018) 044053},
\href{http://arxiv.org/abs/1805.03798}{{\ttfamily arXiv:1805.03798 [gr-qc]}}.

\bibitem{Ayzenberg:2018jip}
D.~Ayzenberg and N.~Yunes, ``{Black Hole Shadow as a Test of General
  Relativity: Quadratic Gravity},''
  \href{http://dx.doi.org/10.1088/1361-6382/aae87b}{{\em Class. Quant. Grav.}
  {\bfseries 35} no.~23, (2018) 235002},
\href{http://arxiv.org/abs/1807.08422}{{\ttfamily arXiv:1807.08422 [gr-qc]}}.

\bibitem{Rahman:2018fgy}
M.~Rahman and A.~A. Sen, ``{Astrophysical Signatures of Black holes in
  Generalized Proca Theories},''
\href{http://arxiv.org/abs/1810.09200}{{\ttfamily arXiv:1810.09200 [gr-qc]}}.

\bibitem{Atamurotov:2013sca}
F.~Atamurotov, A.~Abdujabbarov, and B.~Ahmedov, ``{Shadow of rotating non-Kerr
  black hole},''
\href{http://dx.doi.org/10.1103/PhysRevD.88.064004}{{\em Phys. Rev.} {\bfseries
  D88} no.~6, (2013) 064004}.

\bibitem{Grenzebach:2015oea}
A.~Grenzebach, V.~Perlick, and C.~Lammerzahl, ``{Photon Regions and Shadows of
  Accelerated Black Holes},''
  \href{http://dx.doi.org/10.1142/S0218271815420249}{{\em Int. J. Mod. Phys.}
  {\bfseries D24} no.~09, (2015) 1542024},
\href{http://arxiv.org/abs/1503.03036}{{\ttfamily arXiv:1503.03036 [gr-qc]}}.

\bibitem{Atamurotov:2015xfa}
F.~Atamurotov, S.~G. Ghosh, and B.~Ahmedov, ``{Horizon structure of rotating
  Einstein?Born?Infeld black holes and shadow},''
  \href{http://dx.doi.org/10.1140/epjc/s10052-016-4122-9}{{\em Eur. Phys. J.}
  {\bfseries C76} no.~5, (2016) 273},
\href{http://arxiv.org/abs/1506.03690}{{\ttfamily arXiv:1506.03690 [gr-qc]}}.

\bibitem{Mars:2017jkk}
M.~Mars, C.~F. Paganini, and M.~A. Oancea, ``{The fingerprints of black
  holes-shadows and their degeneracies},''
  \href{http://dx.doi.org/10.1088/1361-6382/aa97ff}{{\em Class. Quant. Grav.}
  {\bfseries 35} no.~2, (2018) 025005},
\href{http://arxiv.org/abs/1710.02402}{{\ttfamily arXiv:1710.02402 [gr-qc]}}.

\bibitem{Einstein:1956zz}
A.~Einstein, ``{Lens-Like Action of a Star by the Deviation of Light in the
  Gravitational Field},''
\href{http://dx.doi.org/10.1126/science.84.2188.506}{{\em Science} {\bfseries
  84} (1936) 506--507}.

\bibitem{Bartelmann:2010fz}
M.~Bartelmann, ``{Gravitational Lensing},''
  \href{http://dx.doi.org/10.1088/0264-9381/27/23/233001}{{\em Class. Quant.
  Grav.} {\bfseries 27} (2010) 233001},
\href{http://arxiv.org/abs/1010.3829}{{\ttfamily arXiv:1010.3829
  [astro-ph.CO]}}.

\bibitem{Cunha:2018acu}
P.~V.~P. Cunha and C.~A.~R. Herdeiro, ``{Shadows and strong gravitational
  lensing: a brief review},''
  \href{http://dx.doi.org/10.1007/s10714-018-2361-9}{{\em Gen. Rel. Grav.}
  {\bfseries 50} no.~4, (2018) 42},
\href{http://arxiv.org/abs/1801.00860}{{\ttfamily arXiv:1801.00860 [gr-qc]}}.

\bibitem{Kochanek2006}
C.~S. Kochanek, {\em Strong Gravitational Lensing},
  \href{http://dx.doi.org/10.1007/978-3-540-30310-7_2}{pp.~91--268}.
\newblock Springer Berlin Heidelberg, Berlin, Heidelberg, 2006.
\newblock \url{https://doi.org/10.1007/978-3-540-30310-7_2}.

\bibitem{PhysRev.133.B835}
S.~Liebes, ``Gravitational lenses,''
  \href{http://dx.doi.org/10.1103/PhysRev.133.B835}{{\em Phys. Rev.} {\bfseries
  133} (Feb, 1964) B835--B844}.
  \url{https://link.aps.org/doi/10.1103/PhysRev.133.B835}.

\bibitem{PhysRevLett.105.251101}
M.~Sereno, A.~Sesana, A.~Bleuler, P.~Jetzer, M.~Volonteri, and M.~C. Begelman,
  ``Strong lensing of gravitational waves as seen by lisa,''
  \href{http://dx.doi.org/10.1103/PhysRevLett.105.251101}{{\em Phys. Rev.
  Lett.} {\bfseries 105} (Dec, 2010) 251101}.
  \url{https://link.aps.org/doi/10.1103/PhysRevLett.105.251101}.

\bibitem{PhysRevD.71.064004}
R.~Whisker, ``Strong gravitational lensing by braneworld black holes,''
  \href{http://dx.doi.org/10.1103/PhysRevD.71.064004}{{\em Phys. Rev. D}
  {\bfseries 71} (Mar, 2005) 064004}.
  \url{https://link.aps.org/doi/10.1103/PhysRevD.71.064004}.

\bibitem{PhysRevD.69.022002}
N.~Seto, ``Strong gravitational lensing and localization of merging massive
  black hole binaries with lisa,''
  \href{http://dx.doi.org/10.1103/PhysRevD.69.022002}{{\em Phys. Rev. D}
  {\bfseries 69} (Jan, 2004) 022002}.
  \url{https://link.aps.org/doi/10.1103/PhysRevD.69.022002}.

\bibitem{Perlick2004}
V.~Perlick, ``Gravitational lensing from a spacetime perspective,''
  \href{http://dx.doi.org/10.12942/lrr-2004-9}{{\em Living Reviews in
  Relativity} {\bfseries 7} no.~1, (Sep, 2004) 9}.
  \url{https://doi.org/10.12942/lrr-2004-9}.

\bibitem{Bozza:2009yw}
V.~Bozza, ``{Gravitational Lensing by Black Holes},''
  \href{http://dx.doi.org/10.1007/s10714-010-0988-2}{{\em Gen. Rel. Grav.}
  {\bfseries 42} (2010) 2269--2300},
\href{http://arxiv.org/abs/0911.2187}{{\ttfamily arXiv:0911.2187 [gr-qc]}}.

\bibitem{PhysRevD.59.124001}
S.~Frittelli and E.~T. Newman, ``Exact universal gravitational lensing
  equation,'' \href{http://dx.doi.org/10.1103/PhysRevD.59.124001}{{\em Phys.
  Rev. D} {\bfseries 59} (Apr, 1999) 124001}.
  \url{https://link.aps.org/doi/10.1103/PhysRevD.59.124001}.

\bibitem{Chakraborty:2016lxo}
S.~Chakraborty and S.~SenGupta, ``{Strong gravitational lensing --- A probe for
  extra dimensions and Kalb-Ramond field},''
  \href{http://dx.doi.org/10.1088/1475-7516/2017/07/045}{{\em JCAP} {\bfseries
  1707} no.~07, (2017) 045},
\href{http://arxiv.org/abs/1611.06936}{{\ttfamily arXiv:1611.06936 [gr-qc]}}.

\bibitem{Doeleman:2008qh}
S.~Doeleman {\em et~al.}, ``{Event-horizon-scale structure in the supermassive
  black hole candidate at the Galactic Centre},''
  \href{http://dx.doi.org/10.1038/nature07245}{{\em Nature} {\bfseries 455}
  (2008) 78},
\href{http://arxiv.org/abs/0809.2442}{{\ttfamily arXiv:0809.2442 [astro-ph]}}.

\bibitem{Dokuchaev:2018kzk}
V.~I. Dokuchaev and N.~O. Nazarova, ``{Event horizon image within black hole
  shadow},''
\href{http://arxiv.org/abs/1804.08030}{{\ttfamily arXiv:1804.08030
  [astro-ph.HE]}}.

\bibitem{Doeleman:2012zc}
S.~S. Doeleman {\em et~al.}, ``{Jet Launching Structure Resolved Near the
  Supermassive Black Hole in M87},''
  \href{http://dx.doi.org/10.1126/science.1224768}{{\em Science} {\bfseries
  338} (2012) 355},
\href{http://arxiv.org/abs/1210.6132}{{\ttfamily arXiv:1210.6132
  [astro-ph.HE]}}.

\bibitem{Guo:2018kis}
M.~Guo, N.~A. Obers, and H.~Yan, ``{Observational signatures of near-extremal
  Kerr-like black holes in a modified gravity theory at the Event Horizon
  Telescope},'' \href{http://dx.doi.org/10.1103/PhysRevD.98.084063}{{\em Phys.
  Rev.} {\bfseries D98} no.~8, (2018) 084063},
\href{http://arxiv.org/abs/1806.05249}{{\ttfamily arXiv:1806.05249 [gr-qc]}}.

\bibitem{Claudel:2000yi}
C.-M. Claudel, K.~S. Virbhadra, and G.~F.~R. Ellis, ``{The Geometry of photon
  surfaces},'' \href{http://dx.doi.org/10.1063/1.1308507}{{\em J. Math. Phys.}
  {\bfseries 42} (2001) 818--838},
\href{http://arxiv.org/abs/gr-qc/0005050}{{\ttfamily arXiv:gr-qc/0005050
  [gr-qc]}}.

\bibitem{Hod:2012ax}
S.~Hod, ``{Spherical null geodesics of rotating Kerr black holes},''
  \href{http://dx.doi.org/10.1016/j.physletb.2012.12.047}{{\em Phys. Lett.}
  {\bfseries B718} (2013) 1552--1556},
\href{http://arxiv.org/abs/1210.2486}{{\ttfamily arXiv:1210.2486 [gr-qc]}}.

\bibitem{Khoo:2016xqv}
F.~S. Khoo and Y.~C. Ong, ``{Lux in obscuro: Photon Orbits of Extremal Black
  Holes Revisited},'' \href{http://dx.doi.org/10.1088/0264-9381/33/23/235002,
  10.1088/1361-6382/aa8706}{{\em Class. Quant. Grav.} {\bfseries 33} no.~23,
  (2016) 235002}, \href{http://arxiv.org/abs/1605.05774}{{\ttfamily
  arXiv:1605.05774 [gr-qc]}}.
[Erratum: Class. Quant. Grav.34,no.21,219501(2017)].

\bibitem{Baldiotti:2014pca}
M.~C. Baldiotti, W.~S. Elias, C.~Molina, and T.~S. Pereira, ``{Thermodynamics
  of quantum photon spheres},''
  \href{http://dx.doi.org/10.1103/PhysRevD.90.104025}{{\em Phys. Rev.}
  {\bfseries D90} no.~10, (2014) 104025},
\href{http://arxiv.org/abs/1410.1894}{{\ttfamily arXiv:1410.1894 [gr-qc]}}.

\bibitem{Decanini:2010fz}
Y.~Decanini, A.~Folacci, and B.~Raffaelli, ``{Unstable circular null geodesics
  of static spherically symmetric black holes, Regge poles and quasinormal
  frequencies},'' \href{http://dx.doi.org/10.1103/PhysRevD.81.104039}{{\em
  Phys. Rev.} {\bfseries D81} (2010) 104039},
\href{http://arxiv.org/abs/1002.0121}{{\ttfamily arXiv:1002.0121 [gr-qc]}}.

\bibitem{Shoom:2017ril}
A.~A. Shoom, ``{Metamorphoses of a photon sphere},''
  \href{http://dx.doi.org/10.1103/PhysRevD.96.084056}{{\em Phys. Rev.}
  {\bfseries D96} no.~8, (2017) 084056},
\href{http://arxiv.org/abs/1708.00019}{{\ttfamily arXiv:1708.00019 [gr-qc]}}.

\bibitem{Cederbaum:2015fra}
C.~Cederbaum and G.~J. Galloway, ``{Uniqueness of photon spheres in
  electro-vacuum spacetimes},''
  \href{http://dx.doi.org/10.1088/0264-9381/33/7/075006}{{\em Class. Quant.
  Grav.} {\bfseries 33} (2016) 075006},
\href{http://arxiv.org/abs/1508.00355}{{\ttfamily arXiv:1508.00355 [math.DG]}}.

\bibitem{Johannsen:2015qca}
T.~Johannsen, ``{Photon Rings around Kerr and Kerr-like Black Holes},''
  \href{http://dx.doi.org/10.1088/0004-637X/777/2/170}{{\em Astrophys. J.}
  {\bfseries 777} (2013) 170},
\href{http://arxiv.org/abs/1501.02814}{{\ttfamily arXiv:1501.02814
  [astro-ph.HE]}}.

\bibitem{Teo2003}
E.~Teo, ``Spherical photon orbits around a kerr black hole,''
  \href{http://dx.doi.org/10.1023/A:1026286607562}{{\em General Relativity and
  Gravitation} {\bfseries 35} no.~11, (Nov, 2003) 1909--1926}.
  \url{https://doi.org/10.1023/A:1026286607562}.

\bibitem{Gallo:2015bda}
E.~Gallo and J.~R. Villanueva, ``{Photon spheres in Einstein and
  Einstein-Gauss-Bonnet theories and circular null geodesics in
  axially-symmetric spacetimes},''
  \href{http://dx.doi.org/10.1103/PhysRevD.92.064048}{{\em Phys. Rev.}
  {\bfseries D92} no.~6, (2015) 064048},
\href{http://arxiv.org/abs/1509.07379}{{\ttfamily arXiv:1509.07379 [gr-qc]}}.

\bibitem{Bhattacharya:2016naa}
S.~Bhattacharya and S.~Chakraborty, ``{Constraining some Horndeski gravity
  theories},'' \href{http://dx.doi.org/10.1103/PhysRevD.95.044037}{{\em Phys.
  Rev.} {\bfseries D95} no.~4, (2017) 044037},
\href{http://arxiv.org/abs/1607.03693}{{\ttfamily arXiv:1607.03693 [gr-qc]}}.

\bibitem{Chakraborty:2012sd}
S.~Chakraborty and S.~SenGupta, ``{Solar system constraints on alternative
  gravity theories},'' \href{http://dx.doi.org/10.1103/PhysRevD.89.026003}{{\em
  Phys. Rev.} {\bfseries D89} no.~2, (2014) 026003},
\href{http://arxiv.org/abs/1208.1433}{{\ttfamily arXiv:1208.1433 [gr-qc]}}.

\bibitem{Rahman:2018oso}
M.~Rahman, S.~Chakraborty, S.~SenGupta, and A.~A. Sen, ``{Fate of Strong Cosmic
  Censorship Conjecture in Presence of Higher Spacetime Dimensions},''
\href{http://arxiv.org/abs/1811.08538}{{\ttfamily arXiv:1811.08538 [gr-qc]}}.

\bibitem{Mukherjee:2018dmm}
S.~Mukherjee, S.~Chakraborty, and N.~Dadhich, ``{On some novel features of the
  Kerr–Newman-NUT spacetime},''
  \href{http://dx.doi.org/10.1140/epjc/s10052-019-6662-2}{{\em Eur. Phys. J.}
  {\bfseries C79} no.~2, (2019) 161},
\href{http://arxiv.org/abs/1807.02216}{{\ttfamily arXiv:1807.02216 [gr-qc]}}.

\bibitem{Abdujabbarov:2015xqa}
A.~A. Abdujabbarov, L.~Rezzolla, and B.~J. Ahmedov, ``{A coordinate-independent
  characterization of a black hole shadow},''
  \href{http://dx.doi.org/10.1093/mnras/stv2079}{{\em Mon. Not. Roy. Astron.
  Soc.} {\bfseries 454} no.~3, (2015) 2423--2435},
\href{http://arxiv.org/abs/1503.09054}{{\ttfamily arXiv:1503.09054 [gr-qc]}}.

\bibitem{Younsi:2016azx}
Z.~Younsi, A.~Zhidenko, L.~Rezzolla, R.~Konoplya, and Y.~Mizuno, ``{New method
  for shadow calculations: Application to parametrized axisymmetric black
  holes},'' \href{http://dx.doi.org/10.1103/PhysRevD.94.084025}{{\em Phys.
  Rev.} {\bfseries D94} no.~8, (2016) 084025},
\href{http://arxiv.org/abs/1607.05767}{{\ttfamily arXiv:1607.05767 [gr-qc]}}.

\bibitem{Goddi:2017pfy}
C.~Goddi {\em et~al.}, ``{BlackHoleCam: Fundamental physics of the galactic
  center},'' \href{http://dx.doi.org/10.1142/S0218271817300014,
  10.1142/9789813226609_0046}{{\em Int. J. Mod. Phys.} {\bfseries D26} no.~02,
  (2016) 1730001}, \href{http://arxiv.org/abs/1606.08879}{{\ttfamily
  arXiv:1606.08879 [astro-ph.HE]}}.
[1,863(2017)].

\bibitem{Lovelock:1971yv}
D.~Lovelock, ``{The Einstein tensor and its generalizations},''
\href{http://dx.doi.org/10.1063/1.1665613}{{\em J. Math. Phys.} {\bfseries 12}
  (1971) 498--501}.

\bibitem{Zwiebach:1985uq}
B.~Zwiebach, ``{Curvature Squared Terms and String Theories},''
\href{http://dx.doi.org/10.1016/0370-2693(85)91616-8}{{\em Phys. Lett.}
  {\bfseries 156B} (1985) 315--317}.

\bibitem{Boulware:1985wk}
D.~G. Boulware and S.~Deser, ``{String Generated Gravity Models},''
\href{http://dx.doi.org/10.1103/PhysRevLett.55.2656}{{\em Phys. Rev. Lett.}
  {\bfseries 55} (1985) 2656}.

\bibitem{Zumino:1985dp}
B.~Zumino, ``{Gravity Theories in More Than Four-Dimensions},''
\href{http://dx.doi.org/10.1016/0370-1573(86)90076-1}{{\em Phys. Rept.}
  {\bfseries 137} (1986) 109}.

\bibitem{Padmanabhan:2013xyr}
T.~Padmanabhan and D.~Kothawala, ``{Lanczos-Lovelock models of gravity},''
  \href{http://dx.doi.org/10.1016/j.physrep.2013.05.007}{{\em Phys. Rept.}
  {\bfseries 531} (2013) 115--171},
\href{http://arxiv.org/abs/1302.2151}{{\ttfamily arXiv:1302.2151 [gr-qc]}}.

\bibitem{Aragone:1987jm}
C.~Aragone, ``{STRINGY CHARACTERISTICS OF EFFECTIVE GRAVITY},'' in {\em {SILARG
  6: 6th Latin American Symposium on Relativity and Gravitation Rio de Janeiro,
  Brazil, July 13-18, 1987}}, pp.~60--69.
\newblock
1987.
\newblock

\bibitem{ChoquetBruhat:1988dw}
Y.~Choquet-Bruhat, ``{The Cauchy Problem for Stringy Gravity},''
\href{http://dx.doi.org/10.1063/1.527841}{{\em J. Math. Phys.} {\bfseries 29}
  (1988) 1891--1895}.

\bibitem{Izumi:2014loa}
K.~Izumi, ``{Causal Structures in Gauss-Bonnet gravity},''
  \href{http://dx.doi.org/10.1103/PhysRevD.90.044037}{{\em Phys. Rev.}
  {\bfseries D90} no.~4, (2014) 044037},
\href{http://arxiv.org/abs/1406.0677}{{\ttfamily arXiv:1406.0677 [gr-qc]}}.

\bibitem{Reall:2014pwa}
H.~Reall, N.~Tanahashi, and B.~Way, ``{Causality and Hyperbolicity of Lovelock
  Theories},'' \href{http://dx.doi.org/10.1088/0264-9381/31/20/205005}{{\em
  Class. Quant. Grav.} {\bfseries 31} (2014) 205005},
\href{http://arxiv.org/abs/1406.3379}{{\ttfamily arXiv:1406.3379 [hep-th]}}.

\bibitem{Papallo:2015rna}
G.~Papallo and H.~S. Reall, ``{Graviton time delay and a speed limit for small
  black holes in Einstein-Gauss-Bonnet theory},''
  \href{http://dx.doi.org/10.1007/JHEP11(2015)109}{{\em JHEP} {\bfseries 11}
  (2015) 109},
\href{http://arxiv.org/abs/1508.05303}{{\ttfamily arXiv:1508.05303 [gr-qc]}}.

\bibitem{Reall:2014sla}
H.~S. Reall, N.~Tanahashi, and B.~Way, ``{Shock Formation in Lovelock
  Theories},'' \href{http://dx.doi.org/10.1103/PhysRevD.91.044013}{{\em Phys.
  Rev.} {\bfseries D91} no.~4, (2015) 044013},
\href{http://arxiv.org/abs/1409.3874}{{\ttfamily arXiv:1409.3874 [hep-th]}}.

\bibitem{Brustein:2017iet}
R.~Brustein and Y.~Sherf, ``{Causality Violations in Lovelock Theories},''
  \href{http://dx.doi.org/10.1103/PhysRevD.97.084019}{{\em Phys. Rev.}
  {\bfseries D97} no.~8, (2018) 084019},
\href{http://arxiv.org/abs/1711.05140}{{\ttfamily arXiv:1711.05140 [hep-th]}}.

\bibitem{Andrade:2016yzc}
T.~Andrade, E.~Caceres, and C.~Keeler, ``{Boundary causality versus
  hyperbolicity for spherical black holes in Gauss?Bonnet gravity},''
  \href{http://dx.doi.org/10.1088/1361-6382/aa7101}{{\em Class. Quant. Grav.}
  {\bfseries 34} no.~13, (2017) 135003},
\href{http://arxiv.org/abs/1610.06078}{{\ttfamily arXiv:1610.06078 [hep-th]}}.

\bibitem{poisson_2004}
E.~Poisson, \href{http://dx.doi.org/10.1017/CBO9780511606601}{{\em A
  Relativist's Toolkit: The Mathematics of Black-Hole Mechanics}}.
\newblock Cambridge University Press, 2004.

\bibitem{Nielsen:2010gm}
A.~B. Nielsen, ``{The Spatial relation between the event horizon and trapping
  horizon},'' \href{http://dx.doi.org/10.1088/0264-9381/27/24/245016}{{\em
  Class. Quant. Grav.} {\bfseries 27} (2010) 245016},
\href{http://arxiv.org/abs/1006.2448}{{\ttfamily arXiv:1006.2448 [gr-qc]}}.

\bibitem{PhysRev.83.10}
P.~C. Vaidya, ``Nonstatic solutions of einstein's field equations for spheres
  of fluids radiating energy,''
  \href{http://dx.doi.org/10.1103/PhysRev.83.10}{{\em Phys. Rev.} {\bfseries
  83} (Jul, 1951) 10--17}.
  \url{https://link.aps.org/doi/10.1103/PhysRev.83.10}.

\bibitem{Caceres:2013dma}
E.~Caceres, A.~Kundu, J.~F. Pedraza, and W.~Tangarife, ``{Strong Subadditivity,
  Null Energy Condition and Charged Black Holes},''
  \href{http://dx.doi.org/10.1007/JHEP01(2014)084}{{\em JHEP} {\bfseries 01}
  (2014) 084},
\href{http://arxiv.org/abs/1304.3398}{{\ttfamily arXiv:1304.3398 [hep-th]}}.

\bibitem{Stuchlik:1999qk}
Z.~Stuchlik and S.~Hledik, ``{Some properties of the Schwarzschild-de Sitter
  and Schwarzschild - anti-de Sitter space-times},''
\href{http://dx.doi.org/10.1103/PhysRevD.60.044006}{{\em Phys. Rev.} {\bfseries
  D60} (1999) 044006}.

\bibitem{Carter:1968rr}
B.~Carter, ``{Global structure of the Kerr family of gravitational fields},''
\href{http://dx.doi.org/10.1103/PhysRev.174.1559}{{\em Phys. Rev.} {\bfseries
  174} (1968) 1559--1571}.

\bibitem{PhysRevD.1.3220}
M.~Murenbeeld and J.~R. Trollope, ``Slowly rotating radiating sphere and a
  kerr-vaidya metric,'' \href{http://dx.doi.org/10.1103/PhysRevD.1.3220}{{\em
  Phys. Rev. D} {\bfseries 1} (Jun, 1970) 3220--3223}.
  \url{https://link.aps.org/doi/10.1103/PhysRevD.1.3220}.

\bibitem{Bardeen:1972fi}
J.~M. Bardeen, W.~H. Press, and S.~A. Teukolsky, ``{Rotating black holes:
  Locally nonrotating frames, energy extraction, and scalar synchrotron
  radiation},''
\href{http://dx.doi.org/10.1086/151796}{{\em Astrophys. J.} {\bfseries 178}
  (1972) 347}.

\bibitem{Misner1973Gravitation}
C.~W. Misner, K.~S. Thorne, and J.~A. Wheeler, {\em Gravitation}.
\newblock Physics Series. W. H. Freeman, San Francisco, first edition~ed.,
  Sept., 1973.
\newblock \url{http://www.worldcat.org/isbn/0716703440}.

\bibitem{Wiltshire:1988uq}
D.~L. Wiltshire, ``{Black Holes in String Generated Gravity Models},''
\href{http://dx.doi.org/10.1103/PhysRevD.38.2445}{{\em Phys. Rev.} {\bfseries
  D38} (1988) 2445}.

\bibitem{PhysRevD.38.2434}
R.~C. Myers and J.~Z. Simon, ``Black-hole thermodynamics in lovelock gravity,''
  \href{http://dx.doi.org/10.1103/PhysRevD.38.2434}{{\em Phys. Rev. D}
  {\bfseries 38} (Oct, 1988) 2434--2444}.
  \url{https://link.aps.org/doi/10.1103/PhysRevD.38.2434}.

\bibitem{Wheeler:1985nh}
J.~T. Wheeler, ``{Symmetric Solutions to the Gauss-Bonnet Extended Einstein
  Equations},''
\href{http://dx.doi.org/10.1016/0550-3213(86)90268-3}{{\em Nucl. Phys.}
  {\bfseries B268} (1986) 737--746}.

\end{thebibliography}\endgroup

\bibliographystyle{./utphys1}
\end{document}